\documentclass[twocolumn]{aastex63}
\usepackage{hyperref}

\usepackage{float}
\usepackage{verbatim}
\usepackage{natbib}
\usepackage{amsmath}

\usepackage{xcolor}

\newcommand{\kms}{km~s$^{-1}$}
\begin{document}

\title{Mysterious Dust-emitting Object Orbiting TIC 400799224}

\correspondingauthor{Brian P. Powell}
\email{brian.p.powell@nasa.gov}

\author[0000-0003-0501-2636]{Brian P. Powell}
\affiliation{NASA Goddard Space Flight Center, 8800 Greenbelt Road, Greenbelt, MD 20771, USA}
%\email{brian.p.powell@nasa.gov}
%
\author[0000-0001-9786-1031]{Veselin~B.~Kostov}
\affiliation{NASA Goddard Space Flight Center, 8800 Greenbelt Road, Greenbelt, MD 20771, USA}
\affiliation{SETI Institute, 189 Bernardo Ave, Suite 200, Mountain View, CA 94043, USA}
\affiliation{GSFC Sellers Exoplanet Environments Collaboration}
%\email{veselin.b.kostov@nasa.gov}
%
%
\author[0000-0003-3182-5569]{Saul A. Rappaport}
\affiliation{Department of Physics, Kavli Institute for Astrophysics and Space Research, M.I.T., Cambridge, MA 02139, USA}
%\email{sar@mit.edu}
%
\author[0000-0002-2084-0782]{Andrei Tokovinin}
\affiliation{Cerro Tololo Inter-American Observatory | NSF's NOIRLab, Casilla 603, La Serena, Chile}
%\email{atokovinin@ctio.noao.edu}
%
\author[0000-0002-1836-3120]{Avi Shporer}
\affil{Department of Physics and Kavli Institute for Astrophysics and Space Research, Massachusetts Institute of Technology, Cambridge, MA 02139, USA}
%\email{shporeravi@gmail.com}
%
\author[0000-0001-6588-9574]{Karen A.\ Collins}
\affiliation{Center for Astrophysics $\vert$ Harvard \& Smithsonian, 60 Garden Street, Cambridge, MA 02138, USA}
%\email{karenacollins@outlook.com}
%
\author[0000-0002-6339-6706]{Hank Corbett}
\affil{University of North Carolina at Chapel Hill, 120 E. Cameron Ave., Chapel Hill, NC 27514, USA}
%\email{htcorbett4@gmail.com}
%
\author[0000-0002-8806-496X]{Tam\'as Borkovits}
\affiliation{Baja Astronomical Observatory of University of Szeged, H-6500 Baja, Szegedi út, Kt. 766, Hungary}
\affiliation{Konkoly Observatory, Research Centre for Astronomy and Earth Sciences, H-1121 Budapest, Konkoly Thege Miklós út 15-17, Hungary}
\affiliation{ELTE Gothard Astrophysical Observatory, H-9700 Szombathely, Szent Imre h. u. 112, Hungary}
%\email{borko@electra.bajaobs.hu}
%
\author[0000-0002-4080-1342]{Bruce L. Gary}
\affiliation{Hereford Arizona Observatory, Hereford, AZ 85615, USA}
%\email{bgary1@cis-broadband.com}
%
\author[0000-0002-6246-2310]{Eugene Chiang}
\affiliation{Department of Astronomy, University of California, Berkeley, Berkeley, CA 94720}
\affiliation{Department of Earth and Planetary Science, University of California, Berkeley, Berkeley, CA 94720}
%\email{echiang@astro.berkeley.edu}
%
\author[0000-0001-8812-0565]{Joseph E. Rodriguez}
\affiliation{Department of Physics and Astronomy, Michigan State University, East Lansing, MI 48824, USA}
%\email{jrod@msu.edu}
%
\author[0000-0001-9380-6457]{Nicholas M. Law}
\affil{University of North Carolina at Chapel Hill, 120 E. Cameron Ave., Chapel Hill, NC 27514, USA}
%\email{nlaw@unc.edu}
%
\author[0000-0001-7139-2724]{Thomas~Barclay}
\affiliation{NASA Goddard Space Flight Center, 8800 Greenbelt Road, Greenbelt, MD 20771, USA}
\affiliation{University of Maryland, Baltimore County, 1000 Hilltop Circle,
Baltimore, MD 21250, USA}
%\email{thomas.barclay@nasa.gov}
%
\author[0000-0002-5665-1879]{Robert Gagliano}
\affiliation{Amateur Astronomer, Glendale, AZ 85308}
%\email{astrowebdoc@gmail.com}
%
\author[0000-0001-7246-5438]{Andrew Vanderburg}
\affiliation{Department of Astronomy, University of Wisconsin-Madison, Madison, WI 53706, USA}
%\email{andrew.m.vanderburg@gmail.com}
%
\author[0000-0001-8472-2219]{Greg Olmschenk}
\affiliation{NASA Goddard Space Flight Center, 8800 Greenbelt Road, Greenbelt, MD 20771, USA}
\affiliation{Universities Space Research Association, 7178 Columbia Gateway Drive, Columbia, MD 21046}
%\email{gregory.olmschenk@nasa.gov}
%
\author[0000-0002-0493-1342]{Ethan Kruse}
\affiliation{NASA Goddard Space Flight Center, 8800 Greenbelt Road, Greenbelt, MD 20771, USA}
\affiliation{Universities Space Research Association, 7178 Columbia Gateway Drive, Columbia, MD 21046}
%\email{ethan.kruse@nasa.gov}
%
\author[0000-0001-5347-7062]{Joshua E. Schlieder}
\affiliation{NASA Goddard Space Flight Center, 8800 Greenbelt Road, Greenbelt, MD 20771, USA}
%\email{joshua.e.schlieder@nasa.gov}
%
\author[0000-0002-1906-1167]{Alan Vasquez Soto}
\affil{University of North Carolina at Chapel Hill, 120 E. Cameron Ave., Chapel Hill, NC 27514, USA}
%\email{vasqua@live.unc.edu}
%
\author[0000-0001-6559-5189]{Erin Goeke}
\affil{University of North Carolina at Chapel Hill, 120 E. Cameron Ave., Chapel Hill, NC 27514, USA}
%\email{eggoeke@live.unc.edu}
%
\author[0000-0003-3988-3245]{Thomas L. Jacobs}
\affiliation{Amateur Astronomer, 12812 SE 69th Place, Bellevue, WA 98006}
%\email{tomjacobs128@gmail.com}
%
\author[0000-0002-2607-138X]{Martti~H.~Kristiansen}
\affil{Brorfelde Observatory, Observator Gyldenkernes Vej 7, DK-4340 T\o{}ll\o{}se, Denmark}
\affil{DTU Space, National Space Institute, Technical University of Denmark, Elektrovej 327, DK-2800 Lyngby, Denmark}
%\email{martti@outinto.space}
%
\author[0000-0002-8527-2114]{Daryll M. LaCourse}
\affiliation{Amateur Astronomer, 7507 52nd Place NE Marysville, WA 98270}
%\email{nhawkb@gmail.com}
%
\author{Mark Omohundro}
\affiliation{Citizen Scientist, c/o Zooniverse, Department of Physics, University of Oxford, Denys Wilkinson Building, Keble Road, Oxford, OX13RH, UK}
%\email{fly.fishing.heaven.pa@gmail.com}
%
\author[0000-0002-1637-2189]{Hans M. Schwengeler}
\affiliation{Citizen Scientist, Planet Hunter, Bottmingen, Switzerland}
%\email{hans.schwengeler@intergga.ch}
%
\author[0000-0002-0654-4442]{Ivan A. Terentev}
\affiliation{Citizen Scientist, Planet Hunter, Petrozavodsk, Russia}
%\email{iterentie@mail.ru}
%
\author[0000-0002-5034-0949]{Allan R. Schmitt}
\affiliation{Citizen Scientist, 616 W. 53rd. St., Apt. 101, Minneapolis, MN 55419, USA}
%\email{aschmitt@comcast.net}
%
%\submitjournal{AAS Journals}
%
\received{2 August 2021}
\revised{13 September 2021}
%\revised{1 October 2021}
\accepted{1 October 2021}

%\section{Abstract}
\begin{abstract}
We report the discovery of a unique object of uncertain nature -- but quite possibly a disintegrating asteroid or minor planet -- orbiting one star of the widely separated binary TIC 400799224. We initially identified the system in data from {\em TESS} Sector 10 via an abnormally-shaped fading event in the light curve (hereafter `dips'). Follow-up speckle imaging determined that TIC 400799224 is actually two stars of similar brightness at $0\farcs62$ separation, forming a likely bound binary with projected separation of $\sim$300 au.  We cannot yet determine which star in the binary is host to the dips in flux.  ASAS-SN and Evryscope archival data show that there is a strong periodicity of the dips at $\sim$19.77 days, leading us to believe that an occulting object is orbiting the host star, though the duration, depth, and shape of the dips vary substantially. Statistical analysis of the ASAS-SN data shows that the dips only occur sporadically at a detectable threshold in approximately one out of every three to five transits, lending credence to the possibility that the occulter is a sporadically-emitted dust cloud.  The cloud is also fairly optically thick, blocking up to 37\% or 75\% of the light from the host star, depending on the true host. Further observations may allow for greater detail to be gleaned as to the origin and composition of the occulter, as well as to a determination of which of the two stars comprising TIC 400799224 is the true host star of the dips.
\end{abstract}


\section{Introduction}\label{sec:intro}

The Full Frame Images (FFI) from The Transiting Exoplanet Survey Satellite ({\em TESS}) \citep{Ricker14} have an instantaneous field of view of $96 \times 24$ degrees (approximately 5\% of the sky), with a cadence as short as 10 min. These data have presented us with an opportunity to thoroughly search the visible sky for variability. 
FFIs have proven valuable for detecting new transient sources \citep{Holoien2019,Fausnaugh2021,Smith2021}, unusual variables \citep{Tajiri2020,Sahoo2020,Payne2021}, and exoplanets \citep{Rodriguez2021,Olmschenk2021,Ikwut2021}.

{\em Kepler} and {\em TESS} have provided us with a truly novel view of the Galaxy and their observations have resulted in the discovery of previously undetected phenomena such as tidally induced stellar pulsations \citep{koi54,heartbeatstars}; shock breakout from supernovae \citep{2016ApJ...820...23G}; circumbinary planets \citep{2020AJ....159..253K,2021arXiv210508614K}; disintegrating planets \citep{2012ApJ...752....1R, 2015ApJ...812..112S}; self-lensing binaries \citep{2014Sci...344..275K}; triply eclipsing triple star systems \citep{2019MNRAS.483.1934B,2020MNRAS.493.5005B,2020MNRAS.498.6034M,2020MNRAS.496.4624B}; disk occultations \citep{2018ApJ...854..109Z,2019MNRAS.485.2681R,2019MNRAS.488.4149C}; the random transiter \citep{2019MNRAS.488.2455R}; and Boyajian's Star \citep{2016MNRAS.457.3988B}.  However, after the completion of the {\em Kepler} mission and three years of the {\em TESS} mission data, discoveries of any truly new behavior in stellar light curves is understandably becoming more infrequent.

Searching for known characteristics in light curves with algorithmic approaches, including machine learning, has been fruitful.  For example, we have identified several hundred thousand eclipsing binaries in {\em TESS} light curves created from the FFIs with the \texttt{eleanor} pipeline \citep{eleanor} using a neural network to search for the feature of the eclipse (Kruse, et al., 2021 in prep). This effort has yielded the discovery of a confirmed sextuple star system \citep{Powell2021}; a confirmed quadruple star system \citep{2021arXiv210512586K}; many additional quadruple star system candidates; numerous triple star system candidates; and candidates for higher-order systems that are currently under investigation.  In an examination of the outputs of an early prototype of an eclipse-finding neural network we identified a particularly unusual source -- TIC 400799224 -- that demonstrates a rapid drop in flux and several sharp variations that could each be weakly interpreted as an eclipse. So, while not intending to find an abnormal or unique shape to a light curve, we fortuitously identified TIC 400799224 through the application of machine learning methods.

As we discuss in detail in Sections \ref{sec:asas-sn} and \ref{sec:Evryscope}, we have found that this dip in flux identified in TIC 400799224 is part of a periodic set of dips, which we postulate is due to an orbiting body episodically emitting dust.  This is in some ways reminiscent of other known orbiting, dust-emitting objects,  such as those hosted by KIC 12557548 \citep{2012ApJ...752....1R}, K2-22 \citep{2015ApJ...812..112S}, WD 1145+017 \citep{2015Natur.526..546V}, ZTF J0328-1219 \citep{2021arXiv210602659V}, and ZTF J0139+5245 \citep{2020ApJ...897..171V}.  Some of the properties of these dusty occulters and the observed periods are listed in Table \ref{tbl:dusty}.

\begin{table*}
\centering
\caption{Properties of Orbiting Bodies with Periodic Dusty Occultations.}
\begin{tabular}{lccccccccccc}
\hline
\hline
  Source  &  $P_{\rm orb}$ & $M_{\rm host}$ & $R_{\rm host}$ & $T_{\rm eff}^a$ & $L_{\rm host}$ & $T_{\rm eq}^b$ & $d/R_{\rm host}$ & $V_{orb}$ & $\beta^c$ & Type$^d$ & Refs  \\
    -     &    days  &  M$_\odot$  &   R$_\odot$ &  K & L$_\odot$ & K & - & km s$^{-1}$ & - & - & -\\  
\hline
    KIC 12557548 & 0.653  & 0.67  & 0.67  & 4500  & 0.17 & 2100 & 4.13 & 215 & 0.19 & MS & 1, 2, 3 \\ % a = 2.77 Rsun
    KOI 2700b  & 0.910  & 0.55   & 0.54  & 4300  & 0.09 & 1350 & 6.00 & 180 & 0.12 & MS & 3, 4 \\  % a = 3.24 Rsun
    K2-22b  & 0.381  & 0.60   & 0.57  & 3830  & 0.063 & 2100 & 3.28 & 250 & 0.07 & MS & 3, 5 \\  % a = 1.87 Rsun
    WD 1145+017  & 0.187 &  0.60  & 0.013 & 15,900 & 0.0094 & 1600 & 90 & 330 & 0.012 & WD & 6 \\  % a = 1.16 Rsun
    ZTF J0328-1219$^e$  & 0.414 &  0.731  & 0.011 & 7630 & 0.0004 & 700 & 192 &  260 & 0.0007 & WD & 7 \\ % a = 2.11 Rsun
    ZTF J0139+5245$^f$  & 107  &  0.52   &  0.014  & 10,500 & 0.0021 & 142 & 5450 & 35 & 0.003 & WD & 8 \\  % a = 76.3 Rsun
    \textbf{TIC 400799224}  & \textbf{19.8} & \textbf{1.5}  & \textbf{2.20} & \textbf{5900} & \textbf{6.3} & \textbf{1525} & \textbf{15} & \textbf{90} & \textbf{$\sim$3} & \textbf{PMS} & \textbf{this work} \\  % a = 33.2 Rsun
\hline
\label{tbl:dusty}
\end{tabular} %Table 1

{Notes. (a) Effective temperature of the host star.  (b) Sub-stellar equilibrium temperature of a body at $P_{\rm orb}$ defined as $T_{\rm eff} \sqrt{d/R_{\rm host}}$, where $d$ is the distance to the orbiting body. (c) Ratio of the radiation pressure force on a dust grain to the force of gravity for a 0.2 $\mu$m dust grain.  (d) WD = white dwarf, MS = main sequence, PMS = pre-MS or post-MS (to be discussed further beginning in Section \ref{sec:sed}). (e) This white dwarf also exhibits a weaker periodicity at 11.2 hours.  (f) The exact underlying period is uncertain by several percent.  References: (1) \citet{2012ApJ...752....1R}; (2) \citet{2013MNRAS.433.2294P}; (3) \citet{2018haex.bookE..15V}; (4) \citet{2014ApJ...784...40R}; (5) \citet{2015ApJ...812..112S}; (6) \citet{2015Natur.526..546V}; (7) \citet{2021arXiv210602659V}; (8) \citet{2020ApJ...897..171V}.}

\end{table*} % Table 1

A planetesimal orbiting KIC 12557548 was the first of the known `disintegrating planets'. In the discovery paper, \citet{2012ApJ...752....1R} identified transits of substantially varying depth with a period of 15.6854 hours and determined that the object was a disintegrating planet emitting a dust cloud.  These authors postulated that the dust, or heavy-element vapor, could be driven off the body at sufficient rates to explain the dusty effluents (see also \citealt{2013MNRAS.433.2294P}). \citet{2012A&A...545L...5B} applied a one-dimensional trailing dust cloud model to the system and concluded that the model explained the system quite well. \citet{2014A&A...561A...3V} improved this model to two dimensions by employing an opaque core and exponential tail, and successfully applied it to many different observed transit shapes.  Multi-wavelength photometry of the transits was analyzed by \citet{2015ApJ...800L..21B}, allowing for improved understanding of the nature of the dust in the tail.  

Another of the known disintegrating planets is K2-22b.  It was originally discovered by \citet{2015ApJ...812..112S}, who provided a robust case for a disintegrating planet and noted variable transit depths in a range from ~0\% to 1.3\% at a period of 9.1457 hours,  quite similar in nature to KIC 12557548b.  \cite{2018AJ....156..227C} observed K2-22b on 45 occasions over the course of seven months, finding varying transit shapes and depths without wavelength dependency. \citet{2019A&A...628A..70R} also observed K2-22b spectrally, finding no signals associated with gas absorption. 

Six periodicities in the white dwarf WD 1145+017 were first identified by \citet{2015Natur.526..546V} and attributed to dust-emitting orbiting bodies.  The periodic dips in flux were $\lesssim 1\%$ deep and the periods were all in the range of 4.5-4.8 hrs. The authors showed that the orbiting bodies had to have masses less than lunar mass in order for the orbits to be stable.  They postulated that the actual transiting events were due to dust given that their durations were much longer than could plausibly be attributed to solid body occulters.  Subsequent ground-based observations showed that the dips in flux (i) could be up to 55\% deep and (ii) the dominant periodicity of 4.5 hours was not strictly periodic (see \citealt{2016MNRAS.458.3904R}; \citealt{gaensicke16}; \citealt{Gary2017}; \citealt{croll17}). \citet{2016ApJ...816L..22X} showed that the dust particles in question had to be larger than a few microns in order to explain the largely colorless dips \citep{2016MNRAS.458.3904R}. The properties of this system are reviewed and summarized by \citet{2018haex.bookE..37V}.  %\ron{I have no idea what the following papers contain, but let's review these later:} \citet{2018AAS...23232001M}, \citet{2019MNRAS.482..999K}, \citet{2020ApJ...888...47F})

ZTF J0139+5245, discovered by \citet{2020ApJ...897..171V}, shows a much different manifestation of the effects of dust.  The authors attribute the up to $\sim$40\% deep and $\sim$25 day duration transits to a stream of dust and planetary debris, possibly created through tidal disruption by the white dwarf.  Means of analysis of the origin of the debris stream are suggested by \citet{2020MNRAS.492.5291V} and \citet{2020MNRAS.493..698M}.  The dips are erratically recurrent with a period of $\sim$107 days.  Most recently, \citet{2021arXiv210602659V} discovered another white dwarf, ZTF J0328-1219, hosting ``debris clumps'', which exhibit periodicities of 9.937 and 11.2 hours.  

TIC 400799224 is analogous to these systems in that (i) there appears to be an orbiting body that shows signs of disintegration, (ii) the resultant transits are variable in depth, shape, and duration, and (iii) the transits may or may not occur at the expected time, only presenting optically measurable evidence of an occultation in one out of every three to five transits. As will be described, its detailed properties differ in significant ways from the other objects listed in Table \ref{tbl:dusty} and, therefore, TIC 400799224 may be in a category of its own.  

The structure of this paper is as follows.  In section \ref{sec:photo}, we present the photometric observations from {\em TESS}, ASAS-SN, and Evryscope.  In section \ref{sec:follow-up}, we describe follow up observations including photometric data from Las Cumbres Observatory (LCO),  spectra from CHIRON, and speckle imaging from SOAR (which resolves TIC 400799224 into two close stellar images).  In section \ref{sec:sed}, we analyze the available spectral energy distribution (SED) data for this object to extract information about the masses of the two stars that we find comprise the image of TIC 400799224.  Lastly, in section \ref{sec:occulter}, we examine the nature of the occulter.

\section{Photometric data}
\label{sec:photo}

\begin{figure}
    \centering
    \includegraphics[width=1.0\linewidth]{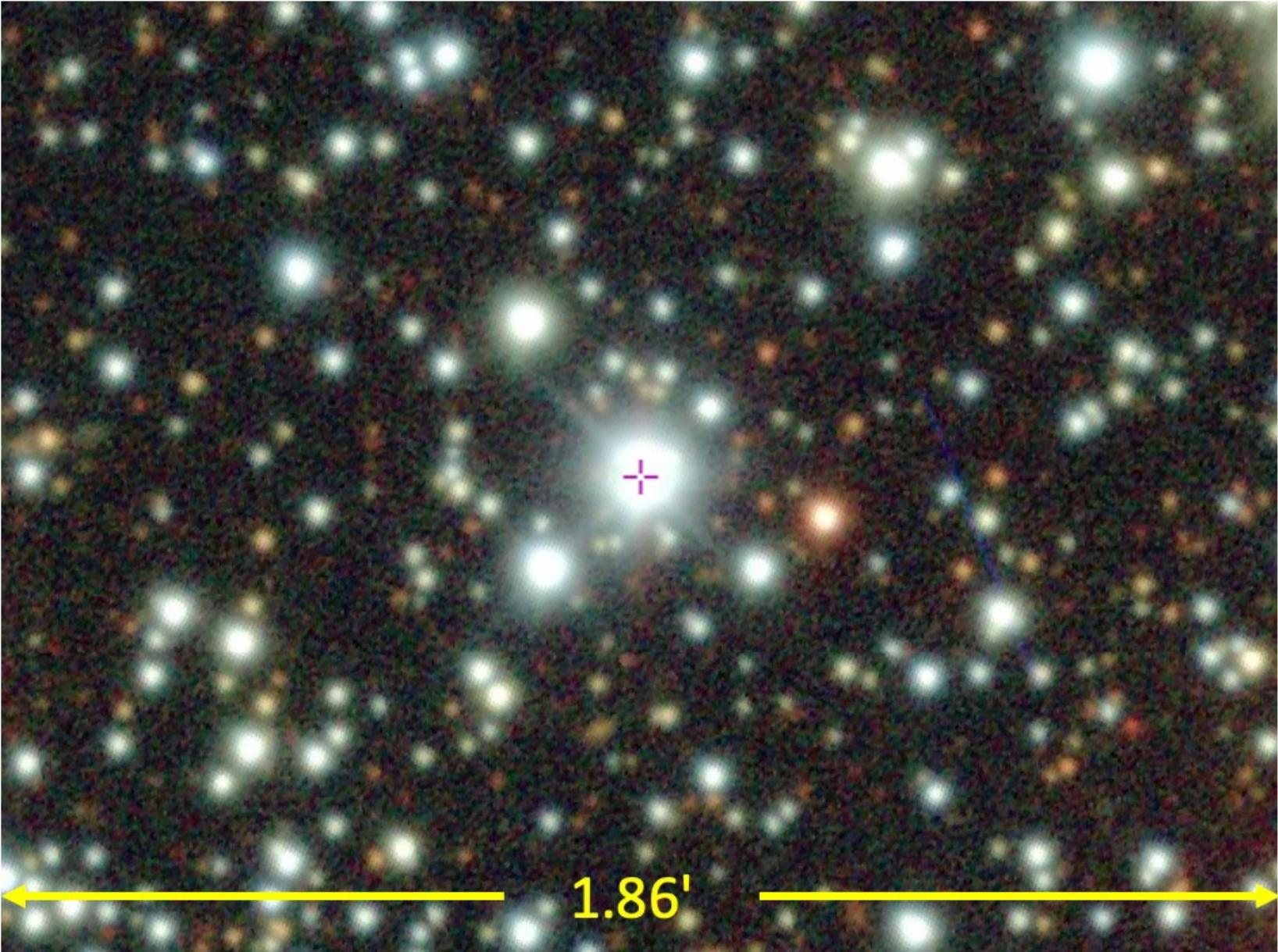}
    \caption{Color image of the field from the DECam Plane Survey (DECaPS; \citealt{schlafly18}), a five-band optical and near-infrared survey of the southern Galactic plane with the Dark Energy Camera at the Cerro Tololo Inter-American Observatory in La Serena, Chile. TIC 400799224 is identified in the crosshairs.}
   \label{fig:field}
\end{figure}

The initial identification of TIC 400799224 as having an unusual dipping feature was made with {\em TESS} data from Sector 10.  Archival data were obtained from both ASAS-SN \citep{2014ApJ...788...48S,2017PASP..129j4502K} and Evryscope \citep{law15}. The known stellar parameters for this system are provided in Table \ref{tab:parameters} and a DECaPS \citep{schlafly18} image of the field containing TIC 400799224 is shown in Figure \ref{fig:field}.

\begin{deluxetable*}{l r r r }
%\tabletypesize{\scriptsize}
\tabletypesize{\small}
\tablecaption{Stellar parameters of TIC 400799224.\label{tab:parameters}}
\tablewidth{0pt}
\tablehead{
\colhead{Parameter} & \colhead{Value} & \colhead{Error} &\colhead{Source}
}
\startdata
%\multicolumn{4}{c}{\em Identifying Information} \\
\multicolumn{4}{l}{\bf Identifying Information} \\
\hline
TIC ID & 400799224 & & TIC \\
{\it Gaia} ID & 5238414793089292160 & & {\it Gaia} EDR3 \\
2MASS ID & 	11095818-6645149 & & TIC \\
ALLWISE ID & J110958.16-664514.8 & & TIC \\
RA  (hh:mm:ss) & 11:09:58.186 &   & TIC \\
Dec (dd:mm:ss) & -66:45:14.91 &   & TIC\\
Distance (pc) & 725 & 140 & {\it Gaia} EDR3 \\
PM (mas/yr) & 11.949 & 0.342 & {\it Gaia} EDR3 \\
PMRA (mas/yr) & -11.895 & 0.320 & {\it Gaia} EDR3 \\
PMDEC (mas/yr) & 1.129 & 0.280 & {\it Gaia} EDR3 \\
\hline
\\
\multicolumn{4}{l}{\bf Photometric Properties} \\
\hline
$T$ (mag) & 11.743 & 0.044 & TIC \\
$B$ (mag) & 13.46 & 0.099 & TIC \\
$V$ (mag) & 12.625 & 0.069 & TIC \\
$Gaia$ (mag) & 	12.710 & 0.003  & {\it Gaia} EDR3 \\
$J$ (mag) & 10.995 & 0.024 & TIC \\
$H$ (mag) & 10.667 & 0.024 & TIC \\
$K$ (mag) & 10.569 & 0.021 & TIC \\
$W1$ (mag) & 10.4 & 0.022 & TIC \\
$W2$ (mag) & 10.387 & 0.02 & TIC \\
$W3$ (mag) & 8.859 & 0.021 & TIC \\
$W4$ (mag) & 8.286 & 0.159 & TIC \\
\enddata
{Notes. -- }
\end{deluxetable*} % Table 2

 % Fig. 1

\subsection{{\em TESS}}
\label{sec:tess}

\begin{figure*}
    \centering
    \includegraphics[width=0.8\linewidth]{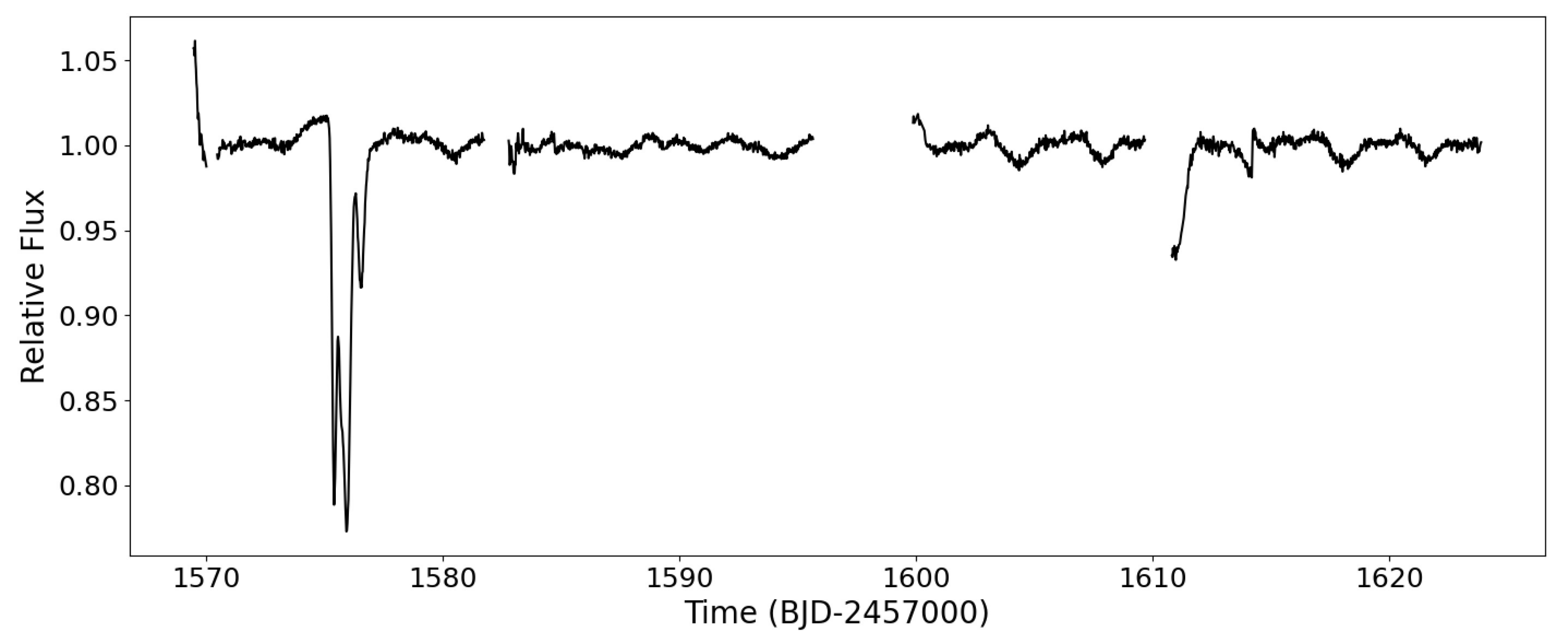}
    \includegraphics[width=0.8\linewidth]{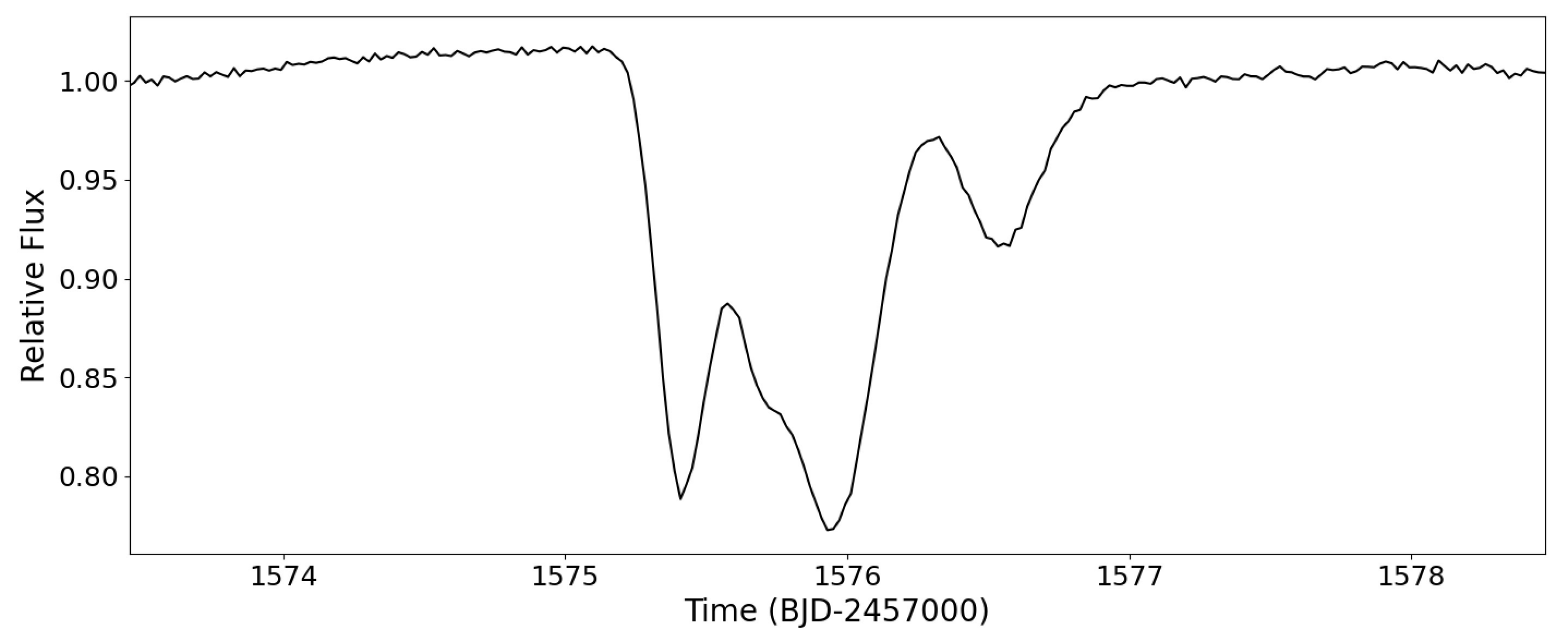}
    \caption{({\em Top panel}) The {\tt eleanor} corrected flux of TIC 400799224 in {\em TESS} Sectors 10 and 11. ({\em Bottom panel}) Close-up view of the the $\sim$1.6-day event demonstrates a substantial occulting of the host star by a highly irregular body.}
   \label{fig:TESS_10_11}
\end{figure*}

{\em TESS} observed TIC 400799224 in Sector 10 (26 March - 22 April 2019), 11 (22 April - 21 May 2019), 37 (2-28 April 2021), and 38 (28 April - 26 May 2021). Figure \ref{fig:TESS_10_11} shows the light curve from Sectors 10 and 11 of the {\em TESS} observation.  The event of interest that drew our attention in the initial discovery of the object occurred at BJTD (BJD-2457000) $\sim$1575 - 1577 of Sector 10, with a close-up view of the event in the bottom panel of the same figure.  The unusual shape of the dip demonstrates three distinct fadings over the course of the event, indicating a highly irregular occulting body, reaching a depth of $\sim$25\%.  We will show in Section \ref{sec:speckle}, that the source light is composite from two stars, so the true fraction of light blocked from the host star is actually much larger.

\begin{figure*}
    \centering
    \includegraphics[width=0.8\linewidth]{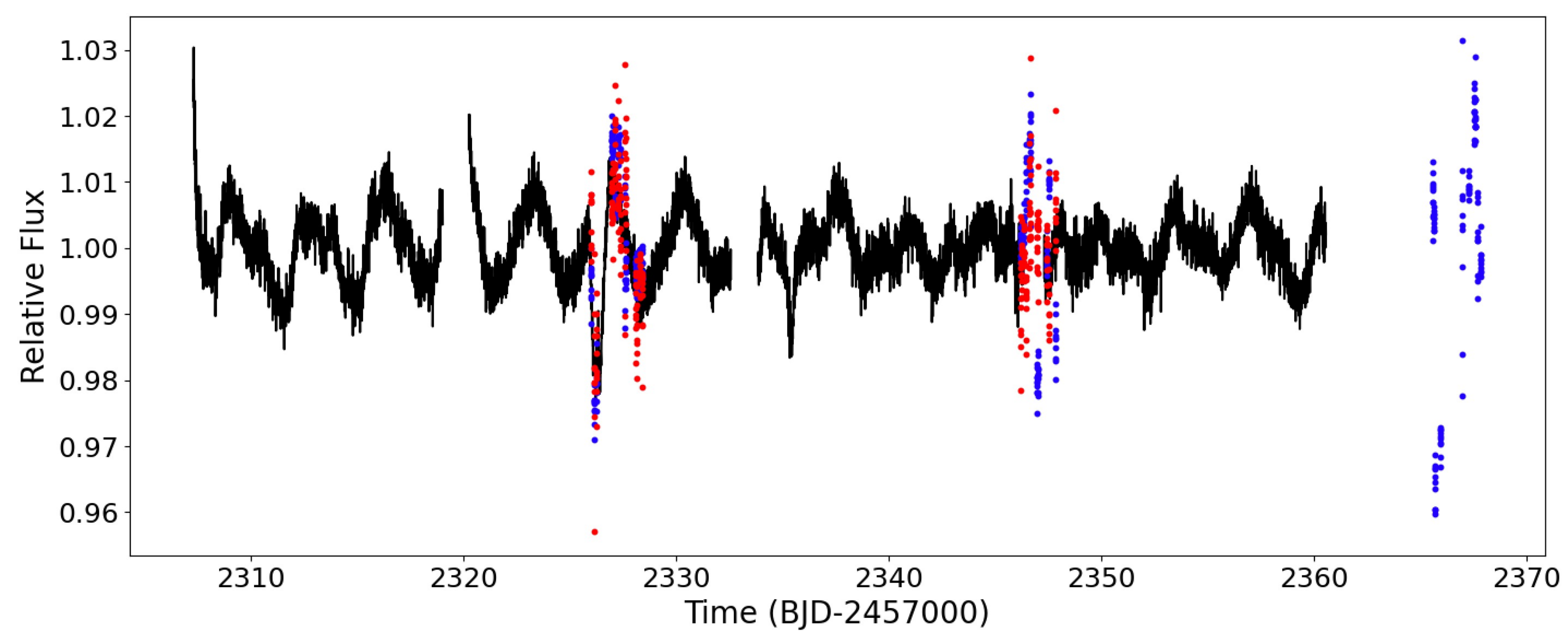}
    \includegraphics[width=0.8\linewidth]{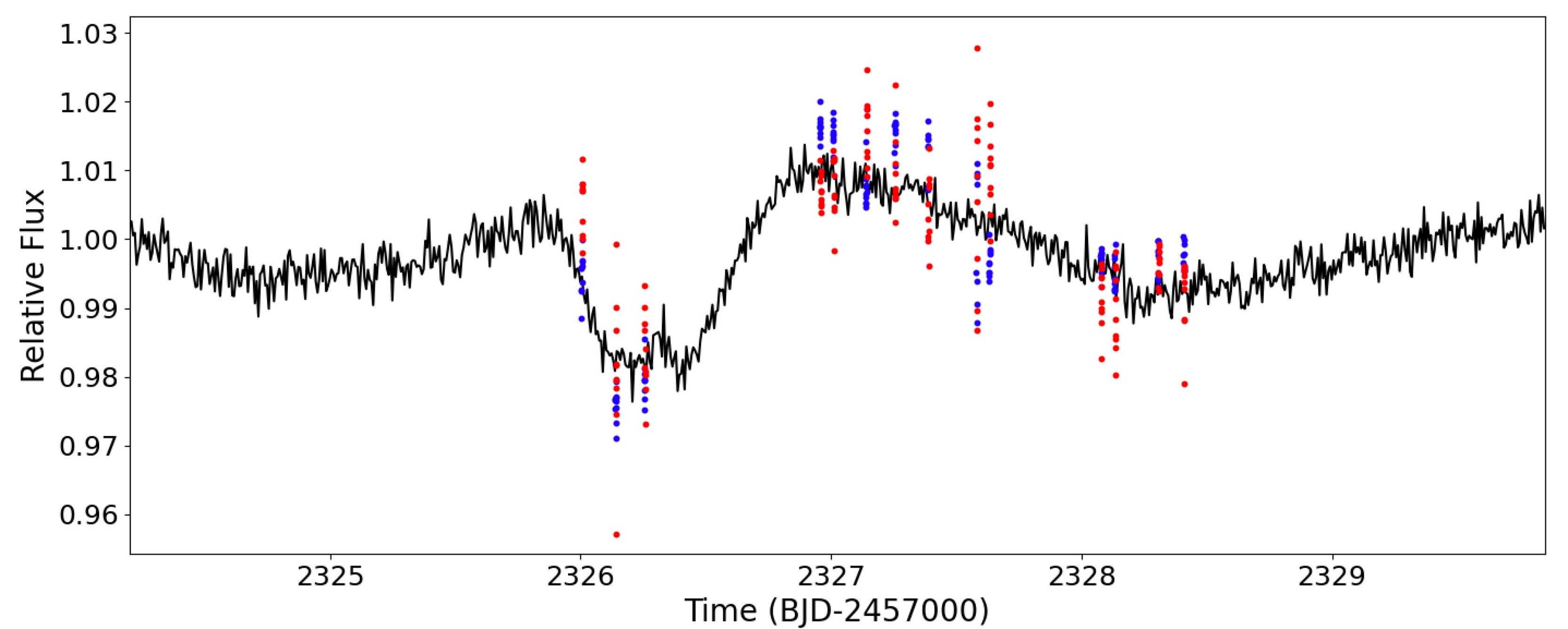}
    \caption{({\em Top panel}) The {\tt eleanor} corrected flux of TIC 400799224 in {\em TESS} Sectors 37 and 38 (black) with LCO G-band (red) and Z-band (blue). ({\em Bottom panel}) Close-up view of the the $\sim$0.9-day event, which was captured by both TESS and LCO.}
   \label{fig:TESS_37_38}
\end{figure*}

After learning of the periodicity of the dips (see Sects.~\ref{sec:asas-sn} and \ref{sec:Evryscope}) we requested follow-up observations from LCO (discussed in Sect.~\ref{sec:photo_follow_up}) to coincide with the TESS sector 37 and 38 observations.  Another dip event was detected at BJTD $\sim$2326 in Sector 37, which is shown with the {\em TESS} Sector 37 and 38 light curve in Fig.~\ref{fig:TESS_37_38}.

 % Fig. 2

 % Fig. 3

\begin{figure}
    \centering
    \includegraphics[width=1.0\linewidth]{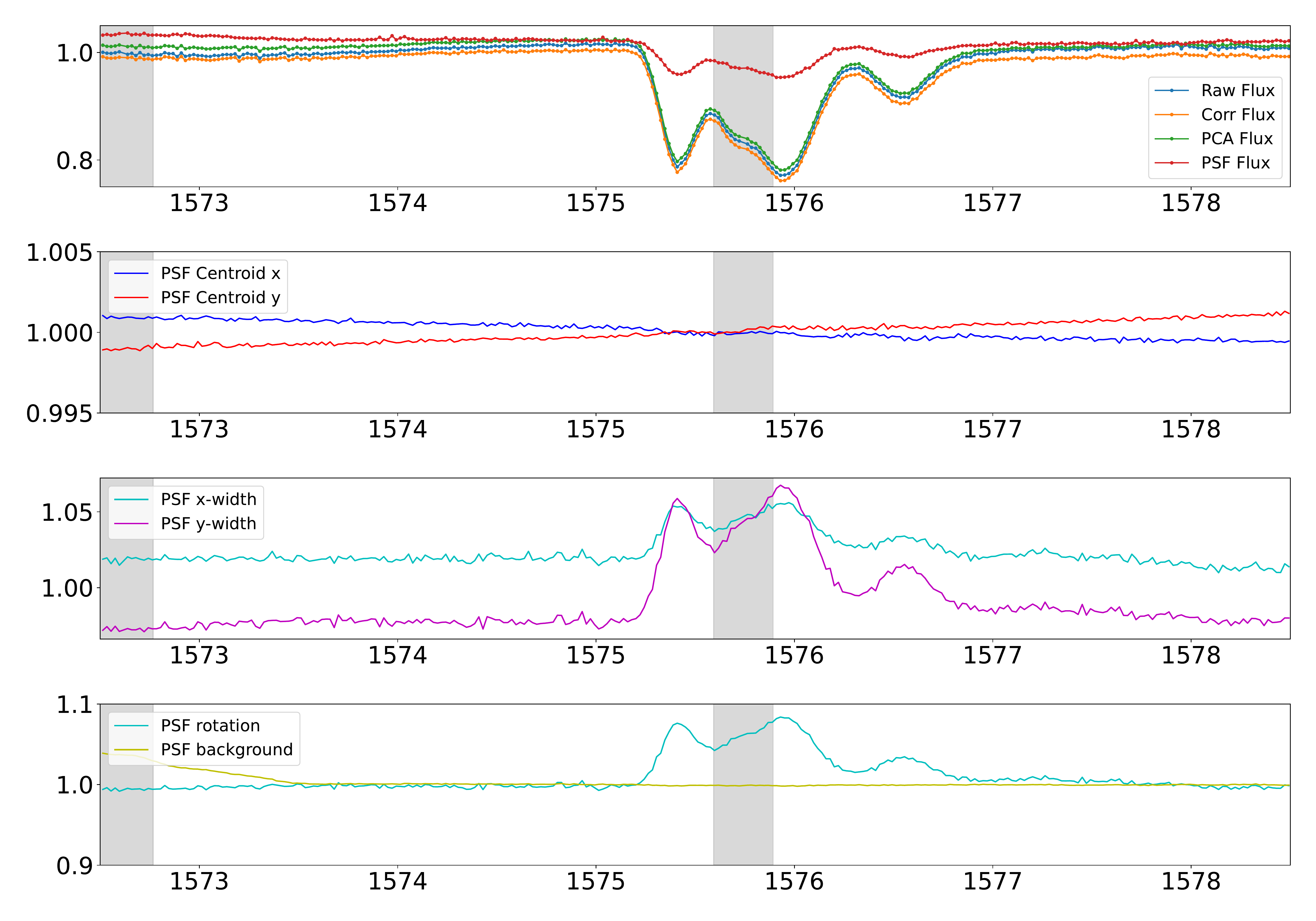}
    \caption{Diagnostic plots from the \texttt{eleanor} pipeline used for ruling out systematic effects as the cause of the Sector 10 event. From top, the panels are 1) segment of the \texttt{eleanor} raw, corrected, PCA, and PSF light curves of the target, centered on the dip; 2) measured normalized PSF x- and y-photocenters; 3) measured normalized PSF x- and y-widths; and 4) measured normalized PSF rotations and backgrounds. While there is a {\em TESS} momentum dump near the center of the event (grey vertical band), there are no significant changes or discontinuities in the measured x- and y-photocenters (panel 2 from top), indicating that the event is produced by either the target star or the nearby star resolved by SOAR ($0\farcs6$ separation). The features seen in panels 3 and 4 (from top) are expected due to the presence of the nearby star (see text for details).}
    \label{fig:vet_S10} 
\end{figure}

\begin{figure}
    \centering
    \includegraphics[width=1.0\linewidth]{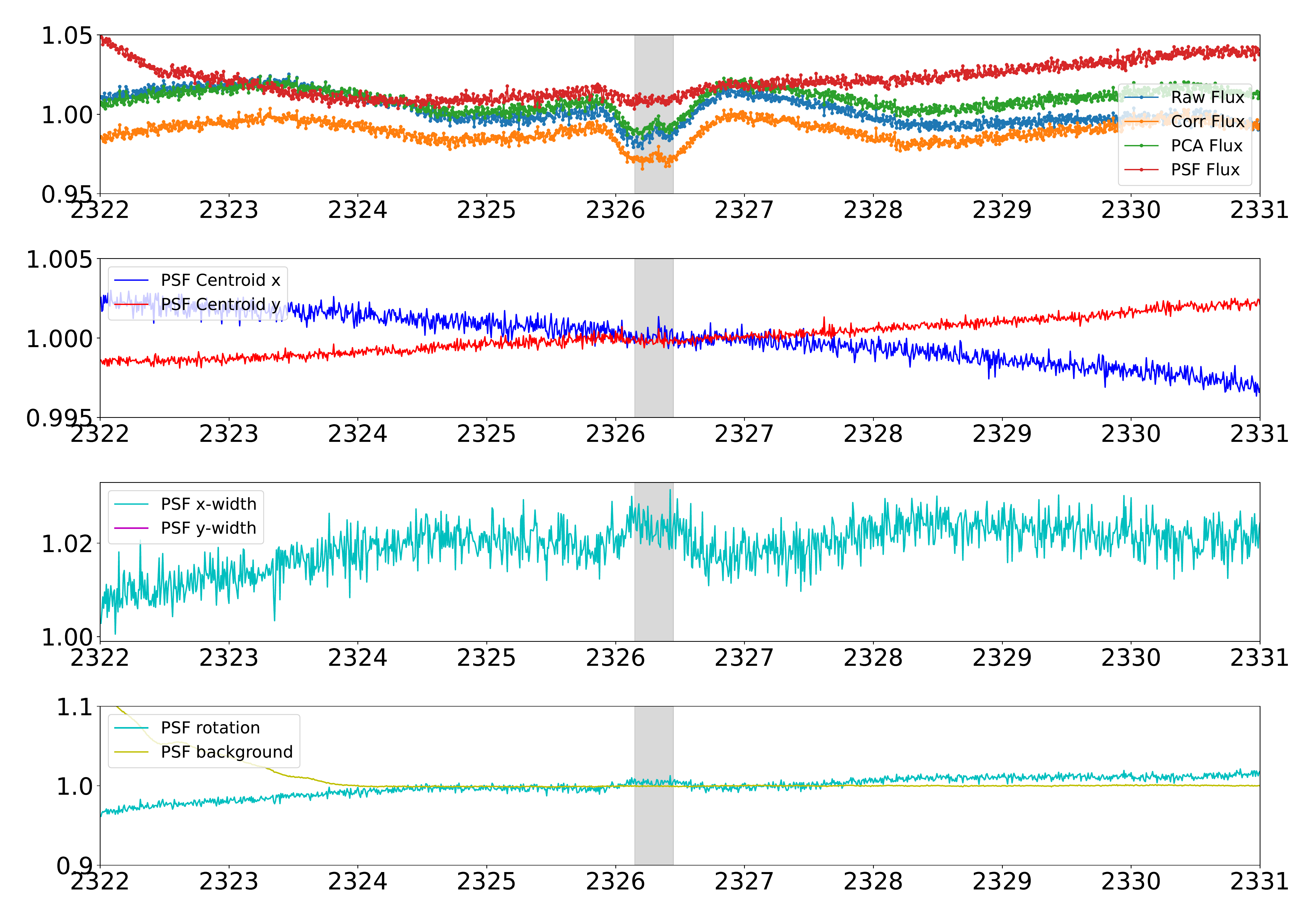}
    \caption{Same as Fig. \ref{fig:vet_S10} but for Sector 37. }
    \label{fig:vet_S37}
\end{figure}

With the dip from Sector 10 demonstrating such a unique shape, we performed thorough pixel-level vetting of the signal and have confirmed that both the Sector 10 and Sector 37 dip events are not due to {\em TESS} systematics and originated from the indicated source, so therefore must be astrophysical in nature. This vetting process is demonstrated in Figures \ref{fig:vet_S10} and \ref{fig:vet_S37}. As seen from the figures, both events coincide with momentum dumps. However, the measured PSF x- and y-photocenters do not show significant changes or discontinuities before, during, or after either event, indicating that their source is either the target star or the $0\farcs62$ separated star resolved by SOAR (see Sect.~\ref{sec:speckle}). The separation between the two stars is too small for resolving which of them is the source of the events based on the photocenter measurements. We note that the PSF x- and y-widths, and orientation do change during the events. This is due to the redistribution of light inside the target’s aperture as the brightness of the source of the dips decreases during the events.

  % Fig. 4

  % Fig. 5

\begin{figure}
    \centering
    \includegraphics[width=1.0\linewidth]{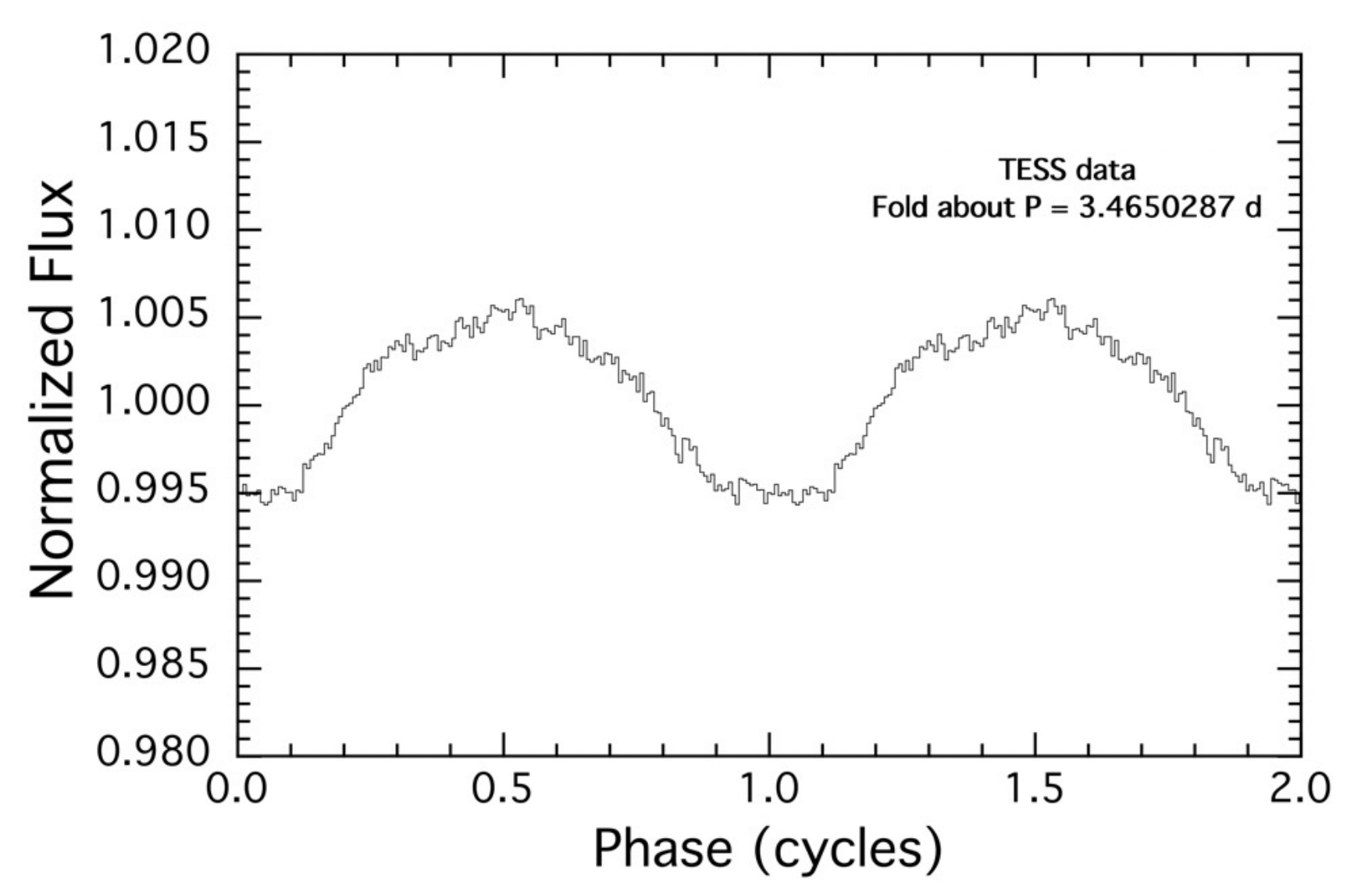}
    \caption{Binned and averaged fold of the {\em TESS} light curve of TIC 400799224, cleaned of the dip events and systematics, shows a modulation at a period of 3.465 days.}
   \label{fig:cleaned_fold}
\end{figure}

We also note from the {\em TESS} light curve that there is a clear modulation in the baseline flux from the system.  Cleaning the light curve of the dip events, we produced the fold shown in Figure \ref{fig:cleaned_fold}.  This modulation is almost certainly related to the rotational period of one of the stars in the binary.  A dip-like feature at BJTD $\sim$2335 is in-phase with the modulation and we assess is most likely to be an anomalous feature on the rotation curve, though we do not rule out the possibility of this being out-of-phase collision debris (see Sect. \ref{sec:occulter}).

 % Fig. 6

\subsection{ASAS-SN}
\label{sec:asas-sn}
\begin{figure*}[h]
\centering
\includegraphics[width=0.45\textwidth]{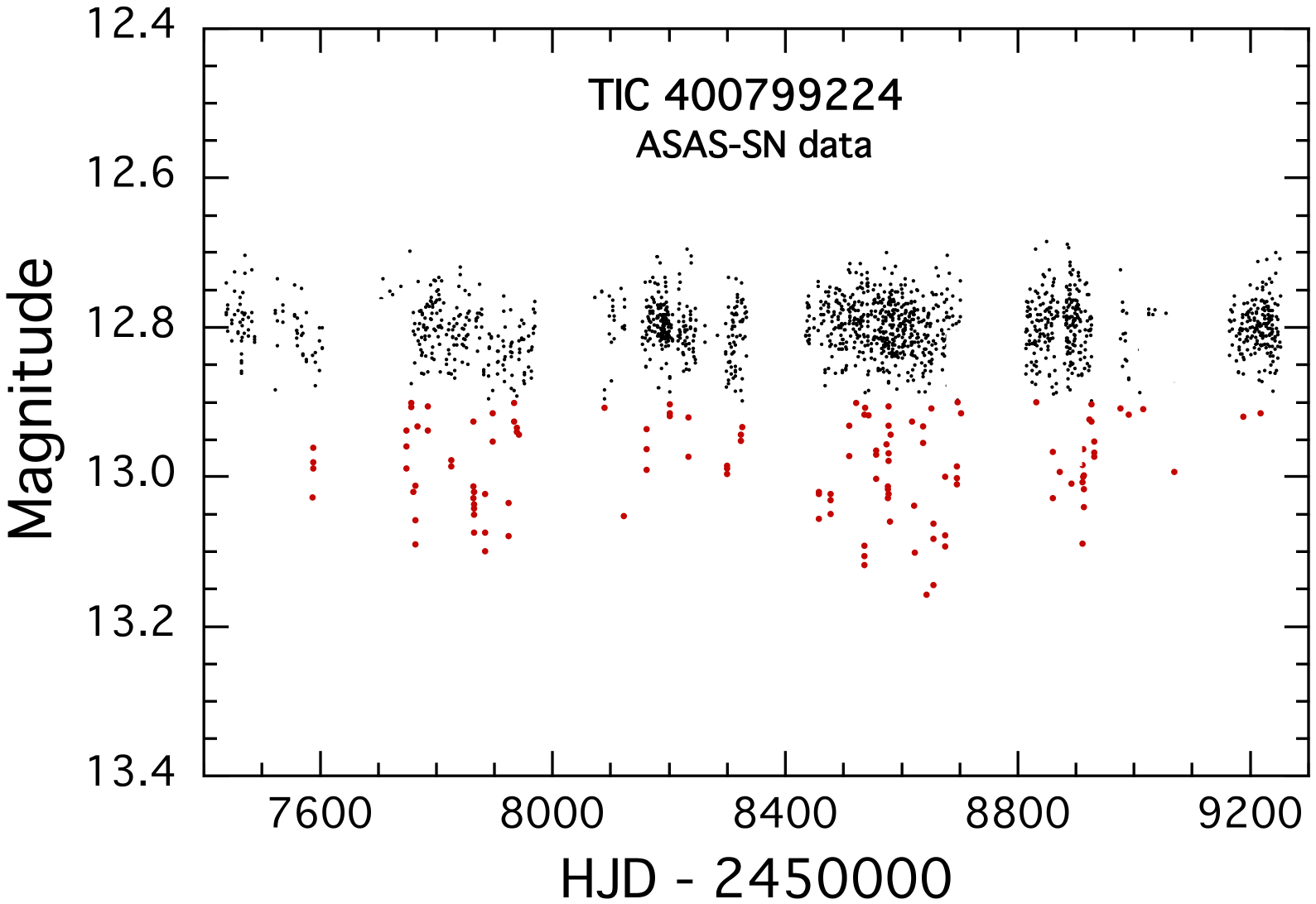}
\includegraphics[width=0.45\textwidth]{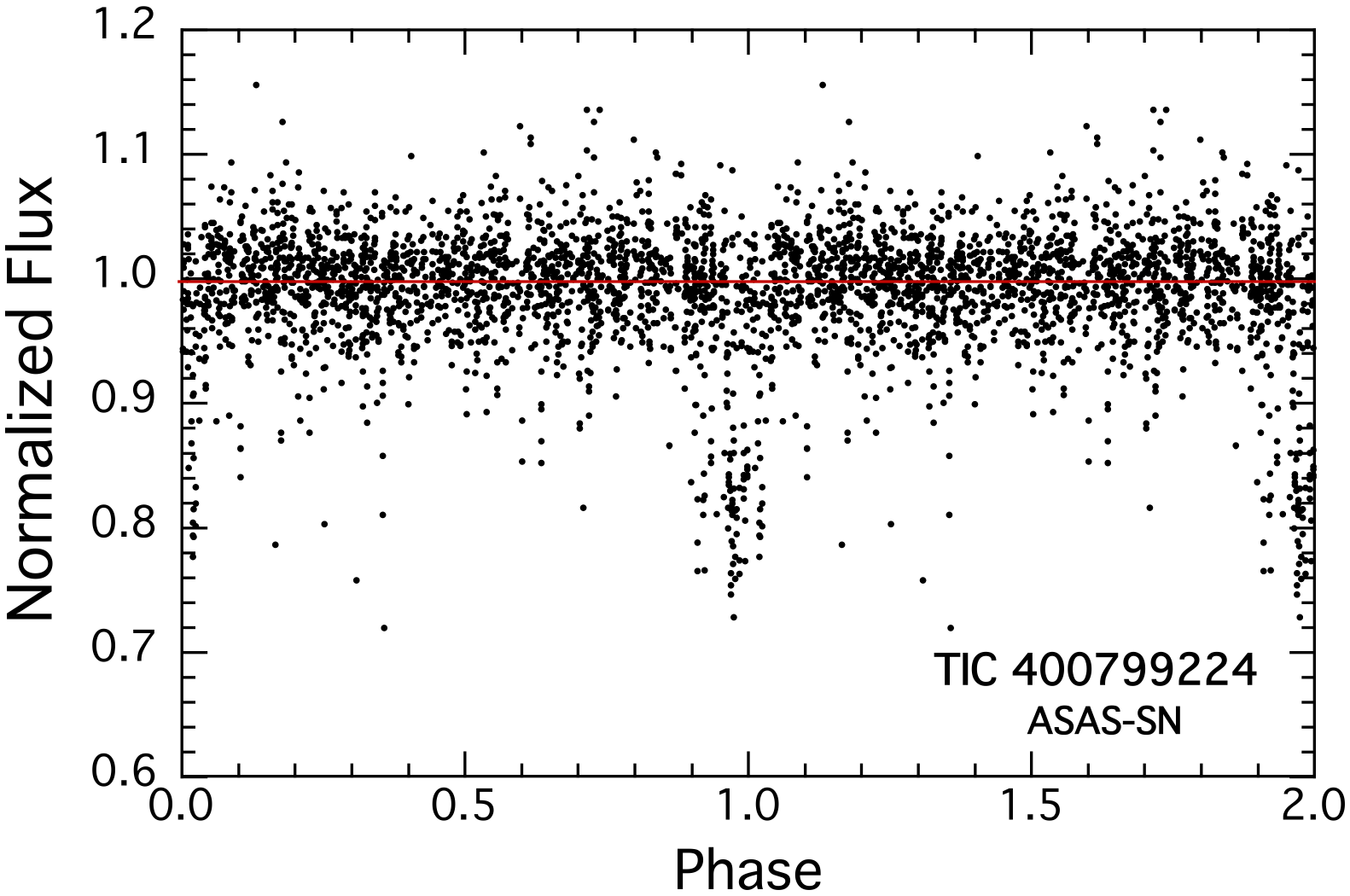}
\caption{({\em Left panel}) The ASAS-SN lightcurve spanning six observing seasons.  The data taken with the green filter were median normalized to those of the V-band data.  Numerous significant individual dips in flux (red) can be seen.  ({\em Right panel}) Fold of the $\sim$1800 ASAS-SN flux measurements (black) about a period of 19.770 days, shown against the median (red). }
\label{fig:LC_fold}
\end{figure*}

More than five years of ASAS-SN archival data (\citealt{Shappee2014}, \citealt{Kochanek2017}) is available for TIC 400799224. This data was of critical importance, as it allowed for the determination of a longer-term periodicity in the dips. The raw ASAS-SN photometric lightcurve is shown in the left panel of Figure \ref{fig:LC_fold}.  Numerous statistically significant dips are readily apparent (these are marked in red).  

In order to extract a possible period for these dips we utilized four different transform algorithms.  These include a standard Box Least Squares (`BLS'; \citealt{Kovacs2002}), Lomb-Scargle (`L-S'; \citealt{Scargle1982}), a less common Plavchan transform \citep{Plavchan2008}, and a custom `Interval Match Transform' (`IMT'; see, e.g., section 7 of \citealt{Gary2017}).  The BLS transform is especially good in searching for periodic signals that involve narrow (i.e., low duty-cycle) features, such as planet transits and narrow eclipses.  The L-S transform is designed to search for sinusoidal like variations in unequally spaced data sets, and somewhat corrects for the window functions in the data set. The Plavchan periodogram \citep{Plavchan2008} is similar to a binless variation of the `phase dispersion minimization' (`PDM') algorithm \citep{Stellingwerf1978}.  Finally, the IMT is a brute force way of searching for common intervals in data sets where a relatively small number of `events' (e.g., dips) can be identified on an individual basis. Basically, it tests a large number of trial periods against all combinations of time differences in the event set.  

  % Fig. 7

\begin{figure*}[h]
\centering
\includegraphics[width=0.45\textwidth]{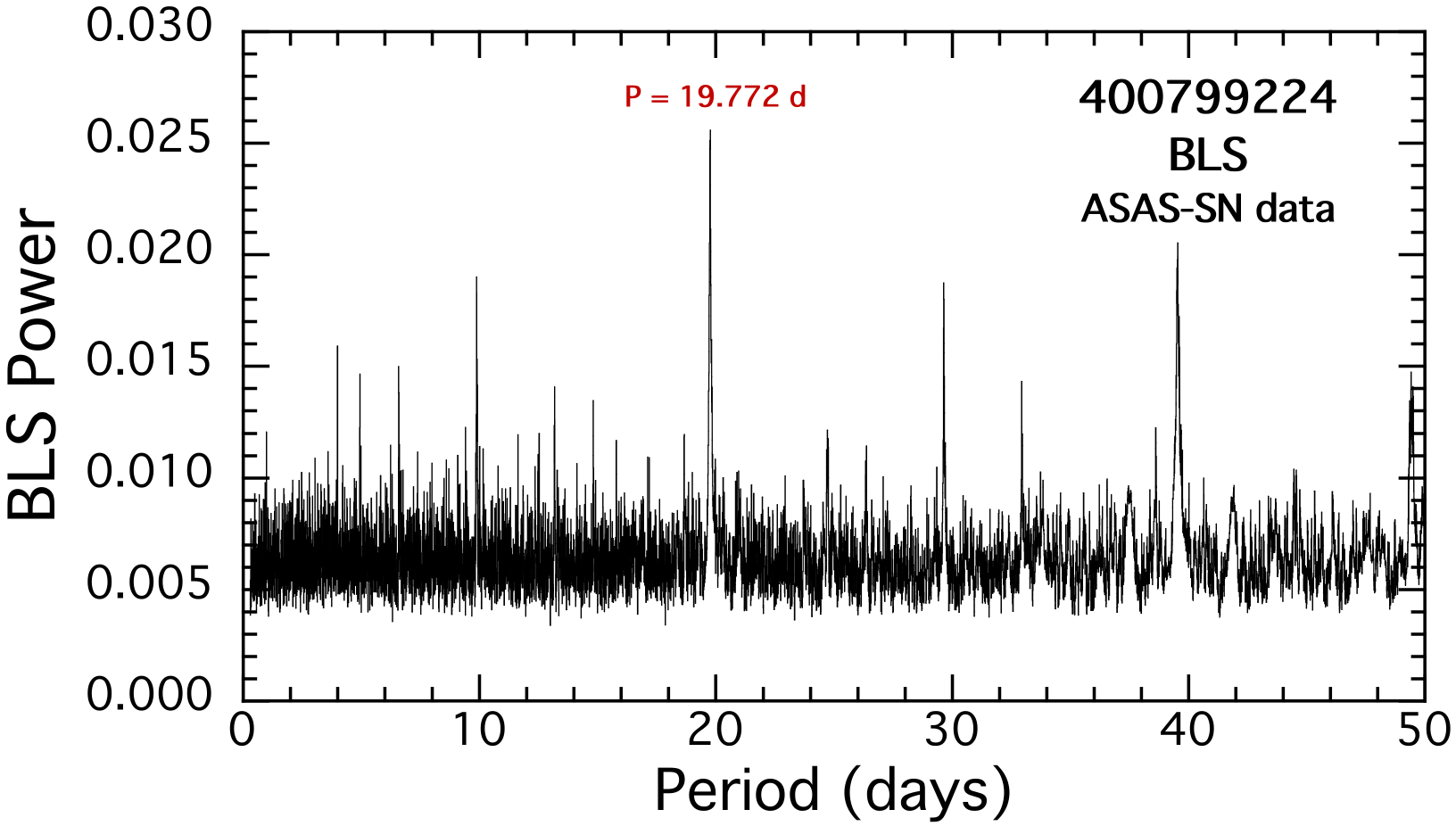} \hglue0.13cm
\includegraphics[width=0.435\textwidth]{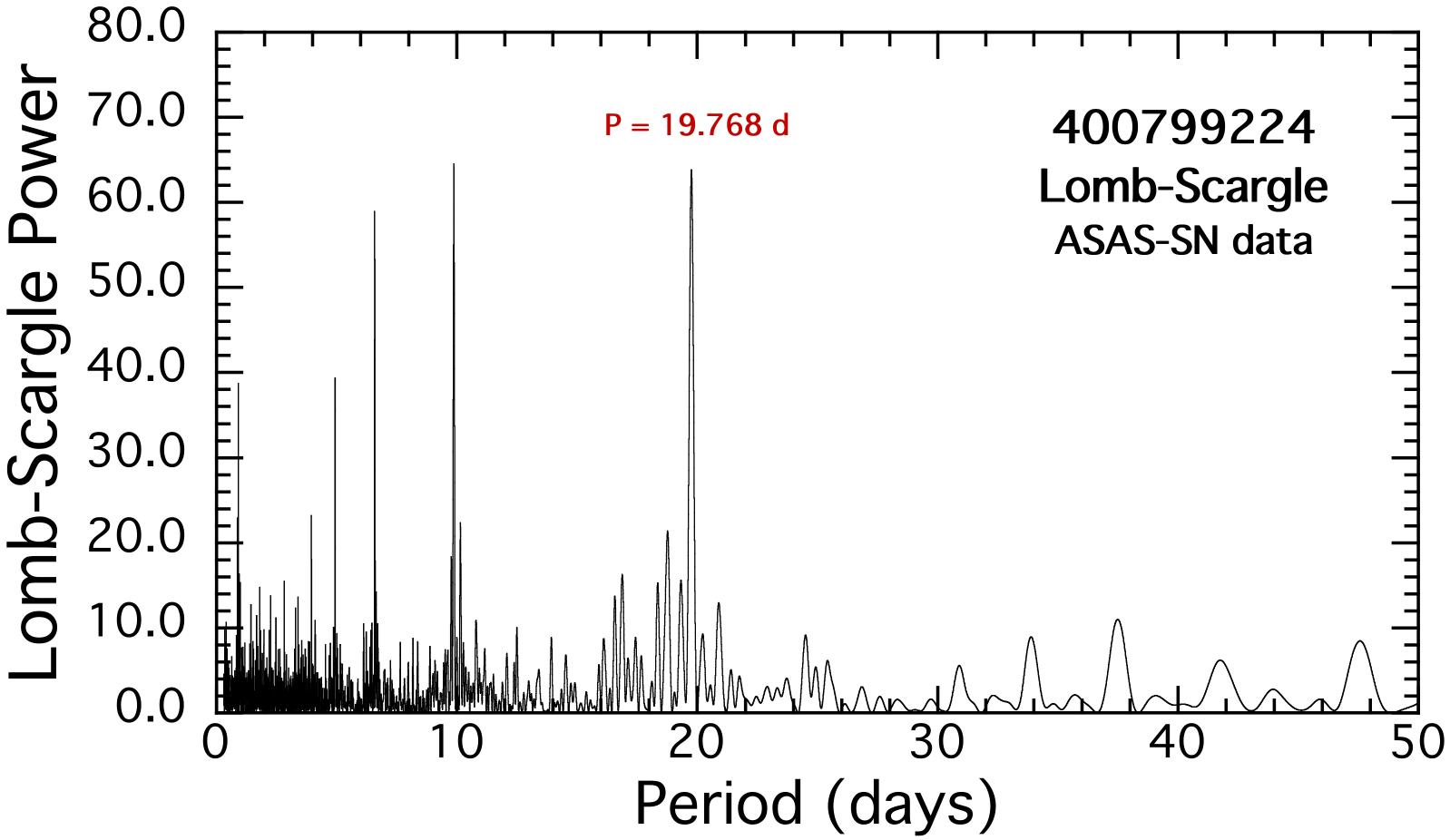}
\includegraphics[width=0.45\textwidth]{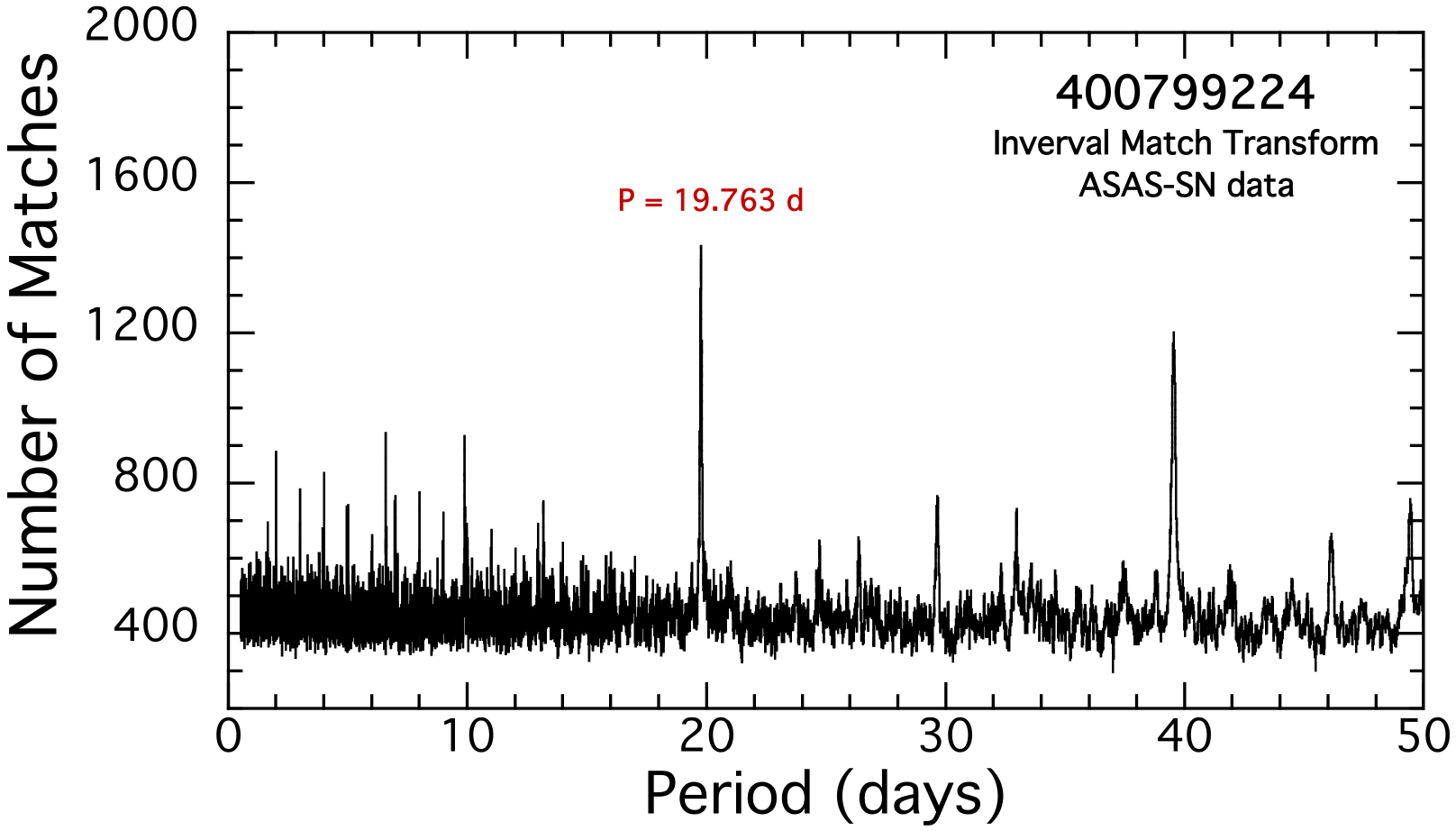}
\includegraphics[width=0.45\textwidth]{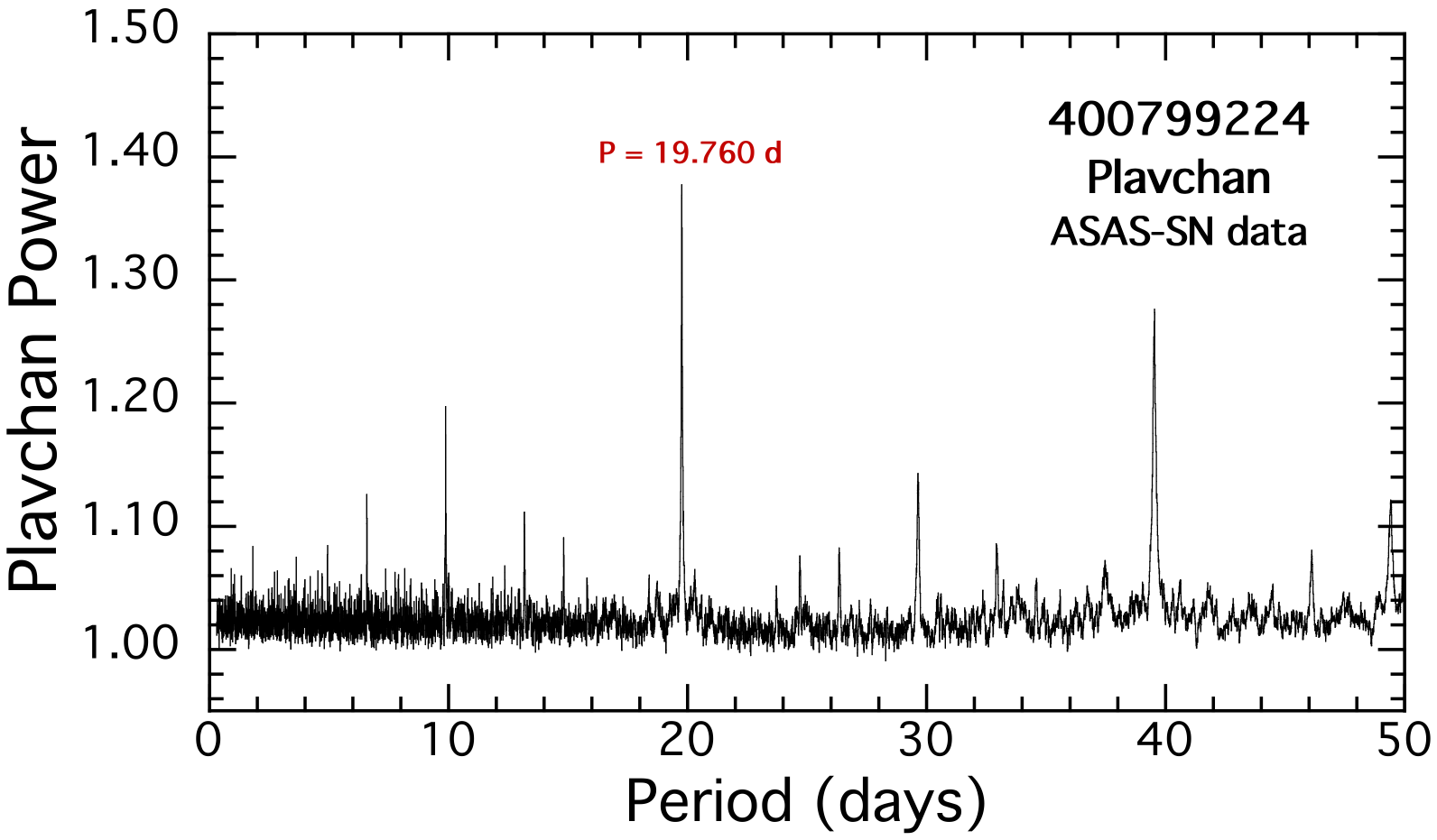}
\caption{Four transforms of the ASAS-SN data.  Clockwise from upper left panel -- Box Least Squares, Lomb-Scarge, Plavchan, and Interval Match Transform.  See text for details.}
\label{fig:transforms}
\end{figure*}

The results of these four transforms applied to the ASAS-SN photometric data are shown in Figure \ref{fig:transforms}. Each transform has its own type of artefacts, e.g., harmonics and subharmonics.  What we see is that the most prominent peak in each transform (except for the L-S) is at a period close to 19.770 days.  Nearly all of the remaining significant peaks are either harmonics or subharmonics of this period.  In the case of the L-S transform, the first harmonic of the 19.77-day period is of comparable height to the base frequency.

   % Fig. 8

From these searches we assess that there is a single unique period in the ASAS-SN data of 19.77 days.  In the right-hand panel of Figure \ref{fig:LC_fold} we show the folded ASAS-SN lightcurve at this period.  A clear dip of up to $\sim$24\% in flux is evident at phase 1.0 on the plot.  However, several other features of this fold are evident that make it quite different from typical profiles of binary eclipses or planet transits. (i) The width of the main dip in flux is $\simeq 15\%$ in phase of the 19.77 d period. (ii) The statistics are not optimum, but the profile of the dip does not look as if there is a monotonic decrease in flux during the ingress, followed by a monotonic rise in flux during the egress. (iii) There are clearly quite a number of low flux points that are not in phase with the main dip.  (iv) There are points with a seemingly unperturbed flux in the phase region where the dip occurs. Therefore, we can conclude immediately that this is not an eclipse or transit of a hard body on a simple Keplerian orbit.  

\begin{figure}
    \centering
    \includegraphics[width=1.0\linewidth]{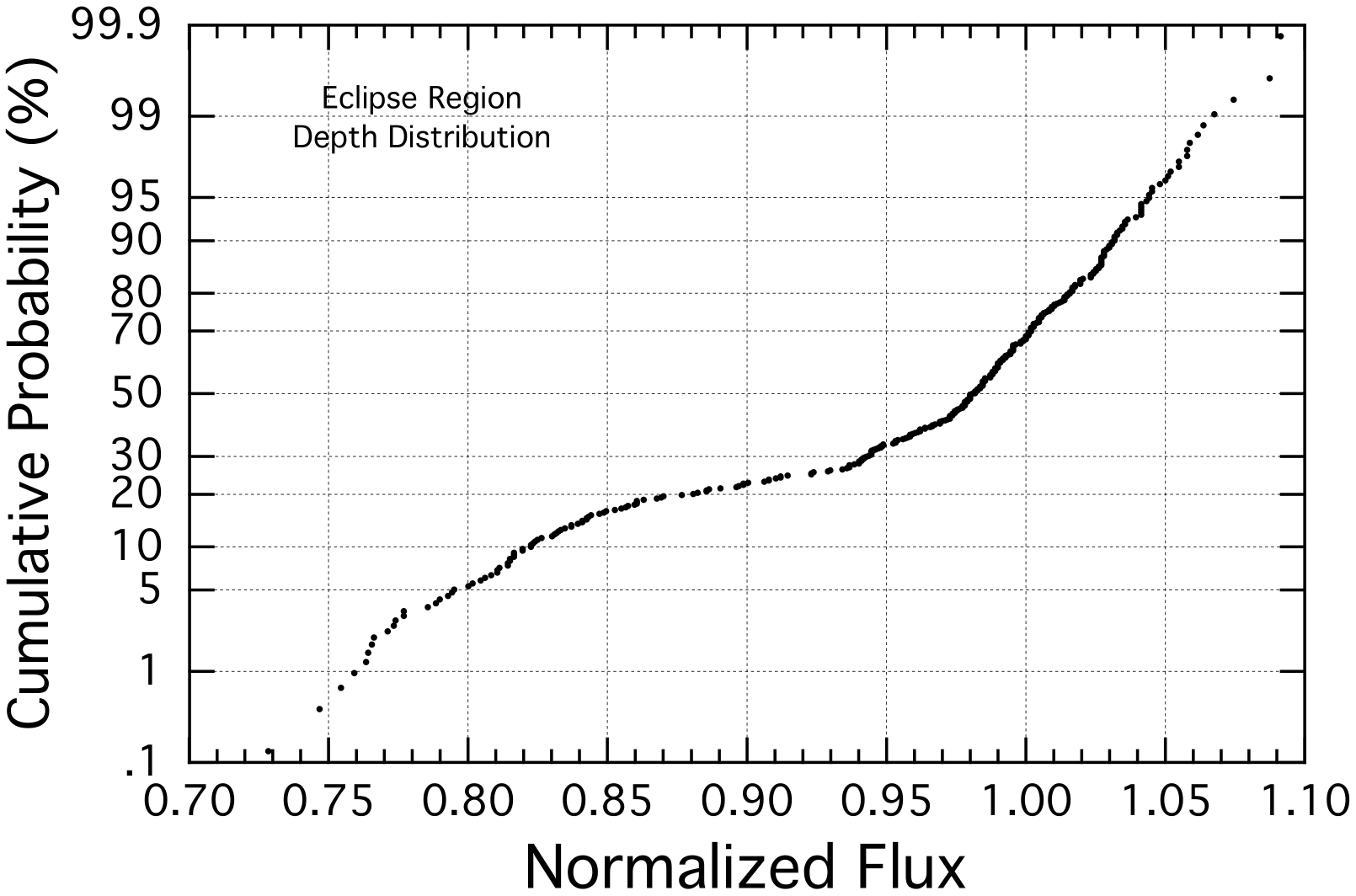}
    \caption{Cumulative distribution of the depth of the ASAS-SN flux in the transit region (phase 0.89 - 1.05 from the right panel of Figure \ref{fig:LC_fold}).}
   \label{fig:cdf}
\end{figure}

Analyzing this further, we isolated the ASAS-SN fluxes within the region of each expected transit (phase 0.89 - 1.05, based on the right panel of Figure \ref{fig:LC_fold}) and measured their depth. The results are shown in Figure \ref{fig:cdf} as a cumulative distribution of flux during the eclipse region. It can be seen from this distribution  that, though the object is clearly periodic at 19.77 days, a measurable reduction in flux will occur in only approximately one out of every three to five transits.  

We can conclude from this that the orbiting object is either too small to cause any noticeable decrease in flux when transiting, or it is itself \textit{not} transiting the star from our perspective.  The former lends to the conclusion that the object is most likely an asteroid rather than a disintegrating planet, while the latter allows for a planet or other orbiting body.  In any case, an occulting cloud of substantial opacity is sporadically present and amorphous, causing a varying depth, duration, and shape of the dips in the light curve.

 % Fig. 9

\subsection{Evryscope}
\label{sec:Evryscope}

\begin{figure*}
    \centering
    \includegraphics[width=0.8\linewidth]{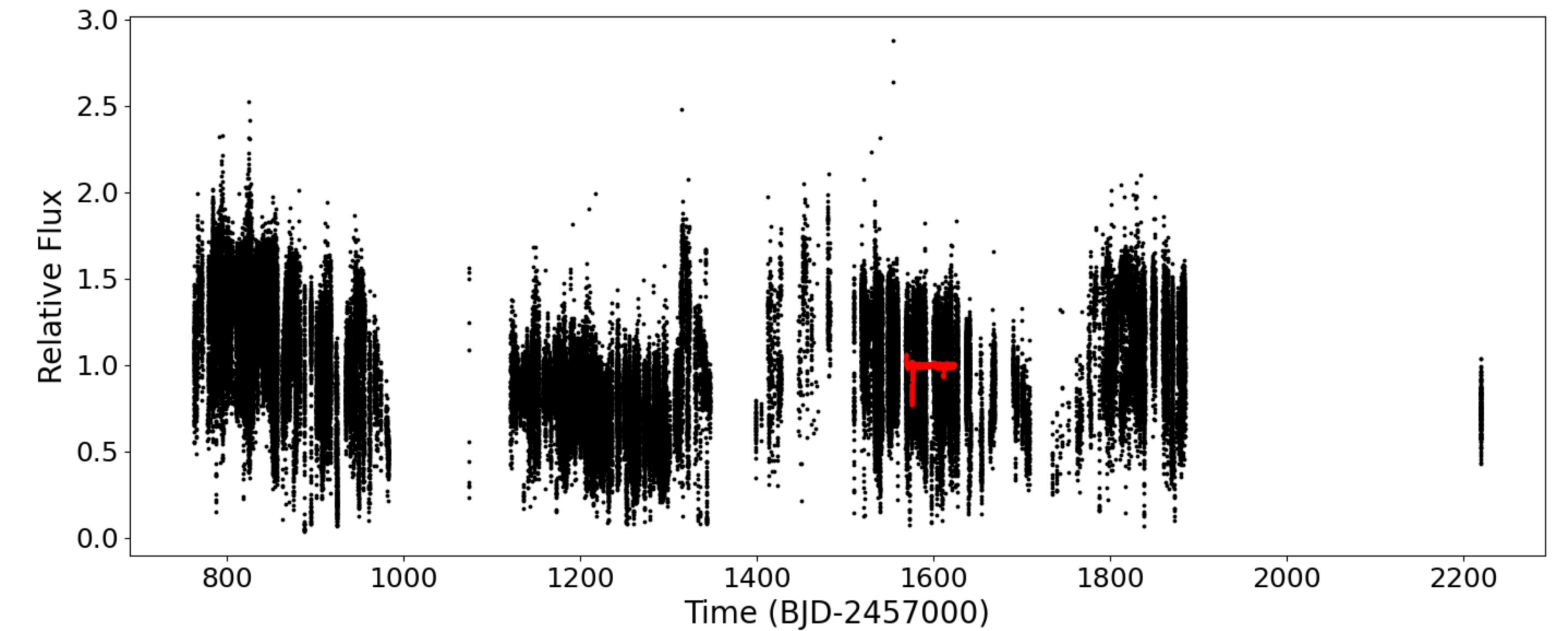}
    \includegraphics[width=0.8\linewidth]{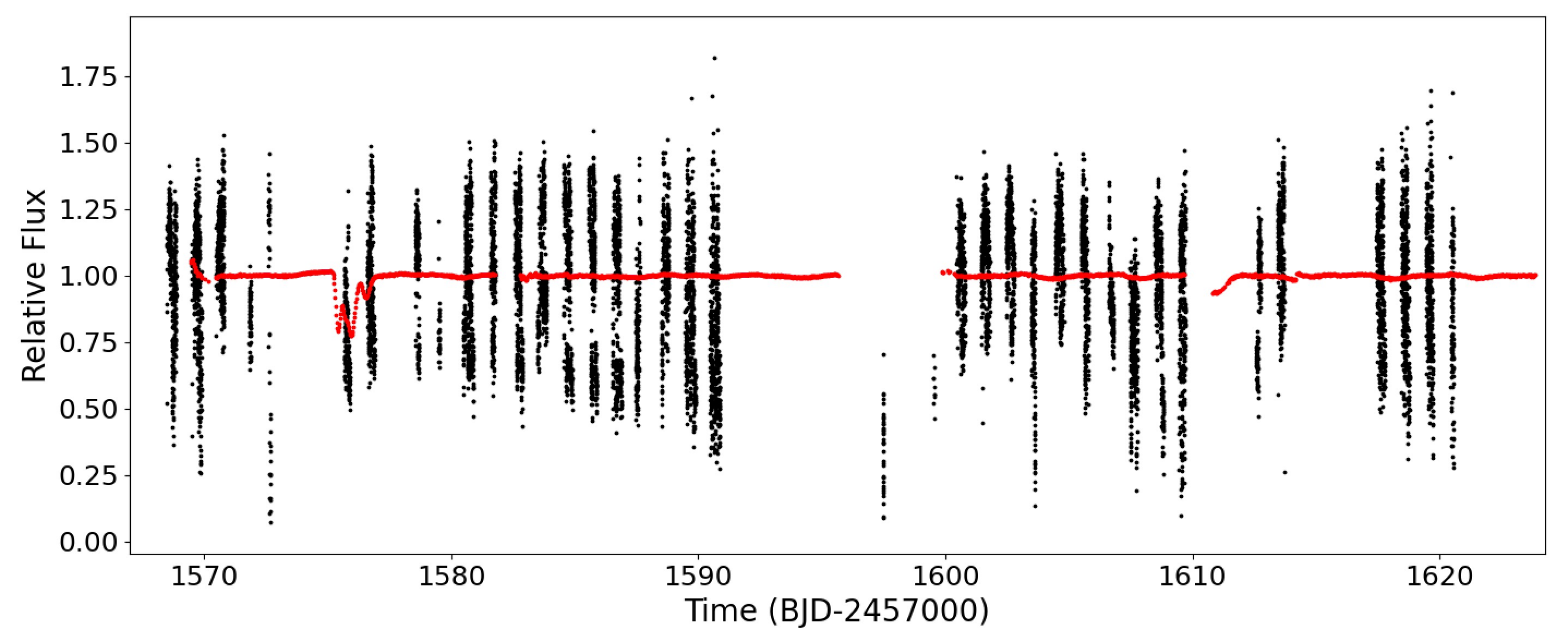}
    \caption{Evryscope light curve (black) for TIC 400799224, {\em top panel}: full extent of Evryscope data, {\em bottom panel}: approximately 50 days of the Evryscope data that overlap with the {\em TESS} Sectors 10 and 11 light curve. In both panels the {\em TESS} lightcurve is superposed (in red) on the Evryscope lightcurve.  It is apparent that the non-astrophysical fluctuations in the Evryscope data make it difficult to see the individual dips; however, the sheer number of observations makes it possible to readily detect the average profile of the 19.77-day periodic dips (see the folded lightcurve in Figure \ref{fig:evryscope_fold}).  In both panels, data points with error bars greater than the flux and with a relative flux greater than 3.0 were excluded.}
   \label{fig:evryscope}
\end{figure*}

We obtained four years of Evryscope \citep{2014SPIE.9145E..0ZL, 2019PASP..131g5001R} archival data for TIC 400799224, ranging from January 2017 to January 2021. Forced aperture photometry was performed on each of 85,394 Sloan $g$-band images using the Evryscope Fast Transient Engine on-demand photometry pipeline as described in \cite{2020ApJ...903L..27C}, using a 26\farcs8 aperture radius. The full light curve is shown in Figure \ref{fig:evryscope}.

\begin{figure*}
    \centering
    \includegraphics[width=0.48\linewidth]{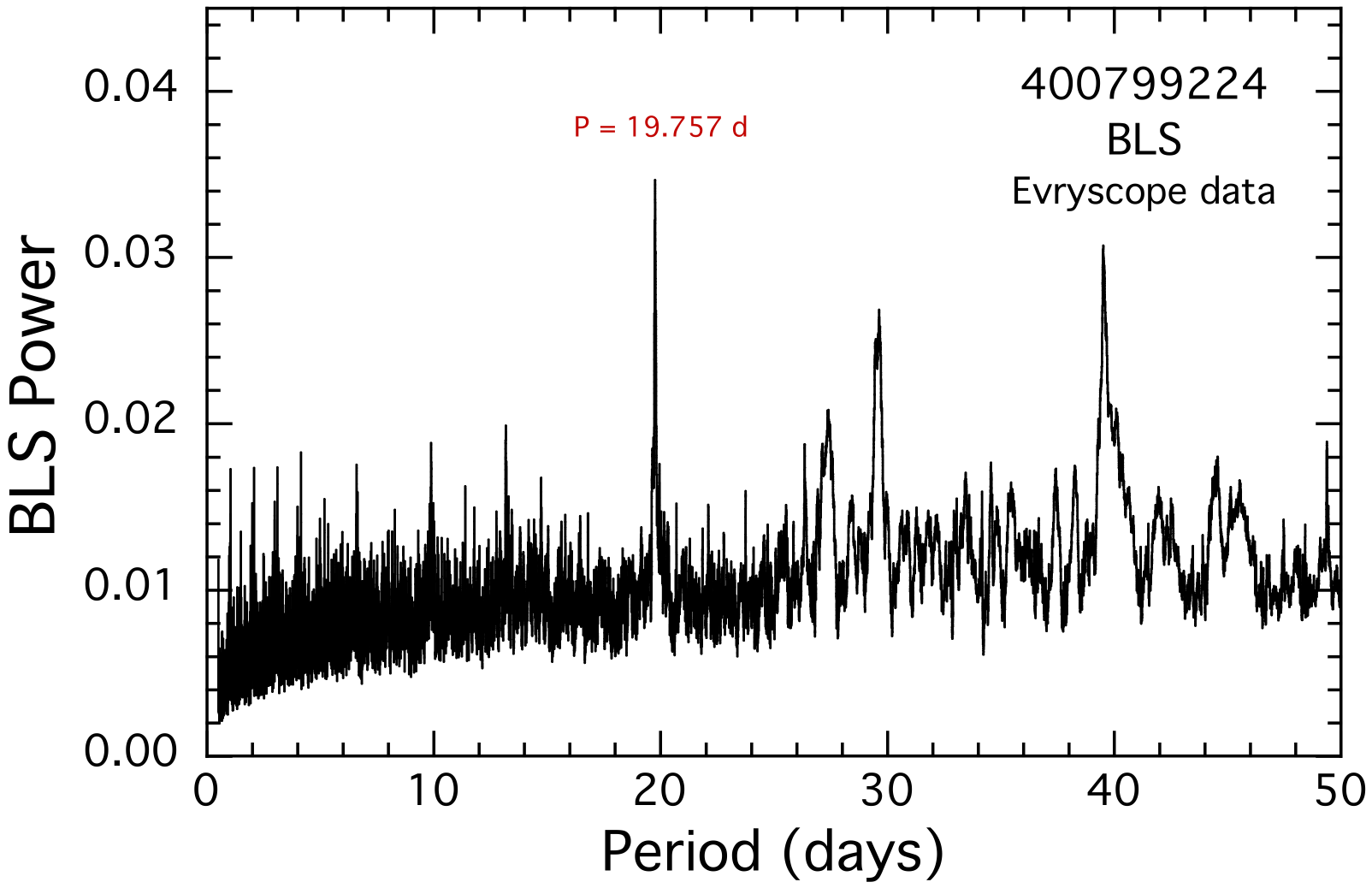} \hglue0.3cm
    \includegraphics[width=0.48\linewidth]{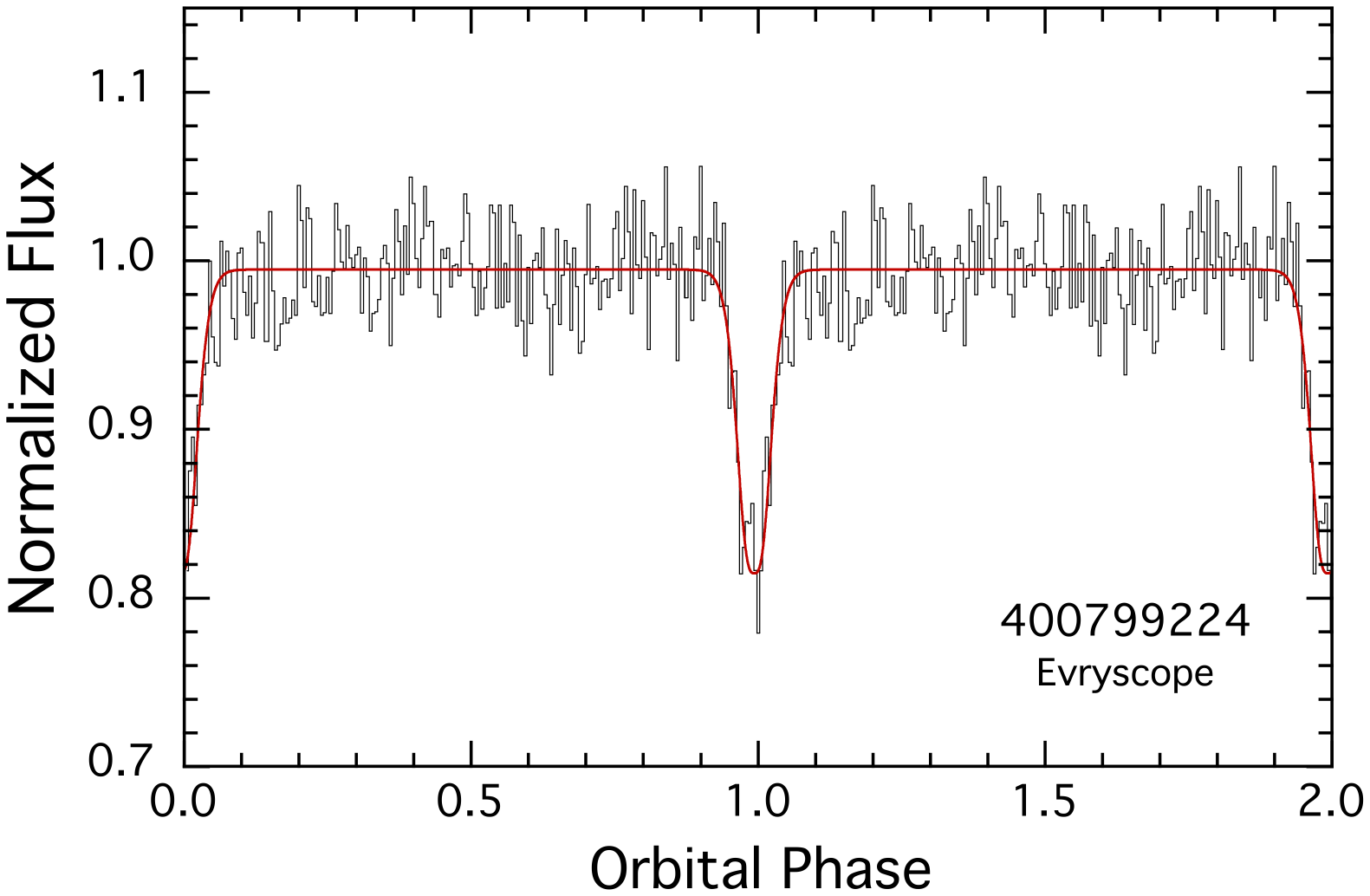}
    \caption{Left: BLS transform of $\sim$85,000 Evryscope photometry data points.  The four highest peaks, in descending order, are the 19.757 day source eclipse period, its lower harmonic near 40 days, the 29.5 day synodic lunar cycle, and the 27.3 day sidereal lunar period.  Right: Binned, averaged fold of the Evryscope data on the 19.757 day period.}
   \label{fig:evryscope_fold}
\end{figure*}

Though these data have a substantial signal due to the Lunar sidereal and synodic periods, the strongest peak in the BLS occurs at 19.757 days and confirms the period found in the ASAS-SN data. The binned and averaged fold of the Evryscope data at this period is shown in the right panel of Fig.~\ref{fig:evryscope_fold}. The epoch time of phase zero on this plot is the same as that of the fold for the ASAS-SN data.  In examination of the folds of the Evryscope and ASAS-SN data, we determined that the areas within the dips differ by approximately 1.6; this could be accounted for by the differences in filters used and seasons sampled.  It is quite possible, even, that Evryscope and ASAS-SN observed an only partially overlapping set of transits.  When considering the variable presence, depth, and duration of the transits, it is strong confirmation of the behavior that the periodicity was even found in both data sets.

 % Fig. 10

 % Fig. 11

\section{Follow-up observations}
\label{sec:follow-up}

\subsection{Photometric}
\label{sec:photo_follow_up}
Following the determination of the periodicity of the dips from ASAS-SN and Evryscope data, we requested photometric follow-up from Las Cumbres Observatory Global Telescope (LCOGT) 1-m network \citep{Brown:2013}. LCOGT attempted to observe the target in Sloan $g'$ and Pan-STARRS $z$-short bands every $3\pm1$ hours during three separate windows of $\pm$1.25 days from a t$_{0}$ corresponding to our calculations of predicted events at 18:17 UT on 22 April 2021, 12:44 UT on 12 May 2021, and 07:12 UT on 1 June 2021. The $4096\times4096$ LCOGT SINISTRO cameras have an image scale of $0\farcs389$ per pixel, resulting in a $26\arcmin\times26\arcmin$ field of view. The images were calibrated by the standard LCOGT {\tt BANZAI} pipeline \citep{McCully:2018}, and photometric data were extracted with {\tt AstroImageJ} \citep{Collins:2017} using apertures with radius $2\farcs7$. 

The 22 April 2021 (BJTD $\sim$2326) dip event, overlapping with {\em TESS} Sector 37, is shown in Figure \ref{fig:TESS_37_38}, with a close-up view in the bottom panel.  While the expected dip in flux at this time is only $\sim$3\% deep in the {\em TESS} data, the LCO photometry seems to follow the same behavior, at least to within its statistical precision.  The 12 May 2021 (BJTD $\sim$2346) observation window coincided with the {\em TESS} Sector 38 data downlink gap, though there does not appear to have been a detectable dip during that time.  The 1 June 2021 (BJTD $\sim$2366) observation window, which occurred after the completion of {\em TESS} Sector 38, shows a $\sim$4\% dip, which we assess to be likely another small event, similar to the 22 April 2021 (BJTD $\sim$2326) dip event.

\subsection{Spectroscopy}
\label{sec:spec}

\begin{figure}
    \centering
    \includegraphics[width=1.0\linewidth]{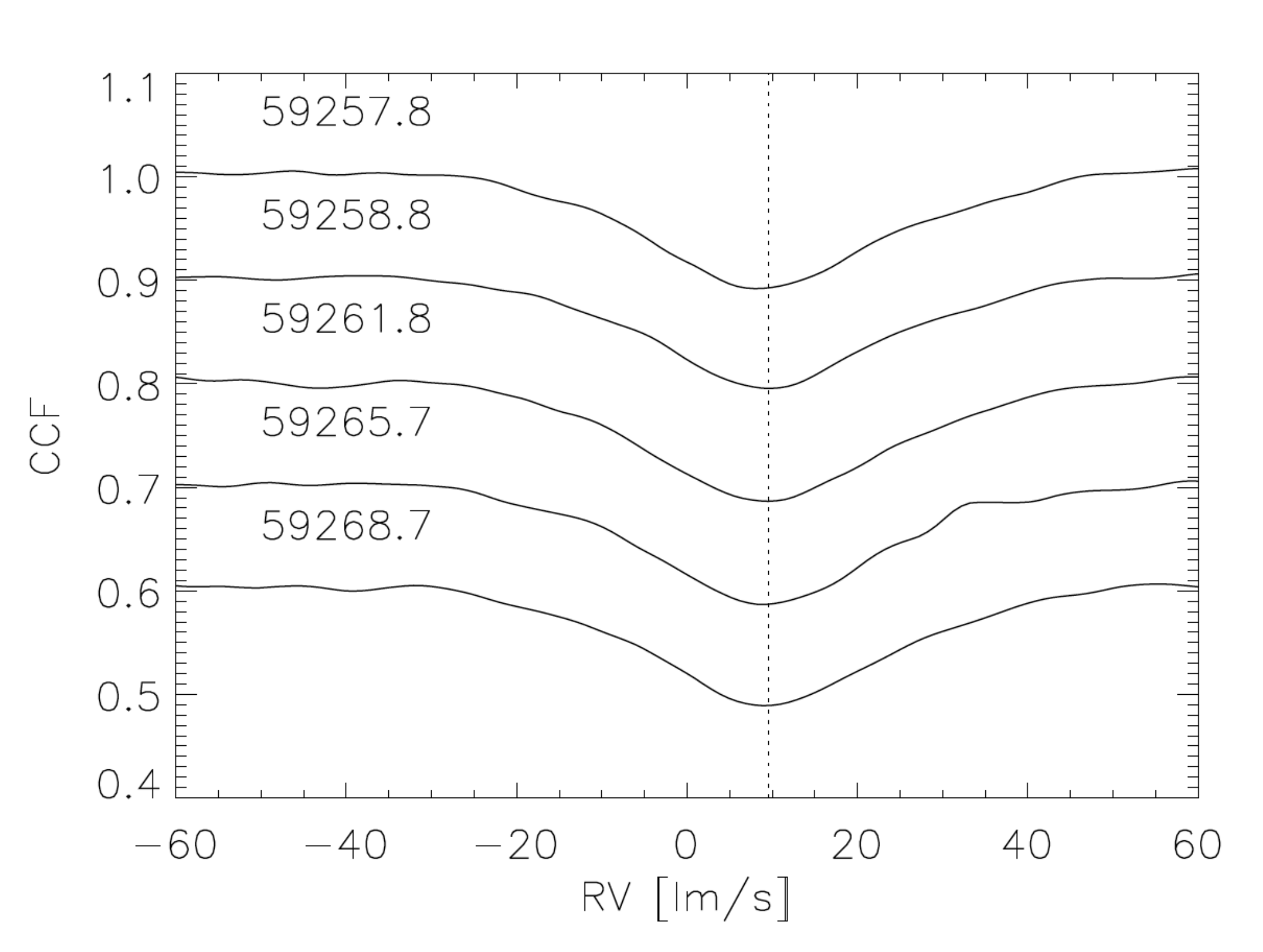}
    \caption{CCFs of the CHIRON spectra with the solar mask observed on five dates. The vertical dotted line marks the average RV. The CCFs are vertically displaced by 0.1, the two 10-min.~CCFs (JD 59261.8) are averaged.}
   \label{fig:ccfsm}
\end{figure}

Optical spectra of TIC 400799224 were recorded using the CHIRON spectrograph at the 1.5 m telescope at Cerro Tololo Inter-American Observatory \citep{chiron} operated by the SMARTS consortium. The observations were made 
in the fiber mode (spectral resolution 25K) on 11, 12, 19, and 22 February 2021 with 15-min. exposure time, accompanied by the ThAr comparison spectra. Moreover, three consecutive 10-min. exposures (also in the fiber mode) were taken on 15 February 2021. The 15-min.~spectra have an average signal of about 600 e$^{-}$ pixel$^{-1}$, or a signal to noise ratio (SNR) of $\sim$25. The cumulative exposure time is 1.5 h. The spectra were reduced using the standard CHIRON pipeline. They were cross-correlated with the binary solar-spectrum mask. All cross-correlation functions (CCFs) have a single dip with a constant radial velocity (RV), having a mean value of 9.54 km~s$^{-1}$ and an rms scatter of 0.24 km~s$^{-1}$. The best fit RVs for each observation are shown in Table \ref{tbl:rvs}.  We assess the RV to be constant within error, implying that the star is not a close binary. If, as we believe, the 19.77-day period is associated with an orbiting body, its mass is too small to cause measurable reflex motion of the host star.   The shape of the CCF profile also does not change, as shown in Figure \ref{fig:ccfsm}. In Sect.~\ref{sec:spectra} we extract other information from the spectra.

\begin{table}
\centering
\caption{CHIRON spectra CCF best fit radial velocities$^a$.}
\begin{tabular}{c c c c}
\hline
\hline
Date & RV & ampl & sigma    \\  
JD-2400000 &  km s$^{-1}$  &   & km s$^{-1}$\\
\hline
59257.8180 &  9.440 &  0.104 & 13.944 \\
59258.7998 &  9.569 &  0.099 & 13.841 \\
59261.8219 &  9.623 &  0.106 & 14.504 \\
59261.8289 &  9.482 &  0.113 & 13.902 \\
59265.7356 &  9.089 &  0.110 & 13.385 \\
59268.7273 &  9.815 &  0.106 & 14.121 \\
\hline 
\label{tbl:rvs} 
\end{tabular} % Table 3

{Notes: (a) Plots of the CCFs are shown in Figure \ref{fig:ccfsm}. }  

\end{table} %table 1

 % Fig. 12

\subsection{Speckle Imaging}
\label{sec:speckle}

\begin{figure}
    \centering
    \includegraphics[width=1.0\linewidth]{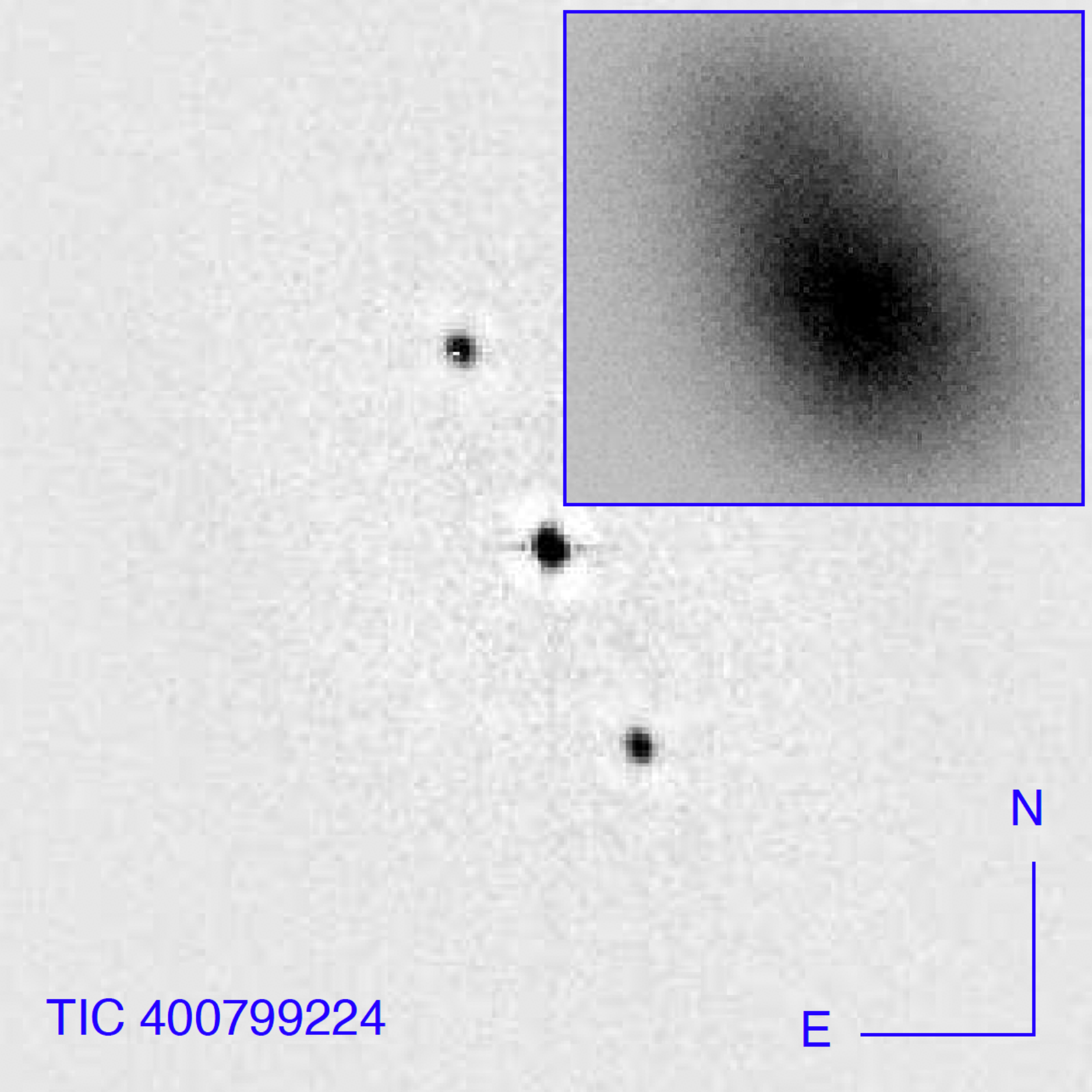}
    \caption{The speckle auto-correlation function of TIC 400799224 in the I filter. The field size is 3\farcs15. The insert shows a semi-resolved long-exposure image produced from the same data cube by re-centering.}
   \label{fig:ccf}
\end{figure}

 TIC 400799224 was observed by the speckle camera at the 4.1 m Southern Astrophysical Research Telescope (SOAR) on 26 February 2021 and resolved as a 0\farcs62 pair with components of comparable flux. The speckle instrument and the data processing are described in \citet{speckle}. It was determined that the photometry and the spectra presented above refer to the sum of two stars (the CHIRON fiber diameter is 2\farcs7). The resulting speckle measurements are shown in Table~\ref{tbl:speckle}. Overall, four data cubes in the I band (with  and  without  binning) and two  data cubes in the V band were recorded. Figure \ref{fig:ccf} illustrates the resolution.
 
 \begin{table}
\centering
\caption{Relative position and photometry of the companion star identified with speckle imaging.}
\begin{tabular}{ccc c c}
\hline
\hline
Date & P.A. & Sep. & $\Delta m$ & Filt.    \\  
(JY)  & (deg) & (arcsec) & (mag) & \\  
\hline
 2021.1573 & 23.91 & 0.6203  &  0.74 &  I \\
 2021.1573 & 24.16 & 0.6206  &  0.82 &  V \\
\hline 
\label{tbl:speckle} 
\end{tabular}
\end{table} %Table 4
 
 The two stars resolved by speckle are most likely mutually bound because the probability of having two unrelated stars of similar brightness so close in the sky is very small. We note that a magnitude difference of $\Delta$G $\sim$1 mag and angular separation of $0\farcs62$ is right at the edge of {\em Gaia}'s contrast sensitivity (see e.g. \citet{2019A&A...621A..86B}). This system is not present in {\em Gaia} DR2, and its absence could be related to its double nature, although DR2 normally provides positions and photometry for such double stars.  The large excess astrometric noise in {\em Gaia} EDR3, where the system {\em is} identified, is also likely caused by the superposition of two stars. To evaluate the probability that the two sources detected in SOAR data are unrelated field stars that just happen to be so close to each other on the sky by coincidence, we use the {\em Gaia} EDR3 catalog as follows.

A query of the catalog shows that there are 13 sources within $\Delta$G $\sim$1 mag of TIC 400799224 inside a $15 \times 15$ {\em TESS} pixel array centered on the target. None of these sources is within an arcmin of the target (corresponding to projected separation of $\sim$28,000 au), or have comparable parallax or proper motion, which indicates that they are unlikely to be gravitationally-bound to TIC 400799224. Assuming that this is the representative density for such sources in the field of view, it follows that there are 13/225 = 0.058 such sources per {\em TESS} pixel, or $\sim$0.00013 sources per square arcsec. Thus the probability that there is a random field star with $\Delta$G $\sim$1 mag from TIC 400799224 and at an angular separation of ~$0\farcs62$ is only $\sim$0.0036.

 % Fig. 13

\subsection{Spectroscopic Analysis}
\label{sec:spectra}

The average parameters of the CCF  are: amplitude $a =0.107$, rms width $\sigma =13.85$ \kms, estimated projected rotation speed $V \sin i = 24.5$ \kms, equivalent width (product of $a \sigma$) equal to 1.49 \kms. According to the calibration of CHIRON CCFs on synthetic spectra, a star of solar metallicity and $T \simeq 6400$ K would yield such a CCF.

Despite the low SNR, the spectra show the lithium doublet at 6707.7\,\AA \, with an equivalent width of 154$\pm$14 m\AA, and the dispersion is 13.8 \kms (same as the CCF width by coincidence: the CCF is broadened by the mask,
while the Li line width is due to blending of its two components). The presence of Lithium will be important for the analysis in Section \ref{sec:sed}, as it leads us to strongly suspect that the system is young. No emission in the H$\alpha$ line is seen.

\begin{figure*}
    \centering
    \includegraphics[width=1.0\linewidth]{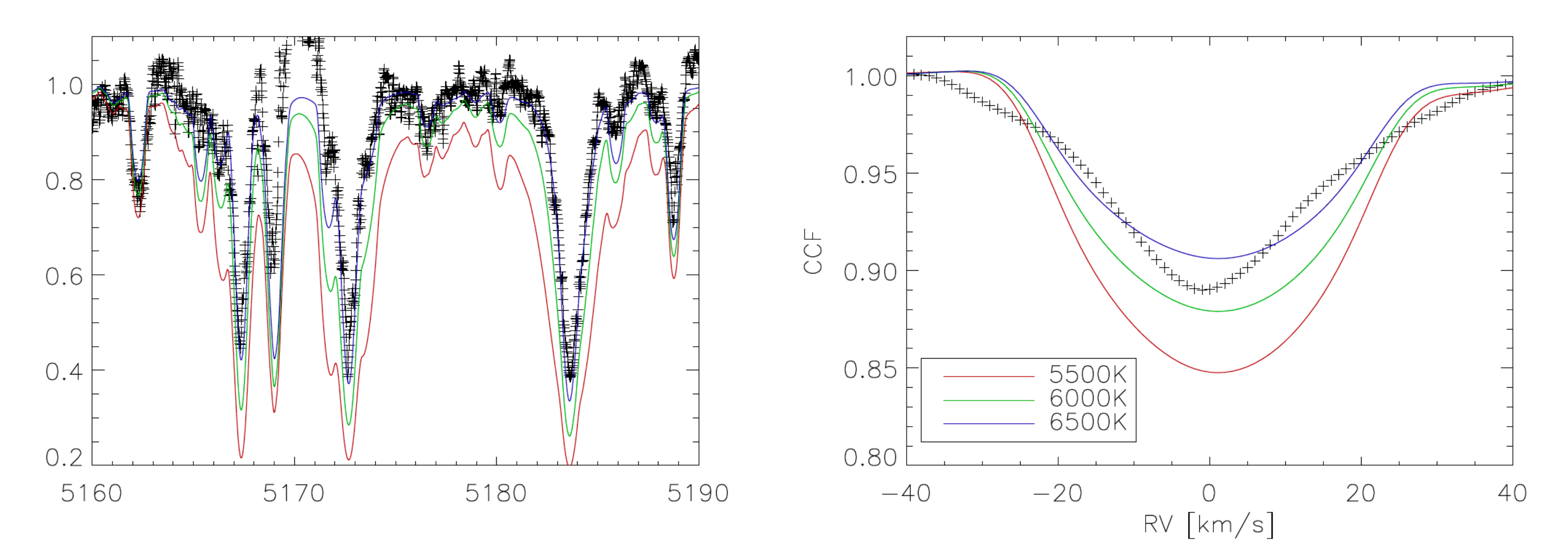}
    \caption{Comparison of the average spectrum of TIC 400799224 (crosses) with the broadened solar-metallicity synthetic spectra with temperatures of 5500, 6000, and 6500 K (red, green, blue lines, respectively). ({\em Left panel}) region around Mg II doublet, ({\em Right panel}) CCFs with solar mask.}
   \label{fig:spectrum}
\end{figure*}

Using known RVs and barycentric corrections, the spectra were order-merged with continuum normalization, shifted to zero velocity, and averaged with weights proportional to fluxes. The first 10-minute spectrum was discarded, as it gives an abnormal CCF, probably due to an un-removed cosmic ray spike. The average spectrum was compared to synthetic spectra with temperatures of 5500, 6000, 6500 K, solar metallicity, and gravity log g = 4.4. The synthetic spectra were rotationally broadened by $V \sin i = 25$ \kms.  The best match between the observed and synthetic spectra, shown in Figure \ref{fig:spectrum}, is found for $T = 6500$ K. The sodium-D doublet in TIC 400799224 is deeper than in the synthetic spectra, possibly because of additional interstellar absorption.

 % Fig. 14

\begin{figure}
    \centering
    \includegraphics[width=1.0\linewidth]{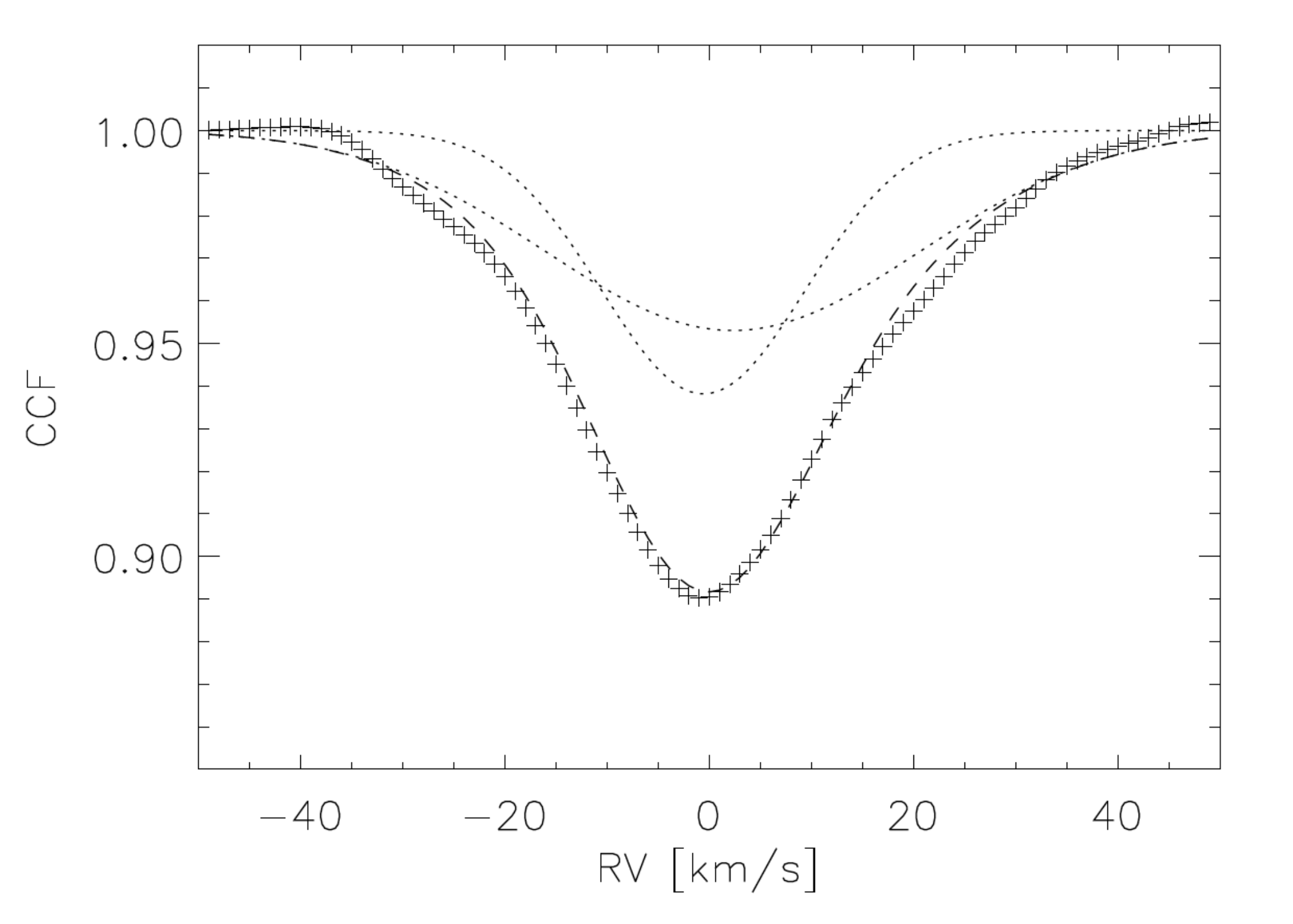}
    \caption{Approximation of the composite CCF (crosses) by the sum of two Gaussians (dashed line). The individual Gaussians are plotted by dotted lines.}
   \label{fig:composite}
\end{figure}

The CCF of TIC 400799224 looks like a superposition of narrow and broad profiles, while the CCFs of the synthetic spectra are closer to a Gaussian. Given that the spectra actually belong to two stars, this is natural. We tentatively identify the two components seen in the spectra with the two stars resolved with SOAR.  Approximating individual CCFs by the sums of two Gaussians does not provide evidence for RV variability of either the broad or narrow component, and the separation of the two heavily blended components is not reliable. Assuming that the RVs of both stars were constant during our observations, the CCF of the {\em average} spectrum was approximated by the double Gaussian. This works better than fitting the individual CCFs, as shown in Figure \ref{fig:composite}, with parameters of the individual Gaussians shown in Table \ref{tbl:gaussians}.

 % Fig. 15

\begin{table}
\centering
\caption{Parameters of Two Gaussians Approximating the CCF.}
\begin{tabular}{ccc c c}
\hline
\hline
RV & $a$ & $\sigma$  & $a \sigma$  & $V\sin i$    \\ 
(\kms) &   & (\kms) & (\kms) & (\kms) \\
%(JY)  & (deg) & (arcsec) & (mag) & \\  
\hline
 2.30 & 0.047 & 18.27  &  0.86 &  34.3 \\
 \hline
$-$0.59 & 0.062 & 9.97 & 0.62 &  16.1 \\
\hline 
\label{tbl:gaussians} 
\end{tabular} % Table 5

{Notes: $a$ -- amplitude, $\sigma$ -- rms width, $V \sin i$ -- projected rotation speed estimated from $\sigma$. The areas of the Gaussians are proportional to $a \sigma$.}

\end{table} %table 3

There is a 3 \kms \,difference between the RVs of the two components. An RV difference of a few \kms \,is expected for a binary with a separation of a few hundred au, supporting  the conclusion that the two stars form a bound system.  The ratio of the areas of the two CCF components  corresponds to $\Delta$m = 0.36 mag. Both the CCF area and the rotation of main-sequence stars depend on their spectral type and change rapidly around F5V owing to transition from radiative to convective envelopes. The brighter star in the TIC 400799224 pair is likely hotter and a faster rotator than the fainter star. Therefore, the CCF areas differ less than the actual magnitude difference measured at SOAR, $\simeq 0.8$ mag in V.

\section{Analysis of the SED Data}
\label{sec:sed}

We fit the 14 available spectral energy distribution (SED) points from VizieR \citep{vizier} and Galex \citep{2005ApJ...619L...1M} with a five-parameter model: the masses of the two resolved stars ($M_1$ and $M_2$), the system age, interstellar extinction $A_V$, and the temperature of a dust component $T_{d}$.  The fit was carried out with a similar MCMC code to the one used previously to fit multi-stellar systems (see, e.g., \citealt{Powell2021,2021arXiv210512586K}).  We take the prior on the distance to be  $725 \pm 140$ pc from {\em Gaia}. The two stars are taken to each be single stars, and to be coeval with no mass having been exchanged between them.  With this assumption, we are able to use {\tt MESA} stellar evolution tracks (see \citealt{paxton11}; \citealt{paxton15}; \citealt{paxton19}; \citealt{dotter16}; \citealt{choi16}) to reduce the number of free parameters in the fit.  Otherwise, we would need to fit independently for two masses, two radii, and two values of $T_{\rm eff}$.  There is insufficient information in the single SED to fit for all those independent parameters in a meaningful way.  Also used to constrain the fitted parameters were the SOAR intensity ratios in the V and I bands.

\begin{figure*}
    \centering
    \includegraphics[width=0.48\linewidth]{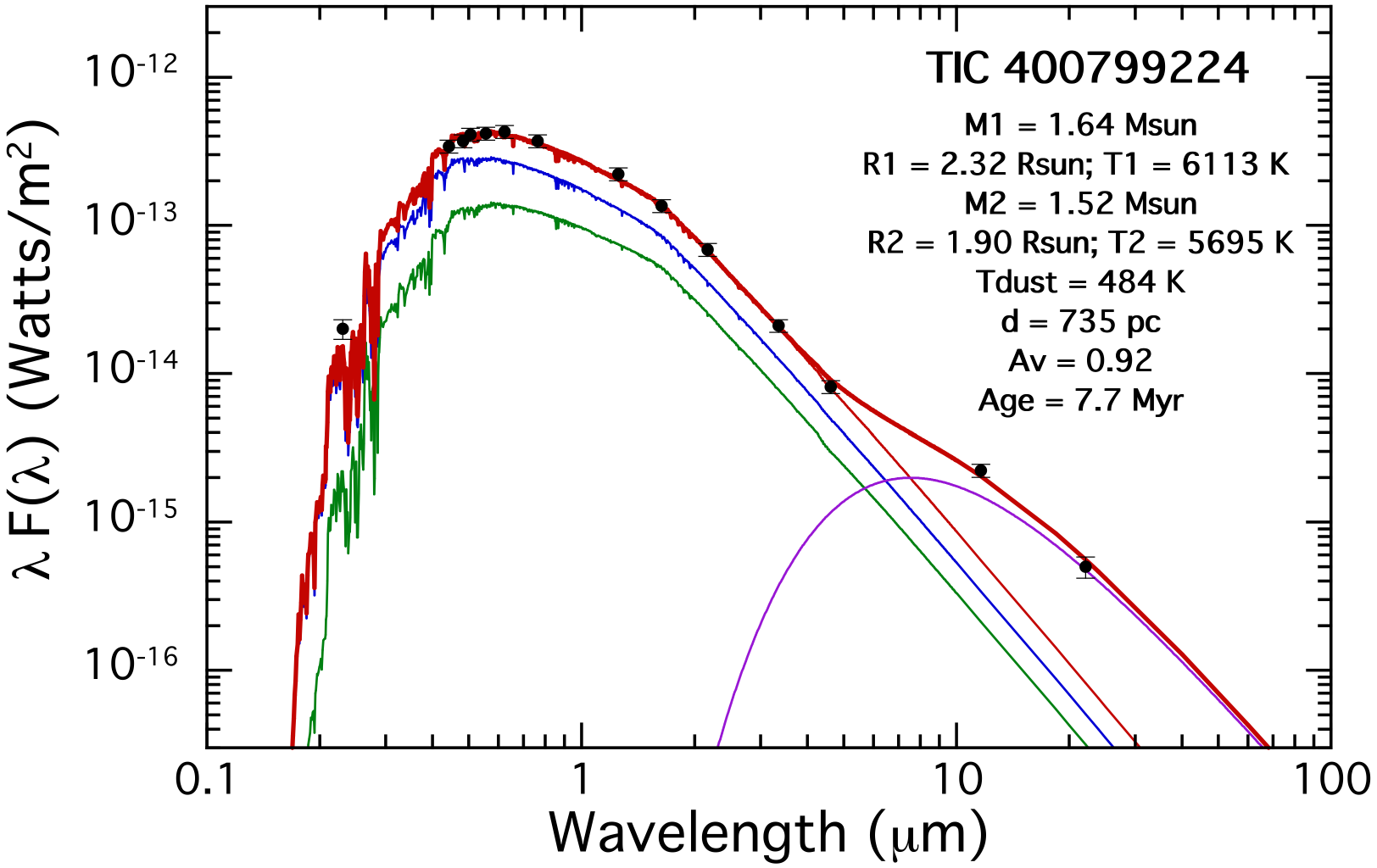}  
    \includegraphics[width=0.48\linewidth]{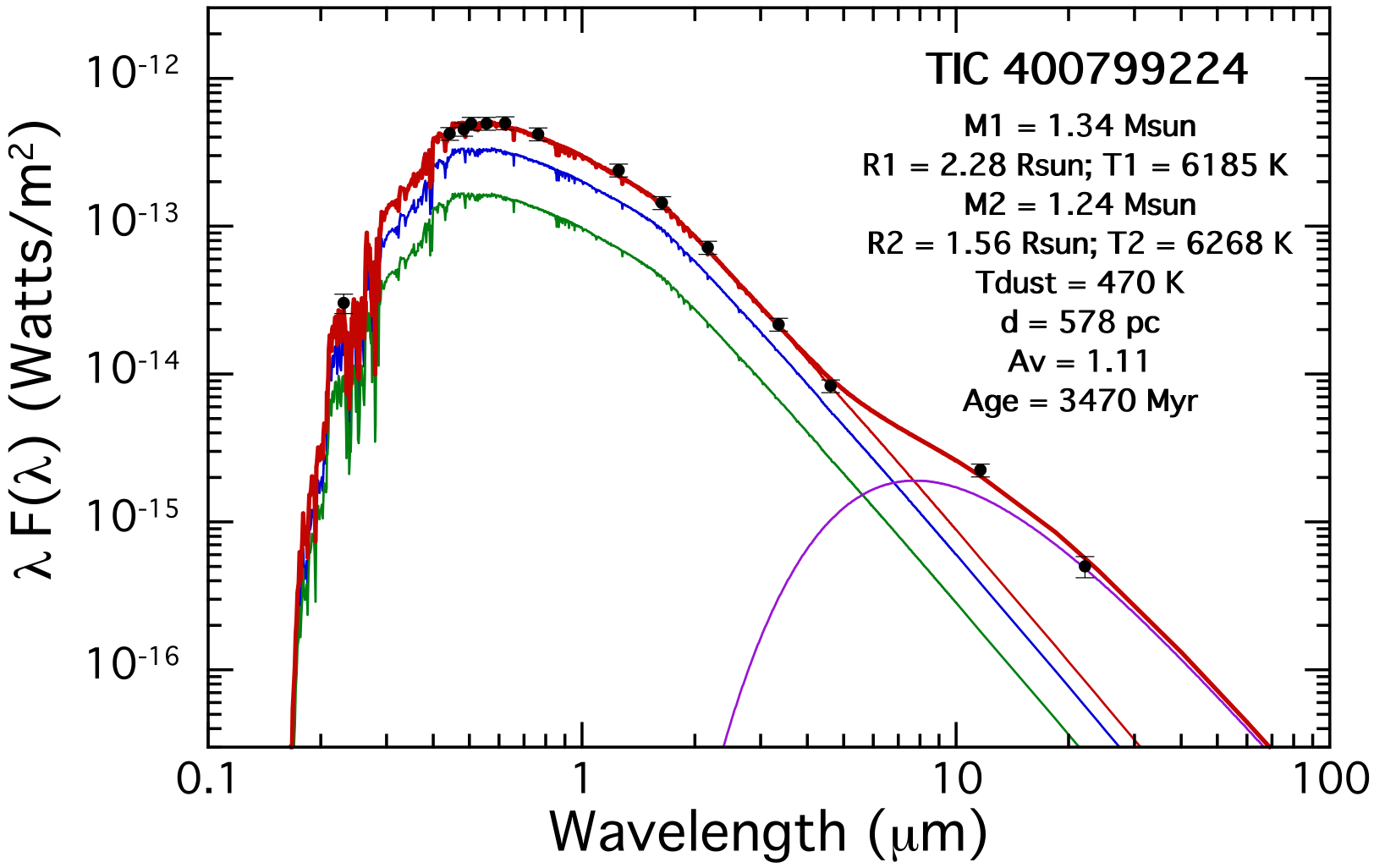}
    \caption{SED fits to 14 spectral flux measurements in different bands, including the Galex NUV. {\em Left panel}: the pre-MS solutions; {\em Right panel}: the post-MS solutions. The lighter green, blue, and red curves are the model spectra for star 2, star 1, and the sum of stars 1 and 2, respectively; the heavy red curve is the model spectrum including both stars and the inferred black body bump from the cool dust component.  Of note, the fit to the NUV point is not quite as good for the pre-MS solution, likely due to the slightly lower temperatures for the two stars. The parameter values listed on the plot are the best fit values which do not necessarily coincide exactly with the median values of the posterior distributions from Table \ref{tbl:stars}.}
   \label{fig:sed}
\end{figure*}

Two independent solutions were found with the SED fit, yielding vastly different system ages (order of Myr vs Gyr).  One solution involves a pair of pre-MS stars, while the other has one of the stars substantially evolved off the MS, with both solutions showing a dust signature at $\sim$450 K. Figure \ref{fig:sed}, left panel, shows the results of the SED fit for the pre-MS solution, while the right panel shows the SED fit for the post-MS solution.  The only significant difference between the fits lies in the Galex NUV point, where the post-MS solution is somewhat better.  We interpret this difference to be a consequence of the higher effective temperatures of the post-MS stars.

\begin{figure}
    \centering
    \includegraphics[width=1.0\linewidth]{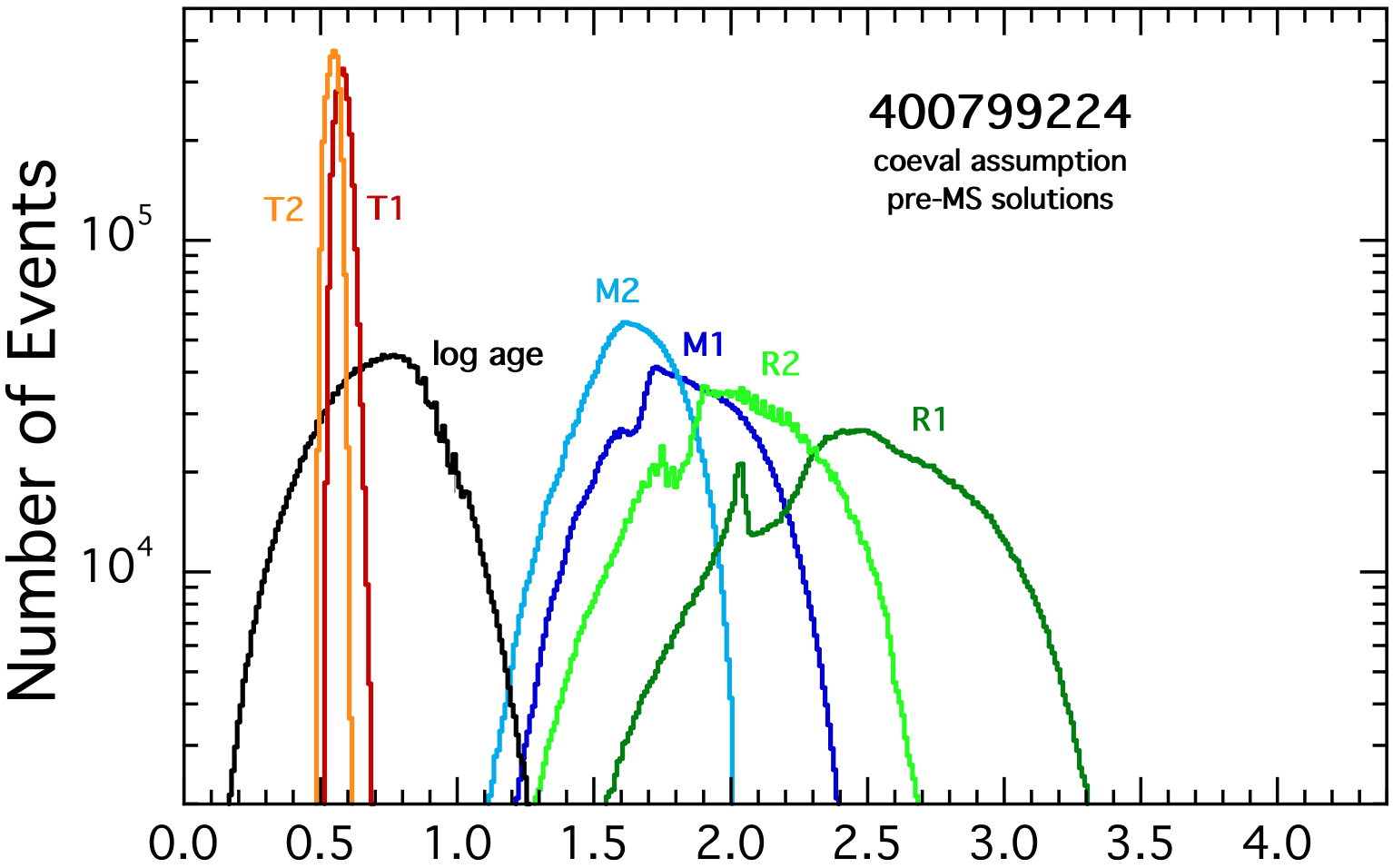}
    \includegraphics[width=1.0\linewidth]{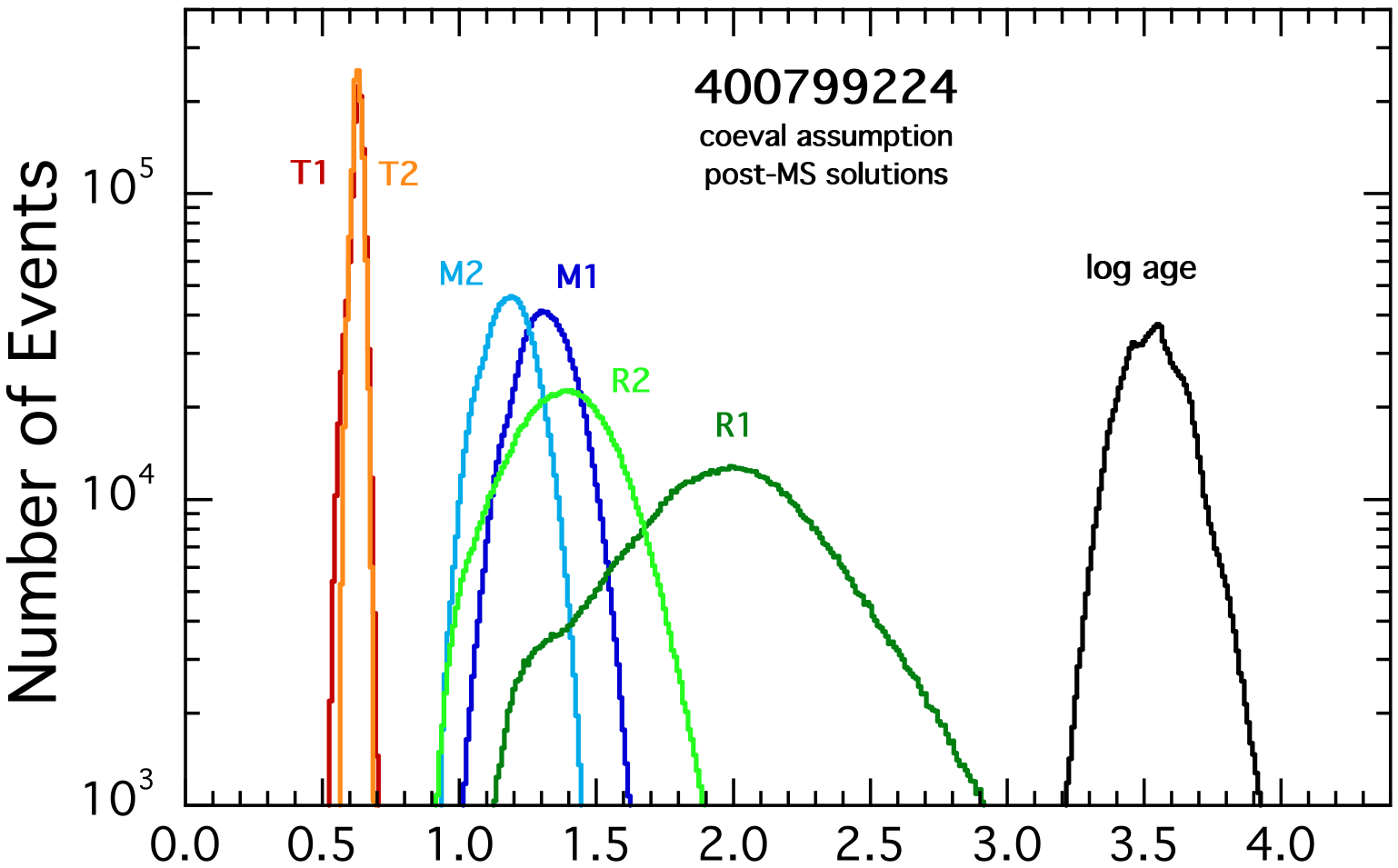}
    \caption{Posterior parameter distributions for the masses, radii, and $T_{\rm eff}$ for the pre-MS ({\em top panel}) and post-MS ({\em bottom panel}) solutions. Units on the x-axis are 10$^4$ K, log(Myr), and solar units for their respective components.}
   \label{fig:global_params}
\end{figure}

We show the posterior distributions of the fitted parameters from the same MCMC procedure that produced the SED fits in Figure \ref{fig:global_params}. In the post-MS solution, the star we have called the primary (i.e., the more massive and the luminous one) seems to have evolved substantially away from the zero age main-sequence, except for cases where the distances to the source are closer than twice the {\em Gaia} uncertainty.  For the most part, the two stars are normal Sun-like stars of masses 1.31 M$_\odot$ and 1.18 M$_\odot$.  By contrast, the two stars in the pre-MS solution are about 0.4 \,M$_\odot$ more massive than the corresponding two stars in the post-MS solution, and are evolving toward the main sequency where they will become somewhat smaller and considerably hotter.    The set of properties of the two stars comprising TIC 400799224 is given in Table \ref{tbl:stars} for both the pre- and post-MS solutions.

\begin{figure}
    \centering
    \includegraphics[width=1.0\linewidth]{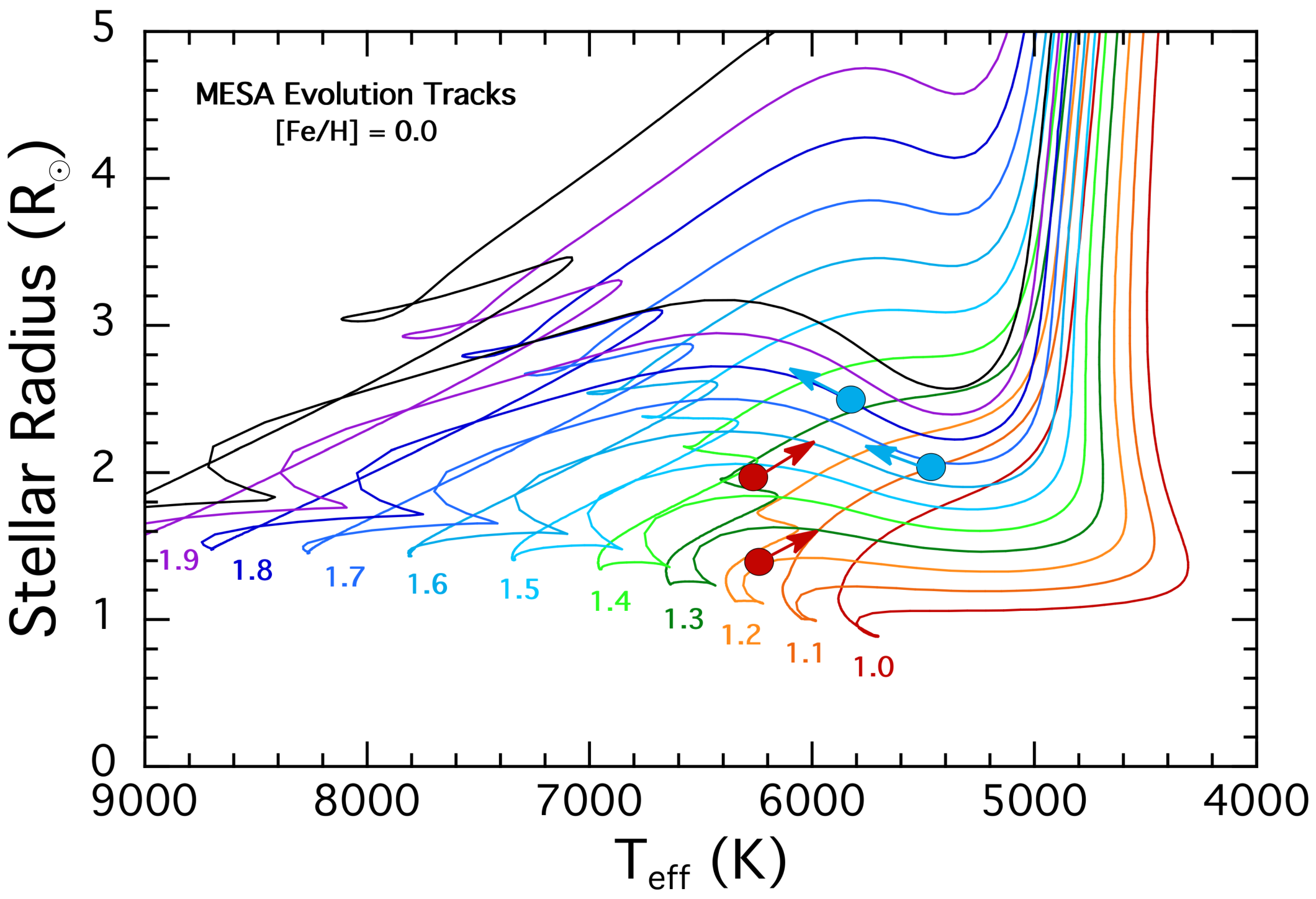}
    \caption{{\tt MESA} tracks covering both pre-MS evolution toward the main sequence and subsequent post-MS evolution away from the main sequence. Each color represents a different stellar mass ranging from 1 to 2\,M$_\odot$ in steps of 0.1\,M$_\odot$. The two filled red circles represent the two stars in the post-MS solution while the filled blue circles are the locations of the two stars in pre-MS solution.  In both cases, the arrows give a sense of the direction of evolution.}
   \label{fig:mist}
\end{figure}

Finally, in Fig.~\ref{fig:mist} we show the locations of the two stars superposed on {\tt MESA} evolution tracks in the radius--$T_{\rm eff}$ plane for both the pre-MS solutions (filled blue circles) and post-MS solutions (filled red circles).  This serves to illustrate visually how two stars descending along coeval tracks to the main-sequence versus two stars on and leaving the main sequence can both match the observed SED for the combined light from the two stars in the system.

\begin{table*}
\centering
\caption{MCMC Fitted Parameters of the Two Stars Comprising TIC 400799224$^{a,b,c}$.}
\begin{tabular}{ccccc}
\hline
 & \multicolumn{2}{c}{Pre-MS Solution} & \multicolumn{2}{c}{Post-MS Solution} \\

Parameter & Star 1 & Star 2 & Star 1 & Star 2  \\ 
\hline
Mass [M$_\odot$]               &             $1.79 \pm 0.24$  &  $1.61 \pm 0.18$      &             $1.31 \pm 0.11$   &   $1.18 \pm 0.10$\\

Radius [R$_\odot$]		&	   $2.47 \pm 0.40$  & $2.02 \pm 0.30$       &            $1.98 \pm 0.40$    &  $1.38 \pm 0.20$\\

Teff  [K]            &                         $5829 \pm 349$  &   $5465 \pm 262$      &           $6250 \pm 265$   &   $6237 \pm 193$\\

Luminosity [L$_\odot$]     &               $6.32 \pm 2.10$     &      $3.25 \pm 1.03$       &            $5.35 \pm 2.04$  &   $2.64 \pm 1.02$\\

Age$^{d}$ [Myr]                    &                     $5.53 \pm 3.20$    &  $5.53 \pm 3.20$    &      $3351 \pm 1237$  &  $3351 \pm 1237$ \\

A$_v$                      &                      $0.84 \pm 0.20$    &     $0.84 \pm 0.20$       &         $0.97 \pm 0.15$   &  $0.97 \pm 0.15$\\

Distance [pc]            &                $787 \pm 107$       &  $787 \pm 107$              &          $ 687  \pm  117 $       &  $ 687  \pm 117 $\\

T$_{d}^{e}$   &   $462 \pm 127$  &   $462 \pm 127$  & $512 \pm 144$ & $512 \pm 144$ \\
\hline 
\label{tbl:stars} 
\end{tabular}

{Notes: (a) See Sect.~\ref{sec:sed} for details of the fit to the SED. (b) These are the median values of the posterior distributions and the cited uncertainties are the rms values. (c) We note that, while the post-MS solution is quite robust, the pre-MS solution can vary widely in its parameters depending on the prior distributions for distance and $A_V$.  The pre-MS solution presented here is constrained by the distance measurement from Gaia and an interstellar extinction limited to $A_V < 0.9$.  For larger, but still somewhat plausible, values of distances (up to 1050 pc) and $A_V$ (up to 1.3), the primary star can become larger (up to 3 R$_\odot$) and hotter (up to 6600 K). (d) The ages of both stars are taken to be the same via the coeval assumption of the SED fit. (e) Temperature of the dust}.   

\end{table*} % Table 6

 % Fig. 18

 % Fig. 19

 % Fig. 20

The pre-MS and post-MS fits offer substantially different interpretations of the system. If the system is indeed old, as found to be 3.4 Gyr in the post-MS solution, then this could be difficult to reconcile with the CHIRON measurement which shows the presence of lithium (see Sect.~\ref{sec:spectra}).  However, the larger primary could still conceivably be a Li-rich giant, which \citet{2021MNRAS.505.5340M} found make up 1.2\% of a sample of giant stars.  On the other hand, the presence of lithium more naturally favors the pre-MS solution, as it is characteristic of young stars.  Young stars could also be expected to host a debris-rich environment containing the occulting object.  The SED fit to the Galex NUV point, however, does still slightly favor the post-MS solution involving older stars, but this is the only evidence supporting this interpretation.

In order to gather more information to robustly choose between the pre-MS and post-MS solutions, we will now examine the celestial region containing TIC 400799244.

\section{Stellar Region Analysis}
\label{sec:region}

The pre-MS solution would be strongly supported by the presence of TIC 400799224 in a stellar formation region or common kinematic group.  So, we examined the {\em Gaia} data in order to check the kinematic properties and the parallax of other stars in the vicinity of TIC 400799224

\begin{figure}
    \centering
    \includegraphics[width=1.0\linewidth]{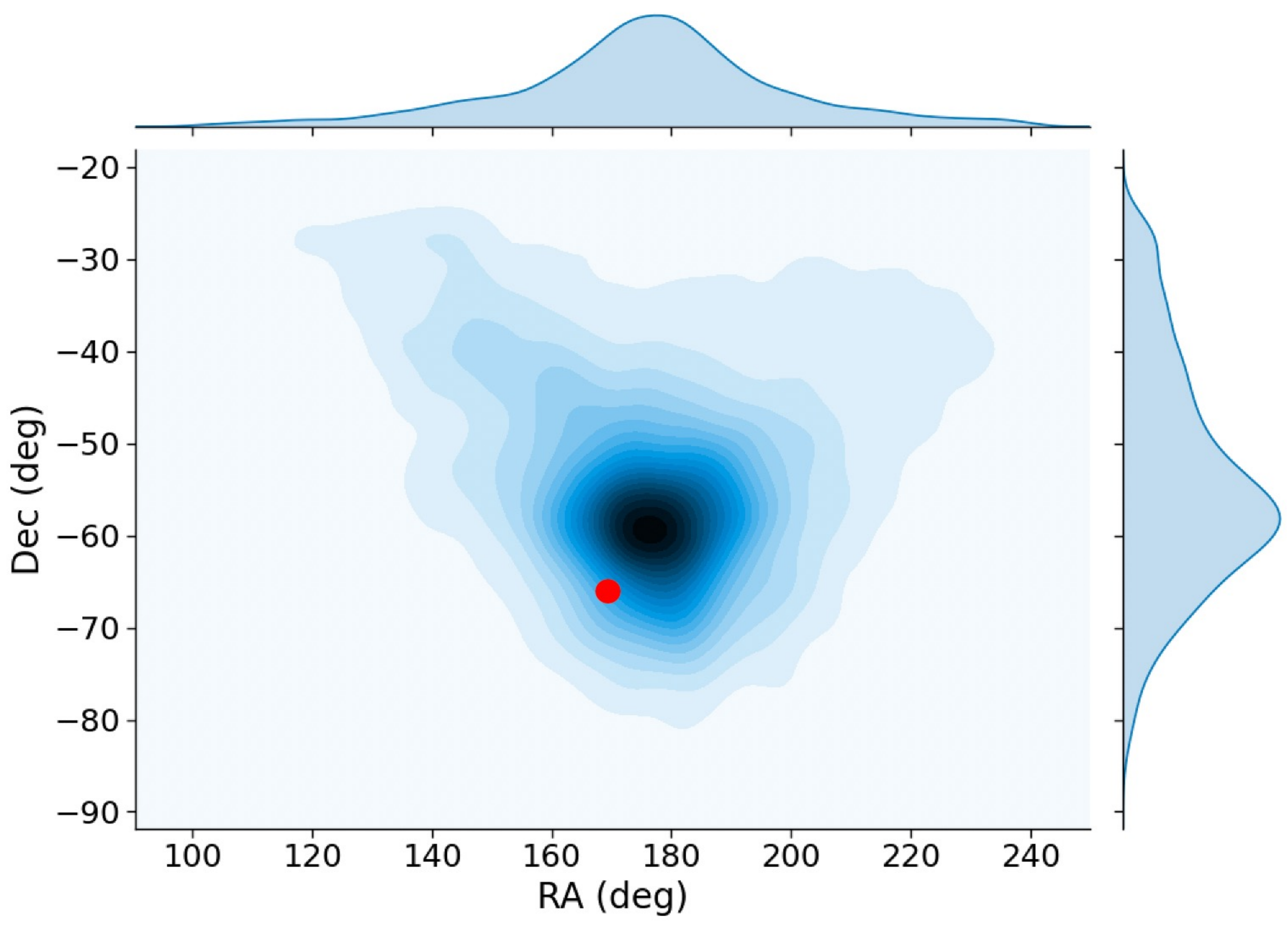}
    \caption{Joint density plot of the RA/Dec coordinates of all objects in the sample within 3$\sigma$ of the parallax, PMRA, and PMDEC of TIC 400799224.  Each level represents and additional 5\% of the total number of objects.  The red dot shows the coordinates of TIC 400799224.}
   \label{fig:kde}
\end{figure}

We used the {\em Gaia} EDR3 archive search tool, available at \url{https://gea.esac.esa.int/archive/}, to extract every source in the right ascension range (100$^{\circ}$, 240$^{\circ}$) and the declination range (-85$^{\circ}$,-25$^{\circ}$), with a {\em Gaia} magnitude $\leq$ 13.0 and parallax $\geq$ 0.0 mas, yielding a total of 1,546,347 objects.  Of these, 5,132 objects have a proper motion in right ascension (PMRA), proper motion in declination (PMDEC), and parallax all within 3$\sigma$ of TIC 400799224.  The coordinates of the 5,132 objects are shown as a joint density plot in Figure \ref{fig:kde}.  This plot indicates that TIC 400799224 may be part of a kinematic group.  The group is rather large to be considered a young stellar association, however, and we have no indication of the age, so we do not accept this to be a certain.

 % Fig 19

\begin{figure}
    \centering
    \includegraphics[width=1.0\linewidth]{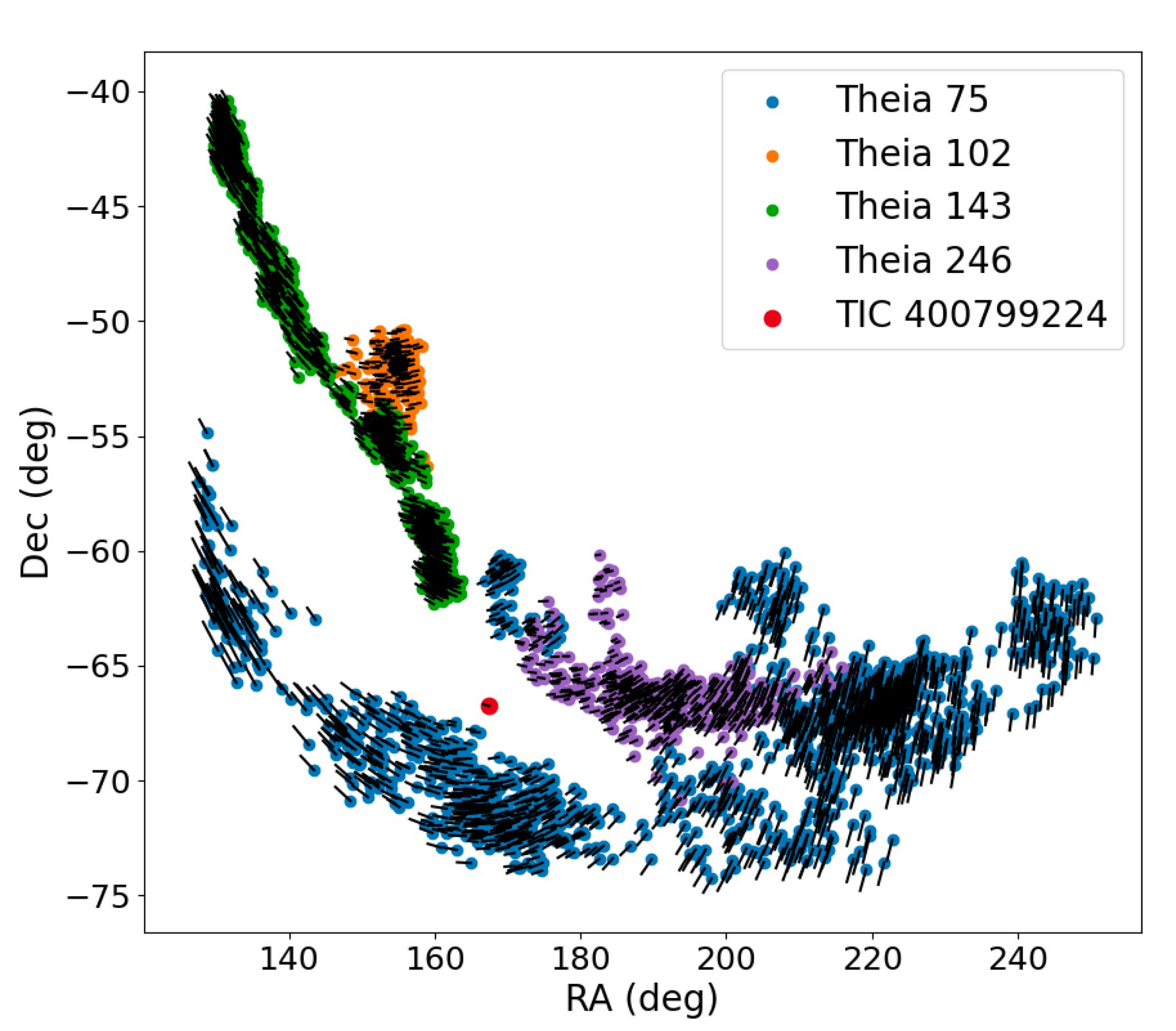}
    \caption{Comparison of TIC 400799224 with all groups (or ``Theias'') identified by \citet{2019AJ....158..122K} that are within 20 degrees, less than 100 Myr old, and have members within 3$\sigma$ of the parallax, PMRA, and PMDEC of TIC 400799224.  The black lines are scaled representations of the proper motion vectors.}
   \label{fig:theia}
\end{figure} % Fig 20

Separately, we checked the celestial region within 20 degrees of TIC 400799224 for young stellar groups identified in \citet{2019AJ....158..122K}.  The four groups (called `Theias') with an age less than 100 Myr and containing members within 3$\sigma$ of the parallax, PMRA, and PMDEC of TIC 400799224 are shown in Figure \ref{fig:theia}, with scaled proper motion vectors represented as the black line originating at each point.  Theias 75, 102, 143, and 246 have ages of 30, 32, 56, and 81 Myr, respectively.  As we mentioned previously, TIC 400799224 was not in the {\em Gaia} DR2 catalog, on which the \citet{2019AJ....158..122K} analysis was based, so it would not have been part of any of the authors' identified groupings.  Although TIC 400799224 is not an obvious fit with any of these groups, it could have been expelled at some point.

In searching the region for stars with similar properties using SIMBAD \citep{2000A&AS..143....9W}, we found a nearby source at $\sim$85$\arcsec$, Gaia DR2 5238414453810235904, identified by \citet{2018A&A...620A.172Z} as a young stellar object candidate.  With PMRA = -13.763 mas yr$^{-1}$ and PMDEC = 2.194 mas yr$^{-1}$, the proper motions of this source are quite similar to those of TIC 400799224.

Although the adoption of either the pre-MS or post-MS solution to the SED fit in no way affects our confidence in the existence of the occulting object, we conclude that the presence of lithium, a debris-rich environment, and possible association with larger kinematic groups all lead strongly toward the adoption of the pre-MS solution.  Further observations, however, should help to better constrain the fit, leading to a more robust conclusion about the age and other parameters of this system.  Additional observations and future study of the stellar system could also help to further refine the hypotheses concerning the processes driving the dust emission, on which we elaborate in the next section.

\section{Nature of the Occulter}
\label{sec:occulter}

Thus far, we have established that one of the two stars in TIC 400799224 has a 19.77-day periodicity, likely due to dusty emissions from an orbiting body of that same period.  While the occultations appear to have a strict underlying period, they are erratic in shape, depth, and duration.  In particular, the occultations seem to be present at a detectable level from the ground only $\sim$1/3rd --1/5th of the time.  The most extreme of the occultations has a depth of $\sim$25\% and a duration of $\sim$2 days. However, given that there are two stars in the system, the actual occultation depth, based on the photometry from SOAR indicating a ratio of $\sim$2 between the luminosities, could be up to 37\% or 75\% depending on which star in the binary is the true host.

The periodic, but erratically occurring occultations in this object are reminiscent of (a) the disintegrating exoplanets (KIC 12557548, KOI 2700b, K2 22b) and (b) the dusty transits of white dwarfs (WD 1145+017, ZTF J0328-1219, ZTF J0139+5245).  The properties of these six objects are given in Table \ref{tbl:dusty} and are compared to those of TIC 400799224.

Among all seven objects from Table \ref{tbl:dusty}, the host star in TIC 400799224 has, by far, the largest luminosity and, we note for later reference, the highest values of $\beta$, the ratio of radiation pressure forces on a dust grain to the corresponding gravitational force.  In all six of the comparison objects, the occultations have been attributed to dust rather than solid bodies, while the sources of the dust are taken to be orbiting bodies ranging from asteroids to lunar size.

\subsection{Dust Required for the Occultations}
\label{sec:dust_required}
If we assume that the occultations in TIC 400799224 are due to dust, then we can estimate the minimum amount of dust required to block 37\% - 75\% of the light from the host star.  Consider a uniform layer of dust with area
\begin{equation}
A_{\rm dust} \simeq 4 \pi d h f \tau
\label{eqn:area}
\end{equation}
where $d$ is the orbital radius of the dust emitting body, $h$ is the height of the dust perpendicular to the orbital plane, $f$ is the fraction of the orbit over which the occultation is observed, and $\tau$ is the optical depth of the dust in the visible band.  In the optically thin limit, the total cross section of all the dust particles, $\sigma_{\rm tot}$,  for a given total mass in dust, $M_{\rm dust}$ is
\begin{equation}
\sigma_{\rm tot} \simeq N_{\rm grain} \sigma_{\rm grain} \simeq \frac{3 M_{\rm dust}\sigma_{\rm grain}}{4 \pi \rho_d s^3}
\label{eqn:sigma}
\end{equation}
where $N_{\rm grain}$ is the total number of dust grains, each of cross section $\sigma_{\rm grain}$, blocking light from the host star, $s$ is the mean effective size of a dust grain, and $\rho_d$ is the mean bulk density of the dust particles. We can equate Eqns.~(\ref{eqn:area}) and (\ref{eqn:sigma}) to find a general expression for the minimum required mass in dust:
\begin{equation}
M_{\rm dust} \gtrsim \frac{16 \pi \, d \, s \, h \, f \, \tau \, \rho_d}{3 (\sigma_{\rm grain}/\sigma_{\rm geom})}
\label{eqn:mdust1}
\end{equation}
where the grain cross section is now normalized to its geometric cross section.  Since the most efficient grain scattering cross sections per unit mass usually occur near fractional micron size particles, where $\sigma_{\rm grain}$ is still close to its geometric cross section in the visible band we find:
\begin{multline}
M_{\rm dust} \gtrsim 16 \, d \, s \, h \, \rho_d \, f \tau \\ ~\simeq 2-4 \times 10^{19}  \left(\frac{d}{33\,R_\odot}\right) \left(\frac{f\tau h/R_*}{0.03}\right)~g
\label{eqn:mdust2}
\end{multline}
and we have taken $s \simeq 0.2 \,\mu$m, $\rho_d \simeq 3$ g/cc, $h\tau \simeq 0.30 - 0.65 R_*$, and $f \simeq 0.15$.  This is a rather substantial amount of dust, e.g., equal to that of an asteroid of radius 10 km.  Given that the dust activity in this source seems to change dramatically on a timescale of about 100 days, if we use that as a proxy for its replenishment lifetime, then the rate at which dust is produced must be of the order of $\dot M \simeq 3 \times 10^{12}$ g s$^{-1}$ (5 M$_\oplus$ Gyr$^{-1}$).  If one dismantled the asteroid Ceres (R = 500 km) at this rate it would last for $\sim$8,000 years.

This raises the question of how such quantities of dust can be produced in the TIC 400799224 system. 

\subsection{Dust Production via Sublimation}

%EC
In the case of the disintegrating planets, dust production has been ascribed to a thermal (Parker) wind generated by surface irradiation and evaporation 
(\citealt{1960ApJ...132..821P}; \citealt{2012ApJ...752....1R}; \citealt{2013MNRAS.433.2294P}; \citealt{vanlieshout18}).  This interpretation is not viable for TIC 400799224. At a temperature $T = 1525$ K (Table \ref{tbl:dusty}), the vapor pressure of silicates is too low for any size object to yield the desired mass loss rate of $\sim10^{12}$ g s$^{-1}$. At fixed temperature, the highest rates of mass loss are obtained in the free-streaming limit, when the thermal speed of molecules $v_{\rm th}$ 
%$ \sim \sqrt{kT/(\mu m_{\rm H})}$ 
exceeds the surface escape velocity $v_{\rm esc}$ from the planet and gravity can be ignored. In this limit $\dot{M} \sim P_{\rm vap} R^2/v_{\rm th}$ where $P_{\rm vap}(T)$ is the equilibrium surface vapor pressure (which is exponentially sensitive to $T$ via the Clausius-Clapeyron relation) and $R$ is the body radius. For $T = 1525$ K, we have $v_{\rm th} \sim 1$ km s$^{-1}$ (assuming a mean molecular weight of $\mu = 30$ appropriate for a silicate gas), $P_{\rm vap} \sim 10^{-4}$ dynes cm$^{-2}$ (\citealt{2013MNRAS.433.2294P}, their figure 1 for olivine), and $R \sim 600$ km as given by the condition $v_{\rm esc} \sim v_{\rm th}$. Then $\dot{M} \sim 4 \times 10^{6}$ g s$^{-1}$. Considering larger $R$ only strengthens surface gravity and decreases $\dot{M}$ (see, e.g., figure 2 of \citealt{2013MNRAS.433.2294P}). Thus thermal mass loss from a single planet fails to explain the inferred $\dot{M}$ by 5--6 orders of magnitude.

\begin{figure}
\centering
\includegraphics[width=1.0\columnwidth]{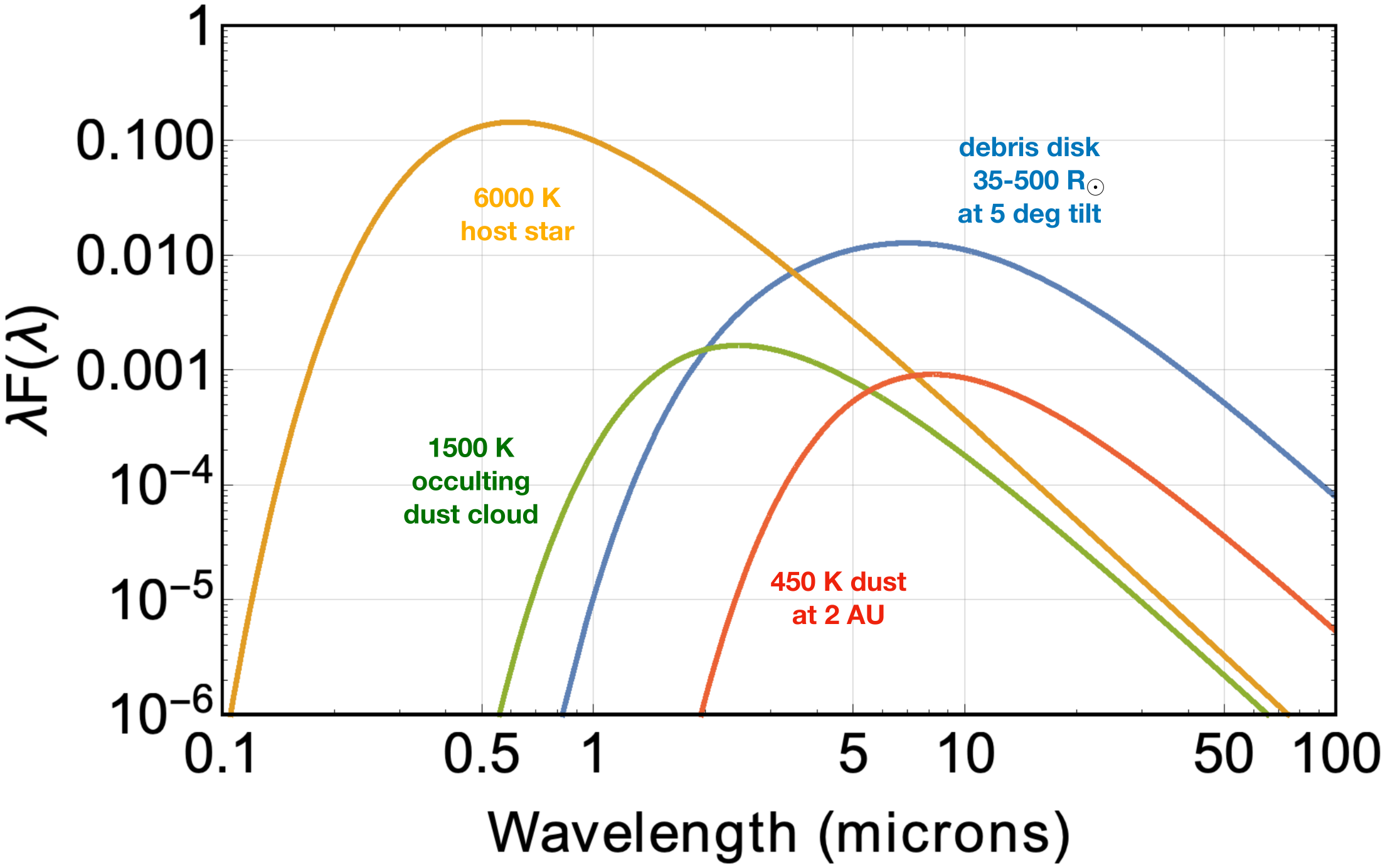}
\caption{Schematic comparison of black body spectra from a host star of $T_{\rm eff} = 6000$ K (orange), a cool dust component with $T_{\rm eq} = 450$ K at 2 AU (red), an occulting dust cloud with $T_{\rm eq} = 1500$ K and the size required to block 37\% of the light from the host star (green), and a debris disk that extends from 35 $R_\odot$ to 500 $R_\odot$ with a 5 degree tilt (blue). $\lambda F(\lambda)$ is in arbitrary units, but the comparisons among the four components are valid.}
\label{fig:composite_blackbody}
\end{figure} % Fig. 21

\subsection{NIR Evidence for Dust}

Regarding the detectability of the occulting dust cloud in the NIR, we first note that the expected equilibrium temperature of dust in a 19.77-d orbit of $\simeq1525$ K does not match the bump seen in the SED (Fig.~\ref{fig:sed}) near 10 $\mu$m.  In Fig.~\ref{fig:composite_blackbody} we show schematically a comparison among the $\lambda F(\lambda)$ curves for black body spectra from (i) a host star of $T_{\rm eff} = 6000$ K, (ii) a cool dust component with $T_{\rm eq} = 450$ K at 2 AU, (iii) an occulting dust cloud with $T_{\rm eq} = 1500$ K and the size required to block 37\% of the light from the host star, and (iv) a debris disk that extends from 35 $R_\odot$ to 500 $R_\odot$ with a 5 degree tilt.  We conclude from this that such an occulting dust cloud at 1500 K would not be readily visible in the SED, especially if it is only present about one quarter of the time and is thereby diluted in the average of the WISE measurements. On the other hand, there is a significant excess in the SED that corresponds to $\sim$450 K dust, which likely would be located at $\sim$2 AU from the host star. The debris disk (blue curve) in Figure \ref{fig:composite_blackbody} will be important in Section \ref{sec:collisions}.

\begin{figure*}[h!]
\vspace{0.0cm}
\begin{center}
\includegraphics[width=0.45 \textwidth]{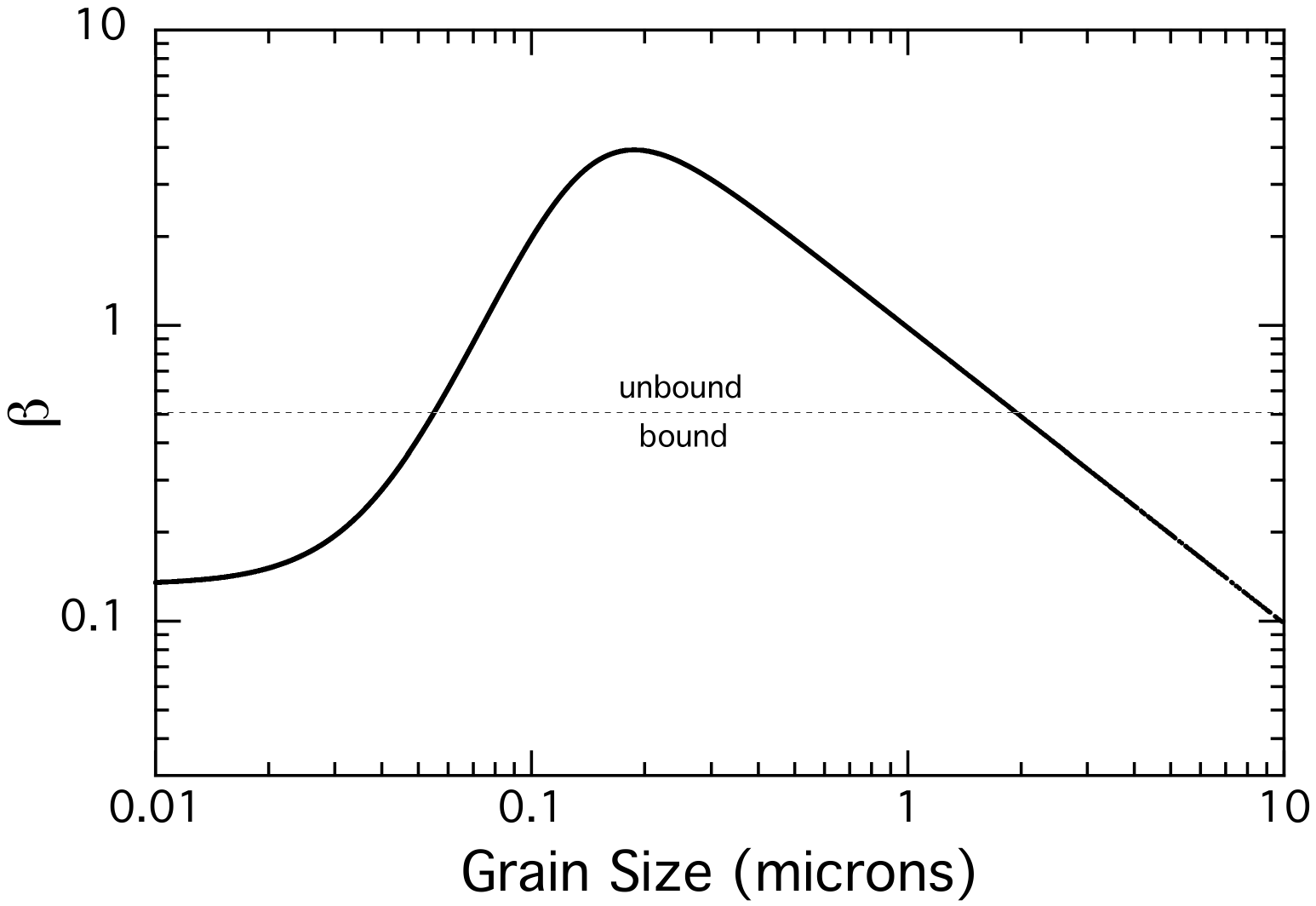}
\includegraphics[width=0.45 \textwidth]{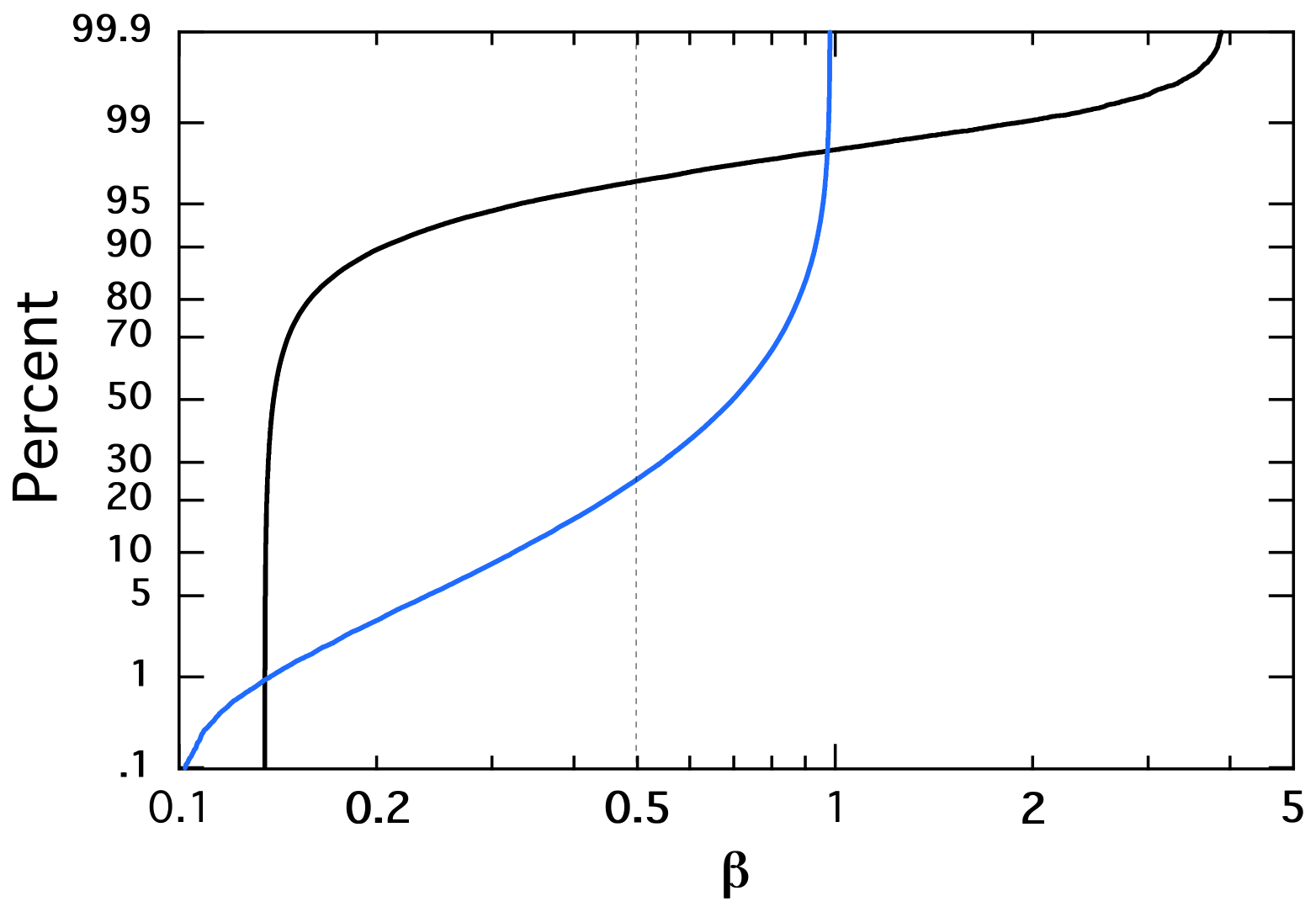}
\caption{{\em Left panel}: Dependence of the parameter $\beta$ on particle size.  Particles with $\beta >1/2$ are unbound from the system.  {\em Right panel}: cumulative distribution of the parameter $\beta$ given the properties of TIC 400799224 and the adopted $s^{-3}$ particle size distribution. Black curve - particle sizes range from 0.01 to 10 $\mu$m. Blue curve - particle sizes range from 1 to 10 $\mu$m.}
\label{fig:beta}
\end{center}
\end{figure*} % Fig. 22

\begin{figure*}[h]
\vspace{0.0cm}
\begin{center}
\includegraphics[width=0.4 \textwidth]{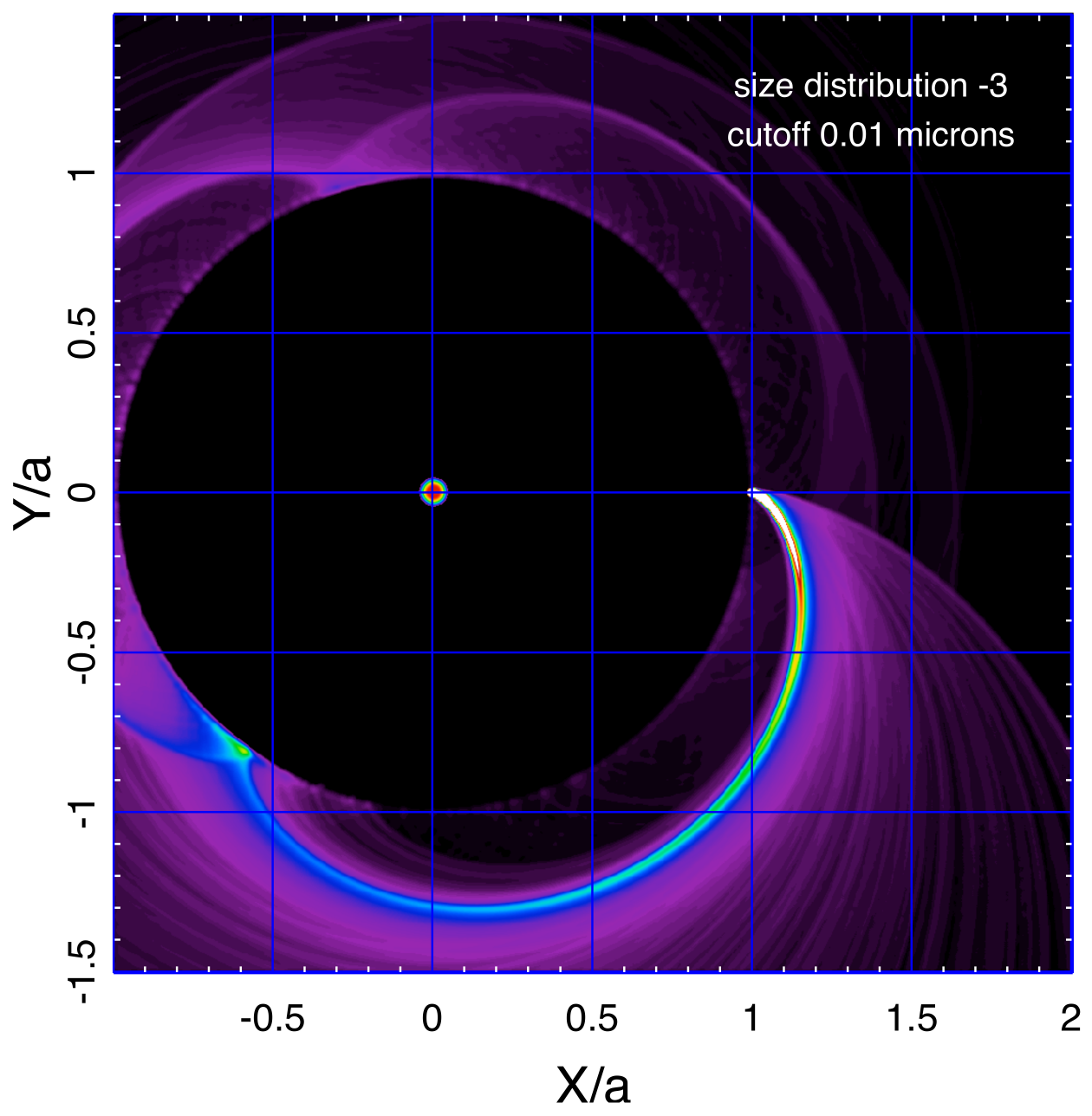}
\includegraphics[width=0.4 \textwidth]{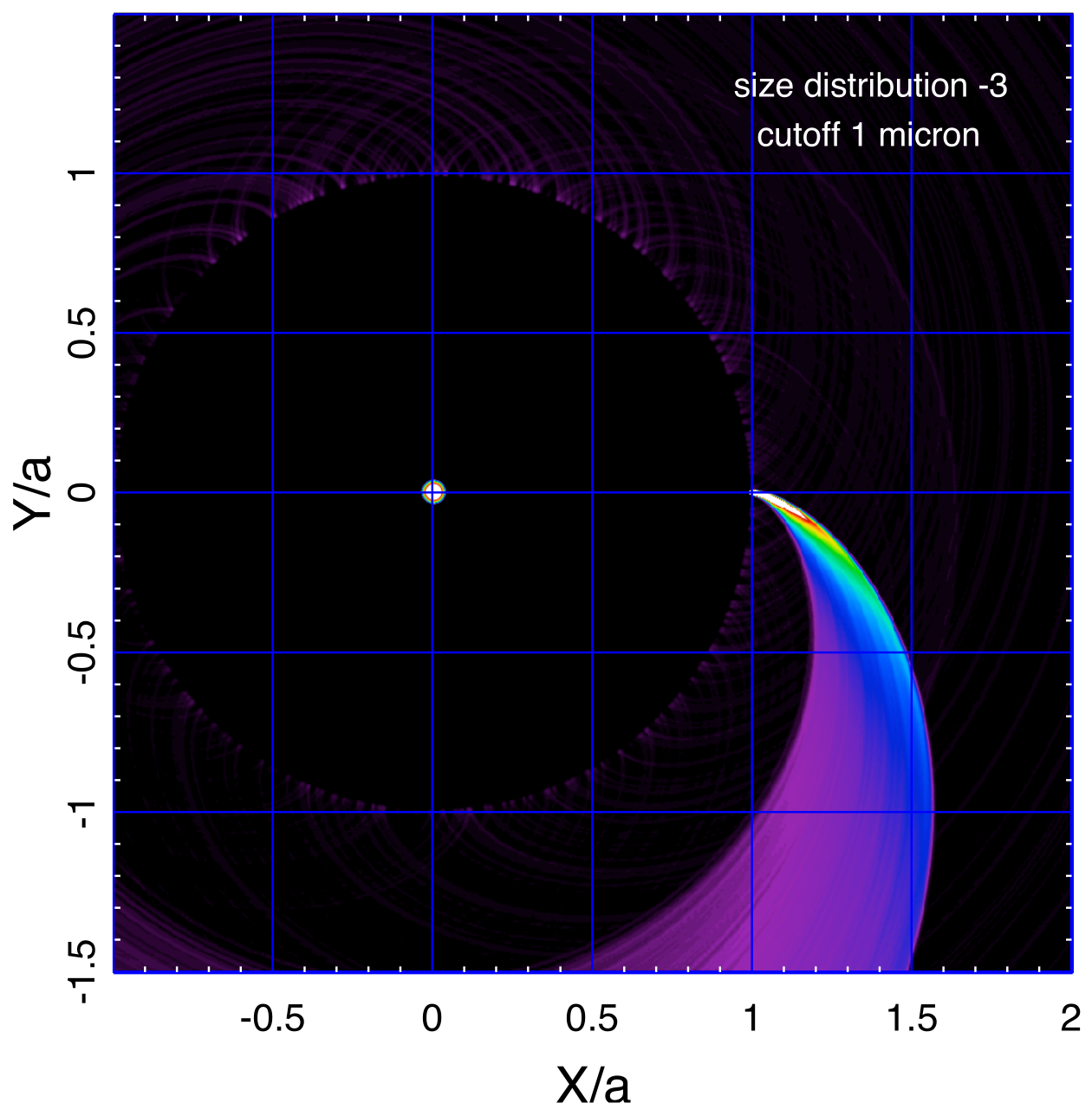}
\caption{Model dust tails for the object hypothesized to be orbiting TIC 400799224. Dust particles with a power-law size distribution of slope $-3$ are ejected from the orbiting body.  The ejection directions are uniform within a 30$^\circ$ cone centered on the host star. {\em Left panel}: The grain sizes range from 0.01 to 10 $\mu$m. {\em Right panel}: Same as for the left panel except that no grains below 1 $\mu$m are included.}
\label{fig:dust_tails}
\end{center}
\end{figure*} % Fig. 23

\subsection{Dust Flows in the System}

Once dust has been released, regardless of the mechanism, there is the question of what radiation pressure subsequently does to it.  Each dust grain that is exposed to the radiation flux of the host star is subjected to a radiation pressure force, $F_{\rm rad}$ equal to $L\sigma_{\rm grain}/(4\pi c d^2)$.  The ratio of $F_{\rm rad}/F_{\rm grav}$ is defined as $\beta$ which is independent of $d$.  $\sigma_{\rm grain}$ in this case is the effective radiation pressure cross section for a photon to impart outward momentum to the dust grain.  Formally, the full expression for $F_{\rm rad}$ is written as an integral over wavelength-dependent Mie scattering and absorption cross sections and the spectral luminosity of the host star (see \citealt{2002Icar..159..529K} for the details).   For the other six systems discussed and listed in Table \ref{tbl:dusty}, the luminosity is lower than for the present star with $L \simeq 6 L_\odot$.  In particular, our estimates for the luminosities of the two stars in the TIC 400799224 system, from Table \ref{tbl:stars} are 6.32 and 3.25 $L_\odot$ (pre-MS solution) and 5.35 and 2.64 $L_\odot$ (post-MS solution).  Thus, compared to the other systems in Table \ref{tbl:dusty}, the highest luminosity of which is 0.17 $L_\odot$, we expect substantial values of $\beta$ for the dust grains in TIC 400799224.

In Fig.~\ref{fig:beta} we show how $\beta$ in this system might vary for a generic dust grain as a function of its size.  The reasons for the uncertainty are that we do not know (i) which of the two stars in TIC 400799224 hosts the dips and (ii) the chemical composition of the dust (in particular the indices of refraction of the dust grains are unknown).  We also show in the right-hand panel of Fig.~\ref{fig:beta} the cumulative distribution of values of $\beta$ for an assumed particle size distribution of $dp/ds \propto s^{-3}$.  For values of $\beta > 1/2$ the dust particles will be unbound from the system. For lower values of $\beta$ the dust particles will go into somewhat eccentric orbits with $\delta P_d/P_{\rm plan} \simeq 2 \beta$ where $\delta P_d$ is the difference between the period of the dust orbit and the emitting planetesimal orbit (for $\beta \ll 1$). In the rest frame of the dust-emitting planetesimal, the dust orbits form rosette patterns with petals marking each time they go all the way around to the azimuth from which they were initially launched (see, e.g., Fig.~7 of \citealt{2014ApJ...784...40R}).  The number of petals in the rosette will be roughly equal to $1/2\beta$.

In Fig.~\ref{fig:dust_tails} we show two illustrative dust-tail images that might exist in the TIC 400799224.  In these simulations, we assume that all the particles are ejected from the vicinity of a lunar-size planetesimal at the escape speed and directed into a 30$^\circ$ cone centered on the direction of the host star.  The size of each particle is chosen at random from a particle size distribution proportional to $s^{-3}$. The particles are assumed, quite arbitrarily, to sublimate on the timescale of the 20-day orbit.  The dust orbits are shown in the reference frame of the orbiting planetesimal.  The left panel shows the paths of 5000 dust particles for the case where the particle sizes are in the range of $10 \,\mu{\rm m} > s > 0.01 \, \mu$m.  Due to the steep power-law size dependence most of the particles are small and $\beta$ is near a constant value of $\sim$0.15.  One can see the formation of something like a rosette pattern, but one where the particles sublimate as time goes on.  In the right panel we truncate the particles below a size of 1 micron.  Because there are now few small particles there is actually a larger range of values for $\beta$. Each different value of $\beta < 1/2$ leads to a different orbit, while grains with $\beta > 1/2$ leave the system.  In spite of the complex nature of the pattern, there is still a clear concentration of dust over a small fraction of the orbit near the planetesimal which could cause dips in the stellar flux if the dust is actively being produced.

\subsection{Dust from Giant Impacts in a Debris Disk}
\label{sec:collisions}

Another possibility for stochastically producing dust in copious quantities is developed by \citet{2014MNRAS.440.3757J} in the context of giant impacts in optically thin, collisional debris disks orbiting young stars. The scenario, which is mentioned by \citet{2021arXiv210602659V} as being possibly
relevant for dust production in ZTF J0328-1219, involves catastrophic collisions among large bodies in a debris disk.   A long-term (at least years) phase coherence in the dips requires a principal body that is undergoing collisions with minor bodies, i.e., ones that (i) do not destroy it, and (ii) do not even change its basic orbital period. The collisions must be fairly regular (at least 20-30 over the last 6 years) and occur at the same orbital phase of the principal body.   Consider, for example, that there is a 100-km asteroid in a 20-day orbit around TIC 400799224.  Further suppose there are  numerous other substantial, but smaller (e.g., $\lesssim 1/10$th the radius), asteroids in near and crossing orbits.  Perhaps this condition was set up in the first place by a massive collision between two larger bodies.  Once there has been such a collision, all the debris returns on the next orbit to nearly the same region in space.  This high concentration of bodies naturally leads to subsequent collisions at the same orbital phase.  Each subsequent collision produces a debris cloud, presumably containing considerable dust and small particles, which expands and contracts vertically, while spreading azimuthally, as time goes on.  This may be sufficient to make one or two dusty transits before the cloud spreads and dissipates.  A new collision is then required to make a new dusty transit.

\begin{figure*}
    %\begin{interactive}{animation}{collision_compressed.mov}
    \centering
    \includegraphics[width=0.46\linewidth]{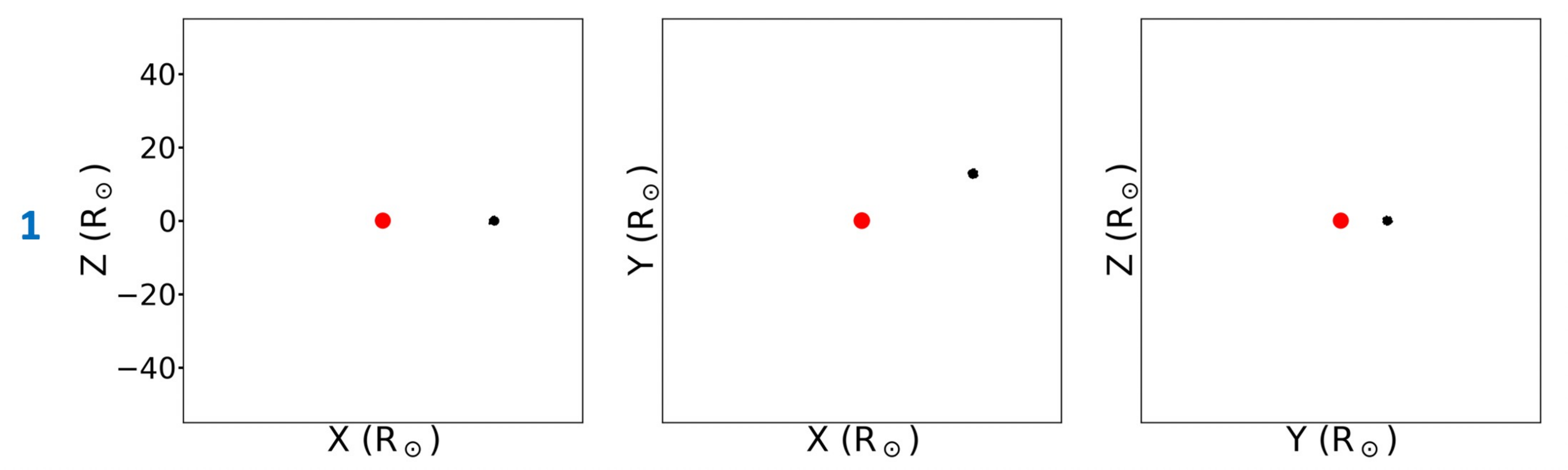} \hspace{.5cm}
    \includegraphics[width=0.46\linewidth]{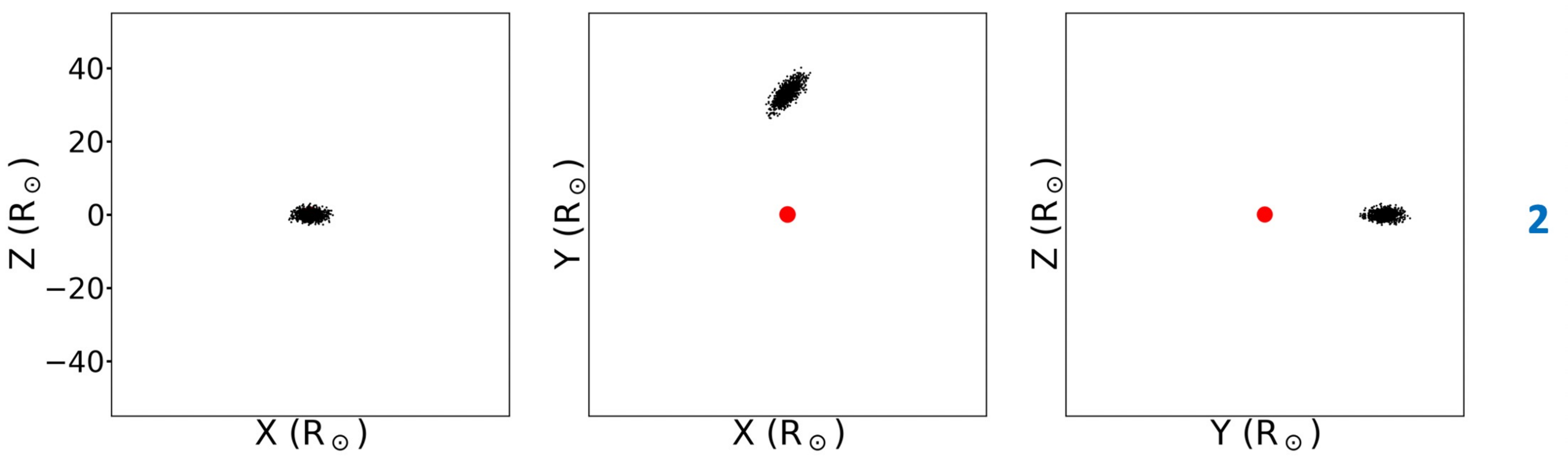}
    \includegraphics[width=0.46\linewidth]{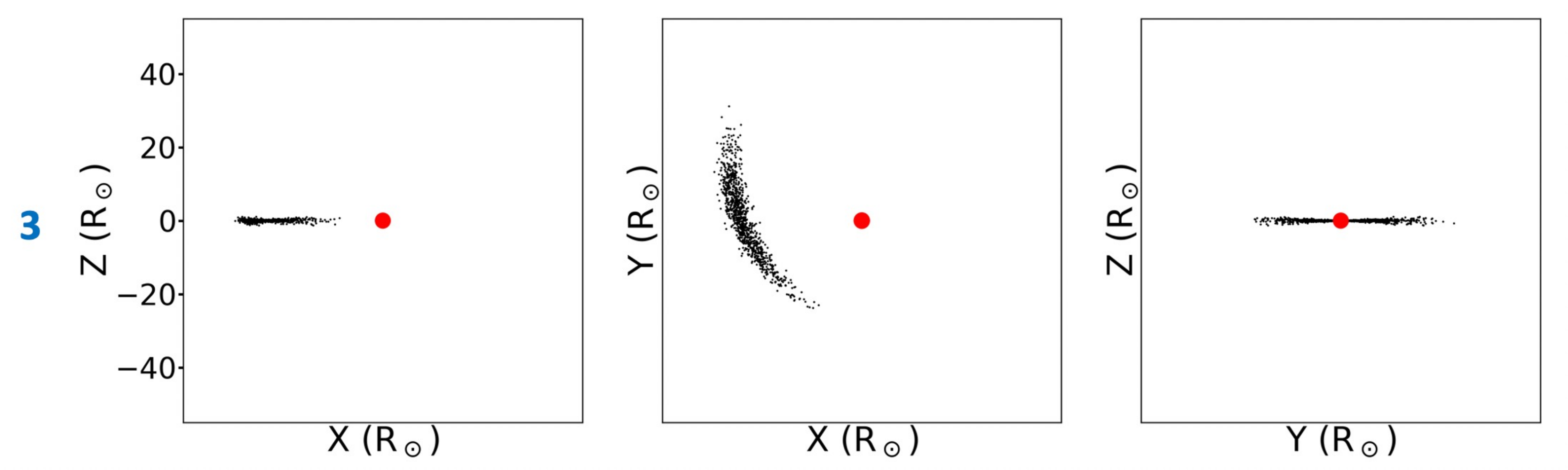}
    \hspace{.5cm}
    \includegraphics[width=0.46\linewidth]{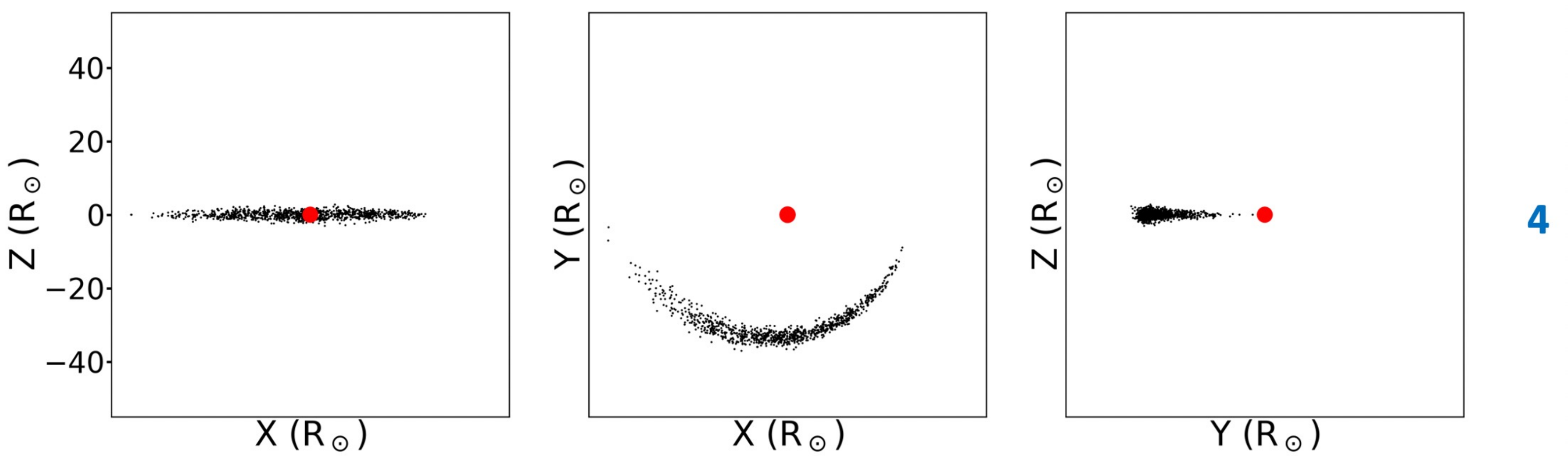}
    \includegraphics[width=0.46\linewidth]{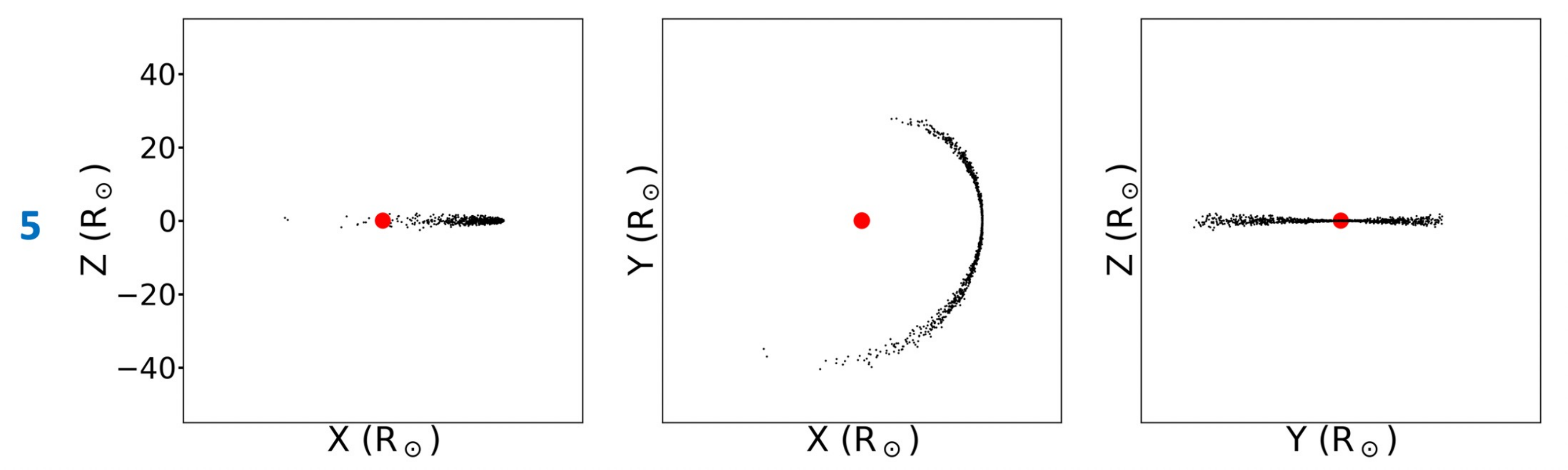}
    \hspace{.5cm}
    \includegraphics[width=0.46\linewidth]{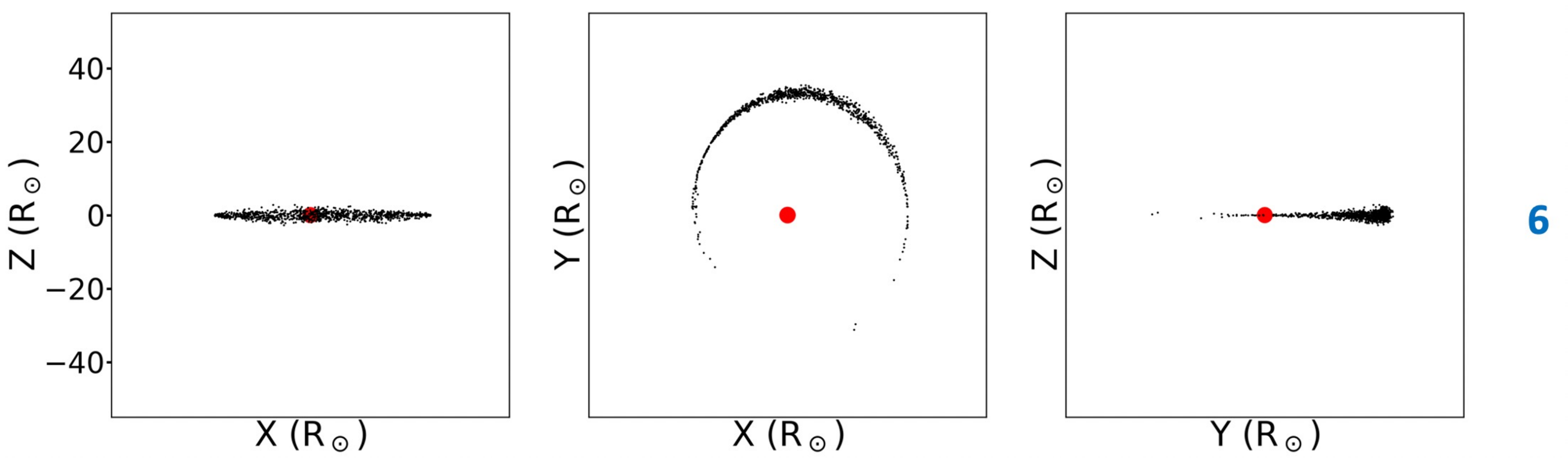}
    \includegraphics[width=0.46\linewidth]{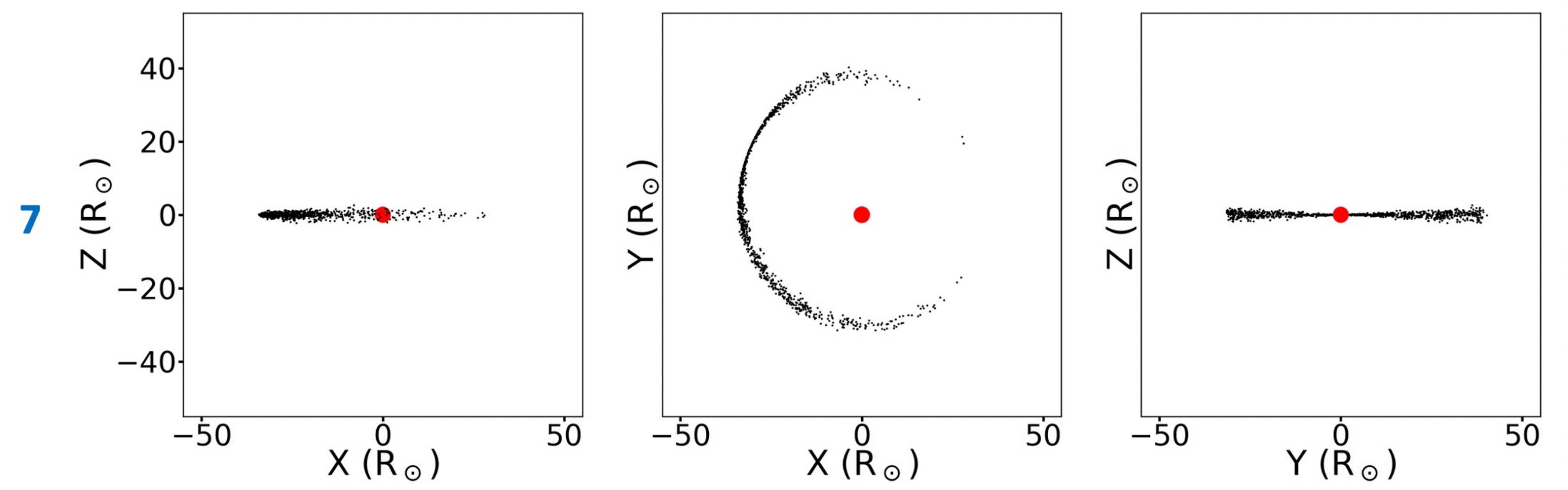}
    \hspace{.5cm}
    \includegraphics[width=0.46\linewidth]{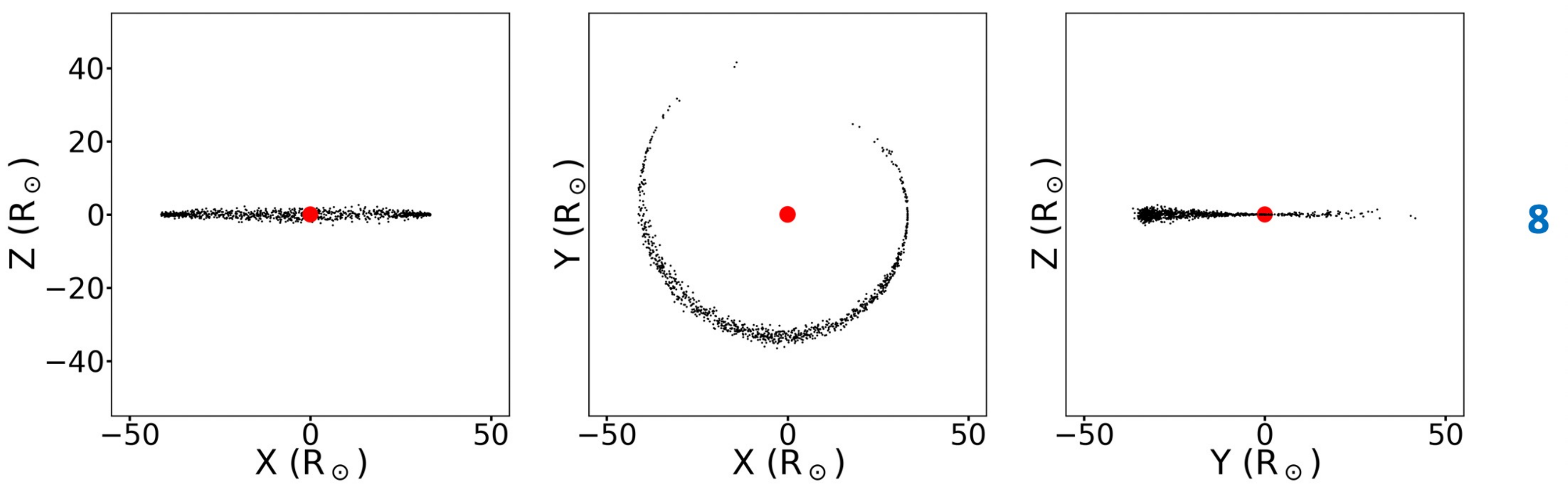}
    \includegraphics[width=1.0\linewidth]{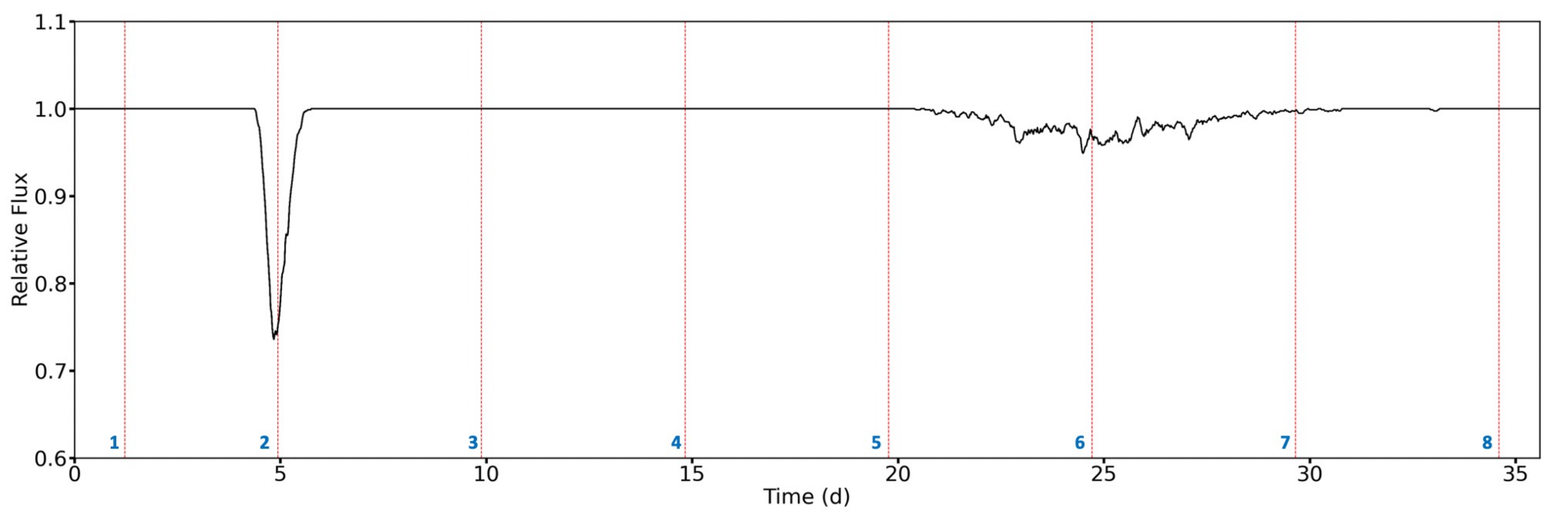}
    \caption{Dust distribution at, from left-to-right and then top-to-bottom, is at 1/16, 1/4, 2/4, 3/4, 4/4, 5/4, 6/4, and 7/4 orbits after the initial collision.  The six panels across are $x-z$, $x-y$, and $y-z$ planes, and then the three repeat, with units in $R_\odot$.  Each individual black point is a modeled dust particle, while the red point at the origin is the host star, here modeled with size $\sim2.2R_\odot$. The bottom-most panel is the net lightcurve observed from the distant $+y$ axis and is produced by assigning an arbitrary small geometrical size to each of the $10^4$ particles.  Red vertical lines correspond to the time of the numbered images above. An animation of this figure is available.}
    %\end{interactive}
   \label{fig:debris_orbits}
\end{figure*} % Fig. 24

We illustrate how this scenario might work in Fig.~\ref{fig:debris_orbits}. In this example there is an asteroid orbiting the host star every 19.77 days when it undergoes a collision that launches 10,000 particles in random directions with a thermal distribution of speeds.  The velocity dispersion, $\sigma$, in units of the orbital speed, is 0.03.  In  Fig.~\ref{fig:debris_orbits} we show 8 snapshots of the full simulation in increments of orbital phase equal to 90$^\circ$.  At each phase we show three panels which represent the views of observers situated at large distances along the $-y$, $+z$, and $+x$ axes. They show how the debris cloud expands both in the azimuthal and the vertical direction (i.e., perpendicular to the orbital plane).  What is significant is that all the particles pass periodically (but not simultaneously), through nearly the same $\{x,y,z\}$ point from which they were launched, and therefore the motion in the $Z$ direction is periodic. The high concentration of debris particles within this `funnel' region could facilitate further disruptive collisions.  In that case, all further debris clouds would be initiated at the same orbit phase as the original `master' asteroid.

A variant of the debris disk scenario involves shepherding of debris by the gravitational force exerted by a master asteroid or planet. We consider this less likely than the above stochastic collision scenario.
Consider for example how Neptune gravitationally sculpts the dusty debris in the Edgeworth-Kuiper belt \citep{1999AJ....118..580L}, or how the Earth traps particles into orbital co-rotation (\citealt{Dermott_etal_1994}). A planet can corral solid particles into its mean-motion resonances (e.g. 1:1, 6:7, 5:6, 4:5, 3:4, 2:3, 1:2, with the orbit of the perturbing planet either inside or outside the orbits of the particles), creating azimuthally localized dust concentrations that occult the Sun when viewed edge-on from without, with wave pattern speeds equal to the planet's orbital frequency. While the occultations would occur with frequencies equal to integer multiples of the planet's orbital frequency,
they would also be highly extended in orbital phase (see e.g.~figures 4 and 5 of \citealt{1999AJ....118..580L}) and produce only gradual, sub-percent modulations in extinction (e.g.~figure 5 of \citealt{Dermott_etal_1994}, and note that the maximum density of the dust clump trailing the Earth is only $\sim$10\% greater than the background density). A fine-tuned viewing geometry could reduce the duration of the transit to better match the event durations observed for TIC 400799224, but at the expense of further muting the transit amplitude. 

Another way to produce shorter-duration events is to situate the planet in an optically thick disk of particles whose viscosity damps away gravitationally driven  perturbations before they travel too far afield. This picture applies to  `propeller' moonlets in Saturn's rings where particle density disturbances are localized to the vicinity of the satellite's Hill sphere (e.g.~\citealt{Tiscareno_etal_2006}). However, an optically thick particle disk is disallowed for TIC 400799224 --- see Figure \ref{fig:composite_blackbody} (blue curve) which shows that such a disk would produce excess emission above 5 $\mu$m wavelengths, which is not observed. A final objection to all of the gravitational shepherding scenarios considered above is that they would predict light curves that would be consistent from transit to transit.

\subsection{Vertical Extent of the Dust}

When launched with a speed $v$ with respect to an orbiting body of orbital speed $V$, the tilt of the new orbit of the small particle with respect to the original orbital plane is
$$\tan \mu = \frac{(v/V) \cos \theta}{1+(v/V) \sin \theta \cos \phi} \approx  (v/V) \cos \theta $$
where $\theta$ is the launch angle from the normal to the orbit and $\phi$ is the azimuthal launch angle (where $\phi = 0$ corresponds to the orbital direction).  Therefore, the maximum vertical height attained by the ejected particle is $h \simeq d \,v/V \cos \theta$ one quarter of an orbit after its release.  Expressing $h$ in units of the host-star radius, we have: $h/R_* \simeq (d/R_*) (v/V) \cos \theta$\footnote{Therefore, completely in general, a deep eclipse from such a debris ejection cannot occur unless $v/V \approx R_*/d$. }.  But, from Table \ref{tbl:dusty} we see that $d/R_*$ for TIC 400799224 is $\simeq$15.  This implies that in order for $h/R_*$ to be as large as 20\% (for dips that deep), dust or particle ejection speeds of $v/V \cos \theta \sim 0.013$ are required.  Finally, for orbital speeds of $V \sim 85$ km s$^{-1}$ corresponding a 20-day orbit about TIC 400799224, launch speeds of $\sim$1 km s$^{-1}$ are implied.   Collision ejecta speeds could be of this order.

\section{Summary}
\label{sec:summary}
We have presented the discovery of a mysterious object orbiting one of the two bound stars comprising TIC 400799224 at a period of $\sim$19.77 days.  The SED fit to the system showed two possible solutions, one pre-MS and the other post-MS.  The presence of lithium and the region kinematics present a strong argument for the former, which also favors the presence of debris and dust.

In our examination of the nature of the orbiting body, we have considered (i) disintegration via sublimation; (ii) collisions with a minor planet-like object producing sporadic dust clouds; and (iii) shepherding of dust by an embedded planet. We conclude that the most likely scenario is (ii) due primarily to mass constraints, the persistent periodicity of the object over the course of six years, and the highly variable occultation depths.

The object appears to sporadically emit a large dust cloud which is able to block up to 37\% or 75\% of the light from its host, depending on which star in the binary is the true host of the object.  The mass of the dust cloud is approximately $10^{19}$ grams.  Remarkably, the dips in flux of the host star's light maintain phase coherence over the 6-year observing interval of the ASAS-SN project.  The three most recent observations by LCO have detected the presence of the dust cloud twice, suggesting that the object remains intact enough for further study in the near term.

The object is bright enough at $V = 12.6$ to be monitored by modest-size backyard telescopes to search for subsequent transits that are up to 25\% deep.  Eventually, when the entire set of DASCH \citep{grindlay17} archival plates are scanned, we should be able to detect these transits in the historical record going back many decades.

\section*{acknowledgments}

We would like to thank the anonymous referee, whose feedback we believe has improved this paper.

Resources supporting this work were provided by the NASA High-End Computing (HEC) Program through the NASA Center for Climate Simulation (NCCS) at Goddard Space Flight Center.  

This paper includes data collected by the {\em TESS} mission, which are publicly available from the Mikulski Archive for Space Telescopes (MAST). Funding for the {\em TESS} mission is provided by NASA's Science Mission directorate.

This work has made use of data from the European Space Agency (ESA) mission
{\it Gaia} (\url{https://www.cosmos.esa.int/gaia}), processed by the {\it Gaia}
Data Processing and Analysis Consortium (DPAC,
\url{https://www.cosmos.esa.int/web/gaia/dpac/consortium}). Funding for the DPAC
has been provided by national institutions, in particular the institutions
participating in the {\it Gaia} Multilateral Agreement.

This research has made use of the SIMBAD database, operated at CDS, Strasbourg, France.

This research has made use of the Exoplanet Follow-up Observation Program website, which is operated by the California Institute of Technology, under contract with the National Aeronautics and Space Administration under the Exoplanet Exploration Program. 

This research is based on observations made with the Galaxy Evolution Explorer, obtained from the MAST data archive at the Space Telescope Science Institute, which is operated by the Association of Universities for Research in Astronomy, Inc., under NASA contract NAS5-26555.

This work makes use of observations from the LCOGT network. Part of the LCOGT telescope time was granted by NOIRLab through the Mid-Scale Innovations Program (MSIP). MSIP is funded by NSF.

\facilities{
\emph{Gaia},
MAST,
TESS,
WASP,
ASAS-SN,
LCOGT,
NCCS,
CTIO:1.5m,
SOAR}

\software{
{\tt Astrocut} \citep{astrocut},
{\tt AstroImageJ} \citep{Collins:2017},
{\tt Astropy} \citep{astropy2013,astropy2018}, 
{\tt Eleanor} \citep{eleanor},
{\tt IPython} \citep{ipython},
{\tt Keras} \citep{keras},
{\tt LcTools} \citep{2021arXiv210310285S},
{\tt Lightkurve} \citep{lightkurve},
{\tt Matplotlib} \citep{matplotlib},
{\tt Mpi4py} \citep{mpi4py2008},
{\tt NumPy} \citep{numpy}, 
{\tt Pandas} \citep{pandas},
{\tt Scikit-learn} \citep{scikit-learn},
{\tt SciPy} \citep{scipy},
{\tt Tensorflow} \citep{tensorflow},
{\tt Tess-point} \citep{tess-point}
}

\bibliography{refs}{}

\begin{thebibliography}{}
\expandafter\ifx\csname natexlab\endcsname\relax\def\natexlab#1{#1}\fi
\providecommand{\url}[1]{\href{#1}{#1}}
\providecommand{\dodoi}[1]{doi:~\href{http://doi.org/#1}{\nolinkurl{#1}}}
\providecommand{\doeprint}[1]{\href{http://ascl.net/#1}{\nolinkurl{http://ascl.net/#1}}}
\providecommand{\doarXiv}[1]{\href{https://arxiv.org/abs/#1}{\nolinkurl{https://arxiv.org/abs/#1}}}

\bibitem[{Abadi {et~al.}(2015)Abadi, Agarwal, Barham, Brevdo, Chen, Citro,
  Corrado, Davis, Dean, Devin, Ghemawat, Goodfellow, Harp, Irving, Isard, Jia,
  Jozefowicz, Kaiser, Kudlur, Levenberg, Man\'{e}, Monga, Moore, Murray, Olah,
  Schuster, Shlens, Steiner, Sutskever, Talwar, Tucker, Vanhoucke, Vasudevan,
  Vi\'{e}gas, Vinyals, Warden, Wattenberg, Wicke, Yu, \& Zheng}]{tensorflow}
Abadi, M., Agarwal, A., Barham, P., {et~al.} 2015, {TensorFlow}: Large-Scale
  Machine Learning on Heterogeneous Systems.
\newblock \url{https://www.tensorflow.org/}

\bibitem[{{Astropy Collaboration} {et~al.}(2013){Astropy Collaboration},
  {Robitaille}, {Tollerud}, {Greenfield}, {Droettboom}, {Bray}, {Aldcroft},
  {Davis}, {Ginsburg}, {Price-Whelan}, {Kerzendorf}, {Conley}, {Crighton},
  {Barbary}, {Muna}, {Ferguson}, {Grollier}, {Parikh}, {Nair}, {Unther},
  {Deil}, {Woillez}, {Conseil}, {Kramer}, {Turner}, {Singer}, {Fox}, {Weaver},
  {Zabalza}, {Edwards}, {Azalee Bostroem}, {Burke}, {Casey}, {Crawford},
  {Dencheva}, {Ely}, {Jenness}, {Labrie}, {Lim}, {Pierfederici}, {Pontzen},
  {Ptak}, {Refsdal}, {Servillat}, \& {Streicher}}]{astropy2013}
{Astropy Collaboration}, {Robitaille}, T.~P., {Tollerud}, E.~J., {et~al.} 2013,
  \aap, 558, A33, \dodoi{10.1051/0004-6361/201322068}

\bibitem[{{Astropy Collaboration} {et~al.}(2018){Astropy Collaboration},
  {Price-Whelan}, {Sip{\H{o}}cz}, {G{\"u}nther}, {Lim}, {Crawford}, {Conseil},
  {Shupe}, {Craig}, {Dencheva}, {Ginsburg}, {Vand erPlas}, {Bradley},
  {P{\'e}rez-Su{\'a}rez}, {de Val-Borro}, {Aldcroft}, {Cruz}, {Robitaille},
  {Tollerud}, {Ardelean}, {Babej}, {Bach}, {Bachetti}, {Bakanov}, {Bamford},
  {Barentsen}, {Barmby}, {Baumbach}, {Berry}, {Biscani}, {Boquien}, {Bostroem},
  {Bouma}, {Brammer}, {Bray}, {Breytenbach}, {Buddelmeijer}, {Burke},
  {Calderone}, {Cano Rodr{\'\i}guez}, {Cara}, {Cardoso}, {Cheedella}, {Copin},
  {Corrales}, {Crichton}, {D'Avella}, {Deil}, {Depagne}, {Dietrich}, {Donath},
  {Droettboom}, {Earl}, {Erben}, {Fabbro}, {Ferreira}, {Finethy}, {Fox},
  {Garrison}, {Gibbons}, {Goldstein}, {Gommers}, {Greco}, {Greenfield},
  {Groener}, {Grollier}, {Hagen}, {Hirst}, {Homeier}, {Horton}, {Hosseinzadeh},
  {Hu}, {Hunkeler}, {Ivezi{\'c}}, {Jain}, {Jenness}, {Kanarek}, {Kendrew},
  {Kern}, {Kerzendorf}, {Khvalko}, {King}, {Kirkby}, {Kulkarni}, {Kumar},
  {Lee}, {Lenz}, {Littlefair}, {Ma}, {Macleod}, {Mastropietro}, {McCully},
  {Montagnac}, {Morris}, {Mueller}, {Mumford}, {Muna}, {Murphy}, {Nelson},
  {Nguyen}, {Ninan}, {N{\"o}the}, {Ogaz}, {Oh}, {Parejko}, {Parley}, {Pascual},
  {Patil}, {Patil}, {Plunkett}, {Prochaska}, {Rastogi}, {Reddy Janga},
  {Sabater}, {Sakurikar}, {Seifert}, {Sherbert}, {Sherwood-Taylor}, {Shih},
  {Sick}, {Silbiger}, {Singanamalla}, {Singer}, {Sladen}, {Sooley},
  {Sornarajah}, {Streicher}, {Teuben}, {Thomas}, {Tremblay}, {Turner},
  {Terr{\'o}n}, {van Kerkwijk}, {de la Vega}, {Watkins}, {Weaver}, {Whitmore},
  {Woillez}, {Zabalza}, \& {Astropy Contributors}}]{astropy2018}
{Astropy Collaboration}, {Price-Whelan}, A.~M., {Sip{\H{o}}cz}, B.~M., {et~al.}
  2018, \aj, 156, 123, \dodoi{10.3847/1538-3881/aabc4f}

\bibitem[{{Bochinski} {et~al.}(2015){Bochinski}, {Haswell}, {Marsh}, {Dhillon},
  \& {Littlefair}}]{2015ApJ...800L..21B}
{Bochinski}, J.~J., {Haswell}, C.~A., {Marsh}, T.~R., {Dhillon}, V.~S., \&
  {Littlefair}, S.~P. 2015, \apjl, 800, L21,
  \dodoi{10.1088/2041-8205/800/2/L21}

\bibitem[{{Borkovits} {et~al.}(2020{\natexlab{a}}){Borkovits}, {Rappaport},
  {Hajdu}, {Maxted}, {P{\'a}l}, {Forg{\'a}cs-Dajka}, {Klagyivik}, \&
  {Mitnyan}}]{2020MNRAS.493.5005B}
{Borkovits}, T., {Rappaport}, S.~A., {Hajdu}, T., {et~al.} 2020{\natexlab{a}},
  \mnras, 493, 5005, \dodoi{10.1093/mnras/staa495}

\bibitem[{{Borkovits} {et~al.}(2019){Borkovits}, {Rappaport}, {Kaye},
  {Isaacson}, {Vanderburg}, {Howard}, {Kristiansen}, {Omohundro},
  {Schwengeler}, {Terentev}, {Shporer}, {Relles}, {Villanueva}, {Tan},
  {Col{\'o}n}, {Blex}, {Haas}, {Cochran}, \& {Endl}}]{2019MNRAS.483.1934B}
{Borkovits}, T., {Rappaport}, S., {Kaye}, T., {et~al.} 2019, \mnras, 483, 1934,
  \dodoi{10.1093/mnras/sty3157}

\bibitem[{{Borkovits} {et~al.}(2020{\natexlab{b}}){Borkovits}, {Rappaport},
  {Tan}, {Gagliano}, {Jacobs}, {Huang}, {Mitnyan}, {Hambsch}, {Kaye}, {Maxted},
  {P{\'a}l}, \& {Schmitt}}]{2020MNRAS.496.4624B}
{Borkovits}, T., {Rappaport}, S.~A., {Tan}, T.~G., {et~al.} 2020{\natexlab{b}},
  \mnras, 496, 4624, \dodoi{10.1093/mnras/staa1817}

\bibitem[{{Boyajian} {et~al.}(2016){Boyajian}, {LaCourse}, {Rappaport},
  {Fabrycky}, {Fischer}, {Gandolfi}, {Kennedy}, {Korhonen}, {Liu}, {Moor},
  {Olah}, {Vida}, {Wyatt}, {Best}, {Brewer}, {Ciesla}, {Cs{\'a}k}, {Deeg},
  {Dupuy}, {Handler}, {Heng}, {Howell}, {Ishikawa}, {Kov{\'a}cs}, {Kozakis},
  {Kriskovics}, {Lehtinen}, {Lintott}, {Lynn}, {Nespral}, {Nikbakhsh},
  {Schawinski}, {Schmitt}, {Smith}, {Szabo}, {Szabo}, {Viuho}, {Wang},
  {Weiksnar}, {Bosch}, {Connors}, {Goodman}, {Green}, {Hoekstra}, {Jebson},
  {Jek}, {Omohundro}, {Schwengeler}, \& {Szewczyk}}]{2016MNRAS.457.3988B}
{Boyajian}, T.~S., {LaCourse}, D.~M., {Rappaport}, S.~A., {et~al.} 2016,
  \mnras, 457, 3988, \dodoi{10.1093/mnras/stw218}

\bibitem[{{Brandeker} \& {Cataldi}(2019)}]{2019A&A...621A..86B}
{Brandeker}, A., \& {Cataldi}, G. 2019, \aap, 621, A86,
  \dodoi{10.1051/0004-6361/201834321}

\bibitem[{{Brasseur} {et~al.}(2019){Brasseur}, {Phillip}, {Fleming},
  {Mullally}, \& {White}}]{astrocut}
{Brasseur}, C.~E., {Phillip}, C., {Fleming}, S.~W., {Mullally}, S.~E., \&
  {White}, R.~L. 2019, {Astrocut: Tools for creating cutouts of TESS images}.
\newblock \doeprint{1905.007}

\bibitem[{{Brogi} {et~al.}(2012){Brogi}, {Keller}, {de Juan Ovelar},
  {Kenworthy}, {de Kok}, {Min}, \& {Snellen}}]{2012A&A...545L...5B}
{Brogi}, M., {Keller}, C.~U., {de Juan Ovelar}, M., {et~al.} 2012, \aap, 545,
  L5, \dodoi{10.1051/0004-6361/201219762}

\bibitem[{{Brown} {et~al.}(2013){Brown}, {Baliber}, {Bianco}, {Bowman},
  {Burleson}, {Conway}, {Crellin}, {Depagne}, {De Vera}, {Dilday}, {Dragomir},
  {Dubberley}, {Eastman}, {Elphick}, {Falarski}, {Foale}, {Ford}, {Fulton},
  {Garza}, {Gomez}, {Graham}, {Greene}, {Haldeman}, {Hawkins}, {Haworth},
  {Haynes}, {Hidas}, {Hjelstrom}, {Howell}, {Hygelund}, {Lister}, {Lobdill},
  {Martinez}, {Mullins}, {Norbury}, {Parrent}, {Paulson}, {Petry}, {Pickles},
  {Posner}, {Rosing}, {Ross}, {Sand}, {Saunders}, {Shobbrook}, {Shporer},
  {Street}, {Thomas}, {Tsapras}, {Tufts}, {Valenti}, {Vander Horst}, {Walker},
  {White}, \& {Willis}}]{Brown:2013}
{Brown}, T.~M., {Baliber}, N., {Bianco}, F.~B., {et~al.} 2013, Publications of
  the Astronomical Society of the Pacific, 125, 1031, \dodoi{10.1086/673168}

\bibitem[{{Burke} {et~al.}(2020){Burke}, {Levine}, {Fausnaugh}, {Vanderspek},
  {Barclay}, {Libby-Roberts}, {Morris}, {Sipocz}, {Owens}, {Feinstein}, \&
  {Camacho}}]{tess-point}
{Burke}, C.~J., {Levine}, A., {Fausnaugh}, M., {et~al.} 2020, {TESS-Point: High
  precision TESS pointing tool}, Astrophysics Source Code Library.
\newblock \doeprint{2003.001}

\bibitem[{{Choi} {et~al.}(2016){Choi}, {Dotter}, {Conroy}, {Cantiello},
  {Paxton}, \& {Johnson}}]{choi16}
{Choi}, J., {Dotter}, A., {Conroy}, C., {et~al.} 2016, \apj, 823, 102,
  \dodoi{10.3847/0004-637X/823/2/102}

\bibitem[{Chollet {et~al.}(2015)}]{keras}
Chollet, F., {et~al.} 2015, Keras, \url{https://keras.io}

\bibitem[{{Collins} {et~al.}(2017){Collins}, {Kielkopf}, {Stassun}, \&
  {Hessman}}]{Collins:2017}
{Collins}, K.~A., {Kielkopf}, J.~F., {Stassun}, K.~G., \& {Hessman}, F.~V.
  2017, \aj, 153, 77, \dodoi{10.3847/1538-3881/153/2/77}

\bibitem[{{Col{\'o}n} {et~al.}(2018){Col{\'o}n}, {Zhou}, {Shporer}, {Collins},
  {Bieryla}, {Espinoza}, {Murgas}, {Pattarakijwanich}, {Awiphan}, {Armstrong},
  {Bailey}, {Barentsen}, {Bayliss}, {Chakpor}, {Cochran}, {Dhillon}, {Horne},
  {Ireland}, {Kedziora-Chudczer}, {Kielkopf}, {Komonjinda}, {Latham}, {Marsh},
  {Mkrtichian}, {Pall{\'e}}, {Ruffolo}, {Sefako}, {Tinney}, {Wannawichian}, \&
  {Yuma}}]{2018AJ....156..227C}
{Col{\'o}n}, K.~D., {Zhou}, G., {Shporer}, A., {et~al.} 2018, \aj, 156, 227,
  \dodoi{10.3847/1538-3881/aae31b}

\bibitem[{{Corbett} {et~al.}(2020){Corbett}, {Law}, {Soto}, {Howard},
  {Glazier}, {Gonzalez}, {Ratzloff}, {Galliher}, {Fors}, \&
  {Quimby}}]{2020ApJ...903L..27C}
{Corbett}, H., {Law}, N.~M., {Soto}, A.~V., {et~al.} 2020, \apjl, 903, L27,
  \dodoi{10.3847/2041-8213/abbee5}

\bibitem[{{Court} {et~al.}(2019){Court}, {Scaringi}, {Rappaport}, {Zhan},
  {Littlefield}, {Castro Segura}, {Knigge}, {Maccarone}, {Kennedy}, {Szkody},
  \& {Garnavich}}]{2019MNRAS.488.4149C}
{Court}, J.~M.~C., {Scaringi}, S., {Rappaport}, S., {et~al.} 2019, \mnras, 488,
  4149, \dodoi{10.1093/mnras/stz2015}

\bibitem[{{Croll} {et~al.}(2017){Croll}, {Dalba}, {Vanderburg}, {Eastman},
  {Rappaport}, {DeVore}, {Bieryla}, {Muirhead}, {Han}, {Latham}, {Beatty},
  {Wittenmyer}, {Wright}, {Johnson}, \& {McCrady}}]{croll17}
{Croll}, B., {Dalba}, P.~A., {Vanderburg}, A., {et~al.} 2017, \apj, 836, 82,
  \dodoi{10.3847/1538-4357/836/1/82}

\bibitem[{Dalcin {et~al.}(2008)Dalcin, Paz, Storti, \& D'Elia}]{mpi4py2008}
Dalcin, L., Paz, R., Storti, M., \& D'Elia, J. 2008, Journal of Parallel and
  Distributed Computing, 68, 655,
  \dodoi{http://dx.doi.org/10.1016/j.jpdc.2007.09.005}

\bibitem[{{Dermott} {et~al.}(1994){Dermott}, {Jayaraman}, {Xu}, {Gustafson}, \&
  {Liou}}]{Dermott_etal_1994}
{Dermott}, S.~F., {Jayaraman}, S., {Xu}, Y.~L., {Gustafson}, B. {\r{A}}.~S., \&
  {Liou}, J.~C. 1994, \nat, 369, 719, \dodoi{10.1038/369719a0}

\bibitem[{{Dotter}(2016)}]{dotter16}
{Dotter}, A. 2016, \apjs, 222, 8, \dodoi{10.3847/0067-0049/222/1/8}

\bibitem[{{Fausnaugh} {et~al.}(2021){Fausnaugh}, {Vallely}, {Kochanek},
  {Shappee}, {Stanek}, {Tucker}, {Ricker}, {Vanderspek}, {Latham}, {Seager},
  {Winn}, {Jenkins}, {Berta-Thompson}, {Daylan}, {Doty}, {F{\H{u}}r{\'e}sz},
  {Levine}, {Morris}, {P{\'a}l}, {Sha}, {Ting}, \& {Wohler}}]{Fausnaugh2021}
{Fausnaugh}, M.~M., {Vallely}, P.~J., {Kochanek}, C.~S., {et~al.} 2021, \apj,
  908, 51, \dodoi{10.3847/1538-4357/abcd42}

\bibitem[{{Feinstein} {et~al.}(2019){Feinstein}, {Montet}, {Foreman-Mackey},
  {Bedell}, {Saunders}, {Bean}, {Christiansen}, {Hedges}, {Luger}, {Scolnic},
  \& {Cardoso}}]{eleanor}
{Feinstein}, A.~D., {Montet}, B.~T., {Foreman-Mackey}, D., {et~al.} 2019,
  \pasp, 131, 094502, \dodoi{10.1088/1538-3873/ab291c}

\bibitem[{{G{\"a}nsicke} {et~al.}(2016){G{\"a}nsicke}, {Aungwerojwit}, {Marsh},
  {Dhillon}, {Sahman}, {Veras}, {Farihi}, {Chote}, {Ashley}, {Arjyotha},
  {Rattanasoon}, {Littlefair}, {Pollacco}, \& {Burleigh}}]{gaensicke16}
{G{\"a}nsicke}, B.~T., {Aungwerojwit}, A., {Marsh}, T.~R., {et~al.} 2016,
  \apjl, 818, L7, \dodoi{10.3847/2041-8205/818/1/L7}

\bibitem[{{Garnavich} {et~al.}(2016){Garnavich}, {Tucker}, {Rest}, {Shaya},
  {Olling}, {Kasen}, \& {Villar}}]{2016ApJ...820...23G}
{Garnavich}, P.~M., {Tucker}, B.~E., {Rest}, A., {et~al.} 2016, \apj, 820, 23,
  \dodoi{10.3847/0004-637X/820/1/23}

\bibitem[{{Gary} {et~al.}(2017){Gary}, {Rappaport}, {Kaye}, {Alonso}, \&
  {Hambschs}}]{Gary2017}
{Gary}, B.~L., {Rappaport}, S., {Kaye}, T.~G., {Alonso}, R., \& {Hambschs},
  F.~J. 2017, \mnras, 465, 3267, \dodoi{10.1093/mnras/stw2921}

\bibitem[{{Grindlay}(2017)}]{grindlay17}
{Grindlay}, J. 2017, in The Science of Time 2016, ed. E.~F. {Arias},
  L.~{Combrinck}, P.~{Gabor}, C.~{Hohenkerk}, \& P.~K. {Seidelmann}, Vol.~50,
  203, \dodoi{10.1007/978-3-319-59909-0\_26}

\bibitem[{Harris {et~al.}(2020)Harris, Millman, van~der Walt, Gommers,
  Virtanen, Cournapeau, Wieser, Taylor, Berg, Smith, Kern, Picus, Hoyer, van
  Kerkwijk, Brett, Haldane, del R{\'{i}}o, Wiebe, Peterson,
  G{\'{e}}rard-Marchant, Sheppard, Reddy, Weckesser, Abbasi, Gohlke, \&
  Oliphant}]{numpy}
Harris, C.~R., Millman, K.~J., van~der Walt, S.~J., {et~al.} 2020, Nature, 585,
  357, \dodoi{10.1038/s41586-020-2649-2}

\bibitem[{{Holoien} {et~al.}(2019){Holoien}, {Vallely}, {Auchettl}, {Stanek},
  {Kochanek}, {French}, {Prieto}, {Shappee}, {Brown}, {Fausnaugh}, {Dong},
  {Thompson}, {Bose}, {Neustadt}, {Cacella}, {Brimacombe}, {Kendurkar},
  {Beaton}, {Boutsia}, {Chomiuk}, {Connor}, {Morrell}, {Newman}, {Rudie},
  {Shishkovksy}, \& {Strader}}]{Holoien2019}
{Holoien}, T. W.~S., {Vallely}, P.~J., {Auchettl}, K., {et~al.} 2019, \apj,
  883, 111, \dodoi{10.3847/1538-4357/ab3c66}

\bibitem[{Hunter(2007)}]{matplotlib}
Hunter, J.~D. 2007, Computing in science \& engineering, 9, 90

\bibitem[{{Ikwut-Ukwa} {et~al.}(2021){Ikwut-Ukwa}, {Rodriguez}, {Quinn},
  {Zhou}, {Vanderburg}, {Ali}, {Bunten}, {Gaudi}, {Latham}, {Howell}, {Huang},
  {Bieryla}, {Collins}, {Carmichael}, {Rabus}, {Eastman}, {Collins}, {Tan},
  {Schwarz}, {Myers}, {Stockdale}, {Kielkopf}, {Radford}, {Oelkers}, {Jenkins},
  {Ricker}, {Seager}, {Vanderspek}, {Winn}, {Burt}, {Butler}, {Calkins},
  {Crane}, {Gnilka}, {Esquerdo}, {Fong}, {Kreidberg}, {Mink}, {Rodriguez},
  {Schlieder}, {Schectman}, {Shporer}, {Teske}, {Ting}, {Villasenor}, \&
  {Yahalomi}}]{Ikwut2021}
{Ikwut-Ukwa}, M., {Rodriguez}, J.~E., {Quinn}, S.~N., {et~al.} 2021, arXiv
  e-prints, arXiv:2102.02222.
\newblock \doarXiv{2102.02222}

\bibitem[{{Jackson} {et~al.}(2014){Jackson}, {Wyatt}, {Bonsor}, \&
  {Veras}}]{2014MNRAS.440.3757J}
{Jackson}, A.~P., {Wyatt}, M.~C., {Bonsor}, A., \& {Veras}, D. 2014, \mnras,
  440, 3757, \dodoi{10.1093/mnras/stu476}

\bibitem[{{Kimura} {et~al.}(2002){Kimura}, {Mann}, {Biesecker}, \&
  {Jessberger}}]{2002Icar..159..529K}
{Kimura}, H., {Mann}, I., {Biesecker}, D.~A., \& {Jessberger}, E.~K. 2002,
  \icarus, 159, 529, \dodoi{10.1006/icar.2002.6940}

\bibitem[{{Kochanek} {et~al.}(2017{\natexlab{a}}){Kochanek}, {Shappee},
  {Stanek}, {Holoien}, {Thompson}, {Prieto}, {Dong}, {Shields}, {Will},
  {Britt}, {Perzanowski}, \& {Pojma{\'n}ski}}]{2017PASP..129j4502K}
{Kochanek}, C.~S., {Shappee}, B.~J., {Stanek}, K.~Z., {et~al.}
  2017{\natexlab{a}}, \pasp, 129, 104502, \dodoi{10.1088/1538-3873/aa80d9}

\bibitem[{{Kochanek} {et~al.}(2017{\natexlab{b}}){Kochanek}, {Shappee},
  {Stanek}, {Holoien}, {Thompson}, {Prieto}, {Dong}, {Shields}, {Will},
  {Britt}, {Perzanowski}, \& {Pojma{\'n}ski}}]{Kochanek2017}
---. 2017{\natexlab{b}}, \pasp, 129, 104502, \dodoi{10.1088/1538-3873/aa80d9}

\bibitem[{{Kostov} {et~al.}(2020){Kostov}, {Orosz}, {Feinstein}, {Welsh},
  {Cukier}, {Haghighipour}, {Quarles}, {Martin}, {Montet}, {Torres}, {Triaud},
  {Barclay}, {Boyd}, {Briceno}, {Cameron}, {Correia}, {Gilbert}, {Gill},
  {Gillon}, {Haqq-Misra}, {Hellier}, {Dressing}, {Fabrycky}, {Furesz},
  {Jenkins}, {Kane}, {Kopparapu}, {Hod{\v{z}}i{\'c}}, {Latham}, {Law},
  {Levine}, {Li}, {Lintott}, {Lissauer}, {Mann}, {Mazeh}, {Mardling}, {Maxted},
  {Eisner}, {Pepe}, {Pepper}, {Pollacco}, {Quinn}, {Quintana}, {Rowe},
  {Ricker}, {Rose}, {Seager}, {Santerne}, {S{\'e}gransan}, {Short}, {Smith},
  {Standing}, {Tokovinin}, {Trifonov}, {Turner}, {Twicken}, {Udry},
  {Vanderspek}, {Winn}, {Wolf}, {Ziegler}, {Ansorge}, {Barnet}, {Bergeron},
  {Huten}, {Pappa}, \& {van der Straeten}}]{2020AJ....159..253K}
{Kostov}, V.~B., {Orosz}, J.~A., {Feinstein}, A.~D., {et~al.} 2020, \aj, 159,
  253, \dodoi{10.3847/1538-3881/ab8a48}

\bibitem[{{Kostov} {et~al.}(2021{\natexlab{a}}){Kostov}, {Powell}, {Orosz},
  {Welsh}, {Cochran}, {Collins}, {Endl}, {Hellier}, {Latham}, {MacQueen},
  {Pepper}, {Quarles}, {Sairam}, {Torres}, {Wilson}, {Bergeron}, {Boyce},
  {Buchheim}, {Ben Christiansen}, {Ciardi}, {Collins}, {Conti}, {Dixon},
  {Guerra}, {Haghighipour}, {Herman}, {Hintz}, {Howard}, {Jensen}, {Kruse},
  {Law}, {Martin}, {Maxted}, {Montet}, {Murgas}, {Nelson}, {Olmschenk},
  {Otero}, {Quimby}, {Richmond}, {Schwarz}, {Shporer}, {Stassun}, {Stephens},
  {Triaud}, {Ulowetz}, {Walter}, {Wiley}, {Wood}, {Yenawine}, {Agol},
  {Barclay}, {Beatty}, {Boisse}, {Caldwell}, {Christiansen}, {Colon},
  {Deleuil}, {Doyle}, {Fabrycky}, {Fausnaugh}, {Furesz}, {Gilbert}, {Hebrard},
  {James}, {Jenkins}, {Kane}, {Kidwell}, {Kopparapu}, {Li}, {Lissauer}, {Lund},
  {Majewski}, {Mazeh}, {Quinn}, {Ricker}, {Rodriguez}, {Rowe}, {Santerne},
  {Schlieder}, {Seager}, {Standing}, {Stevens}, {Ting}, {Vanderspek}, \&
  {Winn}}]{2021arXiv210508614K}
{Kostov}, V.~B., {Powell}, B.~P., {Orosz}, J.~A., {et~al.} 2021{\natexlab{a}},
  arXiv e-prints, arXiv:2105.08614.
\newblock \doarXiv{2105.08614}

\bibitem[{{Kostov} {et~al.}(2021{\natexlab{b}}){Kostov}, {Powell}, {Torres},
  {Borkovits}, {Rappaport}, {Tokovinin}, {Zasche}, {Anderson}, {Barclay},
  {Berlind}, {Brown}, {Calkins}, {Collins}, {Collins}, {Conti}, {Esquerdo},
  {Hellier}, {Jensen}, {Kamler}, {Kruse}, {Latham}, {Masek}, {Murgas},
  {Olmschenk}, {Orosz}, {Pal}, {Palle}, {Schwarz}, {Stockdale}, {Tamayo},
  {Uhlar}, {Welsh}, \& {West}}]{2021arXiv210512586K}
{Kostov}, V.~B., {Powell}, B.~P., {Torres}, G., {et~al.} 2021{\natexlab{b}},
  arXiv e-prints, arXiv:2105.12586.
\newblock \doarXiv{2105.12586}

\bibitem[{{Kounkel} \& {Covey}(2019)}]{2019AJ....158..122K}
{Kounkel}, M., \& {Covey}, K. 2019, \aj, 158, 122,
  \dodoi{10.3847/1538-3881/ab339a}

\bibitem[{{Kov{\'a}cs} {et~al.}(2002){Kov{\'a}cs}, {Zucker}, \&
  {Mazeh}}]{Kovacs2002}
{Kov{\'a}cs}, G., {Zucker}, S., \& {Mazeh}, T. 2002, \aap, 391, 369,
  \dodoi{10.1051/0004-6361:20020802}

\bibitem[{{Kruse} \& {Agol}(2014)}]{2014Sci...344..275K}
{Kruse}, E., \& {Agol}, E. 2014, Science, 344, 275,
  \dodoi{10.1126/science.1251999}

\bibitem[{{Law} {et~al.}(2014){Law}, {Fors}, {Wulfken}, {Ratzloff}, \&
  {Kavanaugh}}]{2014SPIE.9145E..0ZL}
{Law}, N.~M., {Fors}, O., {Wulfken}, P., {Ratzloff}, J., \& {Kavanaugh}, D.
  2014, in Society of Photo-Optical Instrumentation Engineers (SPIE) Conference
  Series, Vol. 9145, Ground-based and Airborne Telescopes V, ed. L.~M. {Stepp},
  R.~{Gilmozzi}, \& H.~J. {Hall}, 91450Z, \dodoi{10.1117/12.2057031}

\bibitem[{{Law} {et~al.}(2015){Law}, {Fors}, {Ratzloff}, {Wulfken},
  {Kavanaugh}, {Sitar}, {Pruett}, {Birchard}, {Barlow}, {Cannon}, {Cenko},
  {Dunlap}, {Kraus}, \& {Maccarone}}]{law15}
{Law}, N.~M., {Fors}, O., {Ratzloff}, J., {et~al.} 2015, \pasp, 127, 234,
  \dodoi{10.1086/680521}

\bibitem[{{Lightkurve Collaboration} {et~al.}(2018){Lightkurve Collaboration},
  {Cardoso}, {Hedges}, {Gully-Santiago}, {Saunders}, {Cody}, {Barclay}, {Hall},
  {Sagear}, {Turtelboom}, {Zhang}, {Tzanidakis}, {Mighell}, {Coughlin}, {Bell},
  {Berta-Thompson}, {Williams}, {Dotson}, \& {Barentsen}}]{lightkurve}
{Lightkurve Collaboration}, {Cardoso}, J.~V.~d.~M., {Hedges}, C., {et~al.}
  2018, {Lightkurve: Kepler and TESS time series analysis in Python},
  Astrophysics Source Code Library.
\newblock \doeprint{1812.013}

\bibitem[{{Liou} \& {Zook}(1999)}]{1999AJ....118..580L}
{Liou}, J.-C., \& {Zook}, H.~A. 1999, \aj, 118, 580, \dodoi{10.1086/300938}

\bibitem[{{Malamud} \& {Perets}(2020)}]{2020MNRAS.493..698M}
{Malamud}, U., \& {Perets}, H.~B. 2020, \mnras, 493, 698,
  \dodoi{10.1093/mnras/staa143}

\bibitem[{{Martell} {et~al.}(2021){Martell}, {Simpson}, {Balasubramaniam},
  {Buder}, {Sharma}, {Hon}, {Stello}, {Ting}, {Asplund}, {Bland-Hawthorn}, {De
  Silva}, {Freeman}, {Hayden}, {Kos}, {Lewis}, {Lind}, {Zucker}, {Zwitter},
  {Campbell}, {{\v{C}}otar}, {Horner}, {Montet}, \&
  {Wittenmyer}}]{2021MNRAS.505.5340M}
{Martell}, S.~L., {Simpson}, J.~D., {Balasubramaniam}, A.~G., {et~al.} 2021,
  \mnras, 505, 5340, \dodoi{10.1093/mnras/stab1356}

\bibitem[{{Martin} {et~al.}(2005){Martin}, {Fanson}, {Schiminovich},
  {Morrissey}, {Friedman}, {Barlow}, {Conrow}, {Grange}, {Jelinsky},
  {Milliard}, {Siegmund}, {Bianchi}, {Byun}, {Donas}, {Forster}, {Heckman},
  {Lee}, {Madore}, {Malina}, {Neff}, {Rich}, {Small}, {Surber}, {Szalay},
  {Welsh}, \& {Wyder}}]{2005ApJ...619L...1M}
{Martin}, D.~C., {Fanson}, J., {Schiminovich}, D., {et~al.} 2005, \apjl, 619,
  L1, \dodoi{10.1086/426387}

\bibitem[{{McCully} {et~al.}(2018){McCully}, {Volgenau}, {Harbeck}, {Lister},
  {Saunders}, {Turner}, {Siiverd}, \& {Bowman}}]{McCully:2018}
{McCully}, C., {Volgenau}, N.~H., {Harbeck}, D.-R., {et~al.} 2018, in Society
  of Photo-Optical Instrumentation Engineers (SPIE) Conference Series, Vol.
  10707, \procspie, 107070K, \dodoi{10.1117/12.2314340}

\bibitem[{McKinney(2010)}]{pandas}
McKinney, W. 2010, in Proceedings of the 9th Python in Science Conference, ed.
  S.~van~der Walt \& J.~Millman, 51 -- 56

\bibitem[{{Mitnyan} {et~al.}(2020){Mitnyan}, {Borkovits}, {Rappaport},
  {P{\'a}l}, \& {Maxted}}]{2020MNRAS.498.6034M}
{Mitnyan}, T., {Borkovits}, T., {Rappaport}, S.~A., {P{\'a}l}, A., \& {Maxted},
  P.~F.~L. 2020, \mnras, 498, 6034, \dodoi{10.1093/mnras/staa2762}

\bibitem[{{Ochsenbein} {et~al.}(2000){Ochsenbein}, {Bauer}, \&
  {Marcout}}]{vizier}
{Ochsenbein}, F., {Bauer}, P., \& {Marcout}, J. 2000, \aaps, 143, 23,
  \dodoi{10.1051/aas:2000169}

\bibitem[{{Olmschenk} {et~al.}(2021){Olmschenk}, {Ishitani Silva}, {Rau},
  {Barry}, {Kruse}, {Cacciapuoti}, {Kostov}, {Powell}, {Wyrwas}, {Schnittman},
  \& {Barclay}}]{Olmschenk2021}
{Olmschenk}, G., {Ishitani Silva}, S., {Rau}, G., {et~al.} 2021, \aj, 161, 273,
  \dodoi{10.3847/1538-3881/abf4c6}

\bibitem[{{Parker}(1960)}]{1960ApJ...132..821P}
{Parker}, E.~N. 1960, \apj, 132, 821, \dodoi{10.1086/146985}

\bibitem[{{Paxton} {et~al.}(2011){Paxton}, {Bildsten}, {Dotter}, {Herwig},
  {Lesaffre}, \& {Timmes}}]{paxton11}
{Paxton}, B., {Bildsten}, L., {Dotter}, A., {et~al.} 2011, \apjs, 192, 3,
  \dodoi{10.1088/0067-0049/192/1/3}

\bibitem[{{Paxton} {et~al.}(2015){Paxton}, {Marchant}, {Schwab}, {Bauer},
  {Bildsten}, {Cantiello}, {Dessart}, {Farmer}, {Hu}, {Langer}, {Townsend},
  {Townsley}, \& {Timmes}}]{paxton15}
{Paxton}, B., {Marchant}, P., {Schwab}, J., {et~al.} 2015, \apjs, 220, 15,
  \dodoi{10.1088/0067-0049/220/1/15}

\bibitem[{{Paxton} {et~al.}(2019){Paxton}, {Smolec}, {Schwab}, {Gautschy},
  {Bildsten}, {Cantiello}, {Dotter}, {Farmer}, {Goldberg}, {Jermyn}, {Kanbur},
  {Marchant}, {Thoul}, {Townsend}, {Wolf}, {Zhang}, \& {Timmes}}]{paxton19}
{Paxton}, B., {Smolec}, R., {Schwab}, J., {et~al.} 2019, \apjs, 243, 10,
  \dodoi{10.3847/1538-4365/ab2241}

\bibitem[{{Payne} {et~al.}(2021){Payne}, {Shappee}, {Hinkle}, {Vallely},
  {Kochanek}, {Holoien}, {Auchettl}, {Stanek}, {Thompson}, {Neustadt},
  {Tucker}, {Armstrong}, {Brimacombe}, {Cacella}, {Cornect}, {Denneau},
  {Fausnaugh}, {Flewelling}, {Grupe}, {Heinze}, {Lopez}, {Monard}, {Prieto},
  {Schneider}, {Sheppard}, {Tonry}, \& {Weiland}}]{Payne2021}
{Payne}, A.~V., {Shappee}, B.~J., {Hinkle}, J.~T., {et~al.} 2021, \apj, 910,
  125, \dodoi{10.3847/1538-4357/abe38d}

\bibitem[{Pedregosa {et~al.}(2011)Pedregosa, Varoquaux, Gramfort, Michel,
  Thirion, Grisel, Blondel, Prettenhofer, Weiss, Dubourg, Vanderplas, Passos,
  Cournapeau, Brucher, Perrot, \& Duchesnay}]{scikit-learn}
Pedregosa, F., Varoquaux, G., Gramfort, A., {et~al.} 2011, Journal of Machine
  Learning Research, 12, 2825

\bibitem[{P\'erez \& Granger(2007)}]{ipython}
P\'erez, F., \& Granger, B.~E. 2007, Computing in Science and Engineering, 9,
  21, \dodoi{10.1109/MCSE.2007.53}

\bibitem[{{Perez-Becker} \& {Chiang}(2013)}]{2013MNRAS.433.2294P}
{Perez-Becker}, D., \& {Chiang}, E. 2013, \mnras, 433, 2294,
  \dodoi{10.1093/mnras/stt895}

\bibitem[{{Plavchan} {et~al.}(2008){Plavchan}, {Jura}, {Kirkpatrick}, {Cutri},
  \& {Gallagher}}]{Plavchan2008}
{Plavchan}, P., {Jura}, M., {Kirkpatrick}, J.~D., {Cutri}, R.~M., \&
  {Gallagher}, S.~C. 2008, \apjs, 175, 191, \dodoi{10.1086/523644}

\bibitem[{{Powell} {et~al.}(2021){Powell}, {Kostov}, {Rappaport}, {Borkovits},
  {Zasche}, {Tokovinin}, {Kruse}, {Latham}, {Montet}, {Jensen}, {Jayaraman},
  {Collins}, {Ma{\v{s}}ek}, {Hellier}, {Evans}, {Tan}, {Schlieder}, {Torres},
  {Smale}, {Friedman}, {Barclay}, {Gagliano}, {Quintana}, {Jacobs}, {Gilbert},
  {Kristiansen}, {Col{\'o}n}, {LaCourse}, {Olmschenk}, {Omohundro},
  {Schnittman}, {Schwengeler}, {Barry}, {Terentev}, {Boyd}, {Schmitt}, {Quinn},
  {Vanderburg}, {Palle}, {Armstrong}, {Ricker}, {Vanderspek}, {Seager}, {Winn},
  {Jenkins}, {Caldwell}, {Wohler}, {Shiao}, {Burke}, {Daylan}, \&
  {Villase{\~n}or}}]{Powell2021}
{Powell}, B.~P., {Kostov}, V.~B., {Rappaport}, S.~A., {et~al.} 2021, \aj, 161,
  162, \dodoi{10.3847/1538-3881/abddb5}

\bibitem[{{Rappaport} {et~al.}(2014){Rappaport}, {Barclay}, {DeVore}, {Rowe},
  {Sanchis-Ojeda}, \& {Still}}]{2014ApJ...784...40R}
{Rappaport}, S., {Barclay}, T., {DeVore}, J., {et~al.} 2014, \apj, 784, 40,
  \dodoi{10.1088/0004-637X/784/1/40}

\bibitem[{{Rappaport} {et~al.}(2016){Rappaport}, {Gary}, {Kaye}, {Vanderburg},
  {Croll}, {Benni}, \& {Foote}}]{2016MNRAS.458.3904R}
{Rappaport}, S., {Gary}, B.~L., {Kaye}, T., {et~al.} 2016, \mnras, 458, 3904,
  \dodoi{10.1093/mnras/stw612}

\bibitem[{{Rappaport} {et~al.}(2012){Rappaport}, {Levine}, {Chiang}, {El
  Mellah}, {Jenkins}, {Kalomeni}, {Kite}, {Kotson}, {Nelson},
  {Rousseau-Nepton}, \& {Tran}}]{2012ApJ...752....1R}
{Rappaport}, S., {Levine}, A., {Chiang}, E., {et~al.} 2012, \apj, 752, 1,
  \dodoi{10.1088/0004-637X/752/1/1}

\bibitem[{{Rappaport} {et~al.}(2019{\natexlab{a}}){Rappaport}, {Zhou},
  {Vanderburg}, {Mann}, {Kristiansen}, {Ol{\'a}h}, {Jacobs}, {Newton},
  {Omohundro}, {LaCourse}, {Schwengeler}, {Terentev}, {Latham}, {Bieryla},
  {Soares-Furtado}, {Bouma}, {Ireland}, \& {Irwin}}]{2019MNRAS.485.2681R}
{Rappaport}, S., {Zhou}, G., {Vanderburg}, A., {et~al.} 2019{\natexlab{a}},
  \mnras, 485, 2681, \dodoi{10.1093/mnras/stz537}

\bibitem[{{Rappaport} {et~al.}(2019{\natexlab{b}}){Rappaport}, {Vanderburg},
  {Kristiansen}, {Omohundro}, {Schwengeler}, {Terentev}, {Dai}, {Masuda},
  {Jacobs}, {LaCourse}, {Latham}, {Bieryla}, {Hedges}, {Dittmann}, {Barentsen},
  {Cochran}, {Endl}, {Jenkins}, \& {Mann}}]{2019MNRAS.488.2455R}
{Rappaport}, S., {Vanderburg}, A., {Kristiansen}, M.~H., {et~al.}
  2019{\natexlab{b}}, \mnras, 488, 2455, \dodoi{10.1093/mnras/stz1772}

\bibitem[{{Ratzloff} {et~al.}(2019){Ratzloff}, {Law}, {Fors}, {Corbett},
  {Howard}, {del Ser}, \& {Haislip}}]{2019PASP..131g5001R}
{Ratzloff}, J.~K., {Law}, N.~M., {Fors}, O., {et~al.} 2019, \pasp, 131, 075001,
  \dodoi{10.1088/1538-3873/ab19d0}

\bibitem[{{Ricker} {et~al.}(2015){Ricker}, {Winn}, {Vanderspek}, {Latham},
  {Bakos}, {Bean}, {Berta-Thompson}, {Brown}, {Buchhave}, {Butler}, {Butler},
  {Chaplin}, {Charbonneau}, {Christensen-Dalsgaard}, {Clampin}, {Deming},
  {Doty}, {De Lee}, {Dressing}, {Dunham}, {Endl}, {Fressin}, {Ge}, {Henning},
  {Holman}, {Howard}, {Ida}, {Jenkins}, {Jernigan}, {Johnson}, {Kaltenegger},
  {Kawai}, {Kjeldsen}, {Laughlin}, {Levine}, {Lin}, {Lissauer}, {MacQueen},
  {Marcy}, {McCullough}, {Morton}, {Narita}, {Paegert}, {Palle}, {Pepe},
  {Pepper}, {Quirrenbach}, {Rinehart}, {Sasselov}, {Sato}, {Seager},
  {Sozzetti}, {Stassun}, {Sullivan}, {Szentgyorgyi}, {Torres}, {Udry}, \&
  {Villasenor}}]{Ricker14}
{Ricker}, G.~R., {Winn}, J.~N., {Vanderspek}, R., {et~al.} 2015, Journal of
  Astronomical Telescopes, Instruments, and Systems, 1, 014003,
  \dodoi{10.1117/1.JATIS.1.1.014003}

\bibitem[{{Ridden-Harper} {et~al.}(2019){Ridden-Harper}, {Snellen}, {Keller},
  \& {Molli{\`e}re}}]{2019A&A...628A..70R}
{Ridden-Harper}, A.~R., {Snellen}, I.~A.~G., {Keller}, C.~U., \&
  {Molli{\`e}re}, P. 2019, \aap, 628, A70, \dodoi{10.1051/0004-6361/201834433}

\bibitem[{{Rodriguez} {et~al.}(2021){Rodriguez}, {Quinn}, {Zhou}, {Vanderburg},
  {Nielsen}, {Wittenmyer}, {Brahm}, {Reed}, {Huang}, {Vach}, {Ciardi},
  {Oelkers}, {Stassun}, {Hellier}, {Gaudi}, {Eastman}, {Collins}, {Bieryla},
  {Christian}, {Latham}, {Carleo}, {Wright}, {Matthews}, {Gonzales}, {Ziegler},
  {Dressing}, {Howell}, {Tan}, {Wittrock}, {Plavchan}, {McLeod}, {Baker},
  {Wang}, {Radford}, {Schwarz}, {Esposito}, {Ricker}, {Vanderspek}, {Seager},
  {Winn}, {Jenkins}, {Addison}, {Anderson}, {Barclay}, {Beatty}, {Berlind},
  {Bouchy}, {Bowen}, {Bowler}, {Brasseur}, {Brice{\~n}o}, {Caldwell},
  {Calkins}, {Cartwright}, {Chaturvedi}, {Chaverot}, {Chimaladinne},
  {Christiansen}, {Collins}, {Crossfield}, {Eastridge}, {Espinoza}, {Esquerdo},
  {Feliz}, {Fenske}, {Fong}, {Gan}, {Giacalone}, {Gill}, {Gordon}, {Granados},
  {Grieves}, {Guenther}, {Guerrero}, {Henning}, {Henze}, {Hesse}, {Hobson},
  {Horner}, {James}, {Jensen}, {Jimenez}, {Jord{\'a}n}, {Kane}, {Kielkopf},
  {Kim}, {Kuhn}, {Latouf}, {Law}, {Levine}, {Lund}, {Mann}, {Mao}, {Matson},
  {Mengel}, {Mink}, {Newman}, {O'Dwyer}, {Okumura}, {Palle}, {Pepper},
  {Quintana}, {Sarkis}, {Savel}, {Schlieder}, {Schnaible}, {Shporer}, {Sefako},
  {Seidel}, {Siverd}, {Skinner}, {Stalport}, {Stevens}, {Stibbards}, {Tinney},
  {West}, {Yahalomi}, \& {Zhang}}]{Rodriguez2021}
{Rodriguez}, J.~E., {Quinn}, S.~N., {Zhou}, G., {et~al.} 2021, \aj, 161, 194,
  \dodoi{10.3847/1538-3881/abe38a}

\bibitem[{{Sahoo} {et~al.}(2020){Sahoo}, {Baran}, {Sanjayan}, \&
  {Ostrowski}}]{Sahoo2020}
{Sahoo}, S.~K., {Baran}, A.~S., {Sanjayan}, S., \& {Ostrowski}, J. 2020,
  \mnras, 499, 5508, \dodoi{10.1093/mnras/staa2991}

\bibitem[{{Sanchis-Ojeda} {et~al.}(2015){Sanchis-Ojeda}, {Rappaport},
  {Pall{\`e}}, {Delrez}, {DeVore}, {Gandolfi}, {Fukui}, {Ribas}, {Stassun},
  {Albrecht}, {Dai}, {Gaidos}, {Gillon}, {Hirano}, {Holman}, {Howard},
  {Isaacson}, {Jehin}, {Kuzuhara}, {Mann}, {Marcy}, {Miles-P{\'a}ez},
  {Monta{\~n}{\'e}s-Rodr{\'\i}guez}, {Murgas}, {Narita}, {Nowak}, {Onitsuka},
  {Paegert}, {Van Eylen}, {Winn}, \& {Yu}}]{2015ApJ...812..112S}
{Sanchis-Ojeda}, R., {Rappaport}, S., {Pall{\`e}}, E., {et~al.} 2015, \apj,
  812, 112, \dodoi{10.1088/0004-637X/812/2/112}

\bibitem[{{Scargle}(1982)}]{Scargle1982}
{Scargle}, J.~D. 1982, \apj, 263, 835, \dodoi{10.1086/160554}

\bibitem[{{Schlafly} {et~al.}(2018){Schlafly}, {Green}, {Lang}, {Daylan},
  {Finkbeiner}, {Lee}, {Meisner}, {Schlegel}, \& {Valdes}}]{schlafly18}
{Schlafly}, E.~F., {Green}, G.~M., {Lang}, D., {et~al.} 2018, \apjs, 234, 39,
  \dodoi{10.3847/1538-4365/aaa3e2}

\bibitem[{{Schmitt} \& {Vanderburg}(2021)}]{2021arXiv210310285S}
{Schmitt}, A., \& {Vanderburg}, A. 2021, arXiv e-prints, arXiv:2103.10285.
\newblock \doarXiv{2103.10285}

\bibitem[{{Shappee} {et~al.}(2014{\natexlab{a}}){Shappee}, {Prieto}, {Grupe},
  {Kochanek}, {Stanek}, {De Rosa}, {Mathur}, {Zu}, {Peterson}, {Pogge},
  {Komossa}, {Im}, {Jencson}, {Holoien}, {Basu}, {Beacom}, {Szczygie{\l}},
  {Brimacombe}, {Adams}, {Campillay}, {Choi}, {Contreras}, {Dietrich},
  {Dubberley}, {Elphick}, {Foale}, {Giustini}, {Gonzalez}, {Hawkins}, {Howell},
  {Hsiao}, {Koss}, {Leighly}, {Morrell}, {Mudd}, {Mullins}, {Nugent},
  {Parrent}, {Phillips}, {Pojmanski}, {Rosing}, {Ross}, {Sand}, {Terndrup},
  {Valenti}, {Walker}, \& {Yoon}}]{2014ApJ...788...48S}
{Shappee}, B.~J., {Prieto}, J.~L., {Grupe}, D., {et~al.} 2014{\natexlab{a}},
  \apj, 788, 48, \dodoi{10.1088/0004-637X/788/1/48}

\bibitem[{{Shappee} {et~al.}(2014{\natexlab{b}}){Shappee}, {Prieto}, {Grupe},
  {Kochanek}, {Stanek}, {De Rosa}, {Mathur}, {Zu}, {Peterson}, {Pogge},
  {Komossa}, {Im}, {Jencson}, {Holoien}, {Basu}, {Beacom}, {Szczygie{\l}},
  {Brimacombe}, {Adams}, {Campillay}, {Choi}, {Contreras}, {Dietrich},
  {Dubberley}, {Elphick}, {Foale}, {Giustini}, {Gonzalez}, {Hawkins}, {Howell},
  {Hsiao}, {Koss}, {Leighly}, {Morrell}, {Mudd}, {Mullins}, {Nugent},
  {Parrent}, {Phillips}, {Pojmanski}, {Rosing}, {Ross}, {Sand}, {Terndrup},
  {Valenti}, {Walker}, \& {Yoon}}]{Shappee2014}
---. 2014{\natexlab{b}}, \apj, 788, 48, \dodoi{10.1088/0004-637X/788/1/48}

\bibitem[{{Smith} {et~al.}(2021){Smith}, {Ridden-Harper}, {Fausnaugh},
  {Daylan}, {Omodei}, {Racusin}, {Weaver}, {Barclay}, {Veres}, {Kann}, \&
  {Arimoto}}]{Smith2021}
{Smith}, K.~L., {Ridden-Harper}, R., {Fausnaugh}, M., {et~al.} 2021, \apj, 911,
  43, \dodoi{10.3847/1538-4357/abe6a2}

\bibitem[{{Stellingwerf}(1978)}]{Stellingwerf1978}
{Stellingwerf}, R.~F. 1978, \apj, 224, 953, \dodoi{10.1086/156444}

\bibitem[{{Tajiri} {et~al.}(2020){Tajiri}, {Kawahara}, {Aizawa}, {Fujii},
  {Hattori}, {Kasagi}, {Kotani}, {Masuda}, {Momose}, {Muto}, {Ohsawa}, \&
  {Takita}}]{Tajiri2020}
{Tajiri}, T., {Kawahara}, H., {Aizawa}, M., {et~al.} 2020, \apjs, 251, 18,
  \dodoi{10.3847/1538-4365/abbc17}

\bibitem[{{Thompson} {et~al.}(2012){Thompson}, {Everett}, {Mullally},
  {Barclay}, {Howell}, {Still}, {Rowe}, {Christiansen}, {Kurtz}, {Hambleton},
  {Twicken}, {Ibrahim}, \& {Clarke}}]{heartbeatstars}
{Thompson}, S.~E., {Everett}, M., {Mullally}, F., {et~al.} 2012, \apj, 753, 86,
  \dodoi{10.1088/0004-637X/753/1/86}

\bibitem[{{Tiscareno} {et~al.}(2006){Tiscareno}, {Burns}, {Hedman}, {Porco},
  {Weiss}, {Dones}, {Richardson}, \& {Murray}}]{Tiscareno_etal_2006}
{Tiscareno}, M.~S., {Burns}, J.~A., {Hedman}, M.~M., {et~al.} 2006, \nat, 440,
  648, \dodoi{10.1038/nature04581}

\bibitem[{{Tokovinin}(2018)}]{speckle}
{Tokovinin}, A. 2018, \pasp, 130, 035002, \dodoi{10.1088/1538-3873/aaa7d9}

\bibitem[{{Tokovinin} {et~al.}(2013){Tokovinin}, {Fischer}, {Bonati},
  {Giguere}, {Moore}, {Schwab}, {Spronck}, \& {Szymkowiak}}]{chiron}
{Tokovinin}, A., {Fischer}, D.~A., {Bonati}, M., {et~al.} 2013, \pasp, 125,
  1336, \dodoi{10.1086/674012}

\bibitem[{{van Lieshout} \&
  {Rappaport}(2018{\natexlab{a}})}]{2018haex.bookE..15V}
{van Lieshout}, R., \& {Rappaport}, S.~A. 2018{\natexlab{a}}, {Disintegrating
  Rocky Exoplanets}, ed. H.~J. {Deeg} \& J.~A. {Belmonte}, 15,
  \dodoi{10.1007/978-3-319-55333-7\_15}

\bibitem[{{van Lieshout} \& {Rappaport}(2018{\natexlab{b}})}]{vanlieshout18}
---. 2018{\natexlab{b}}, {Disintegrating Rocky Exoplanets}, ed. H.~J. {Deeg} \&
  J.~A. {Belmonte}, 15, \dodoi{10.1007/978-3-319-55333-7\_15}

\bibitem[{{van Werkhoven} {et~al.}(2014){van Werkhoven}, {Brogi}, {Snellen}, \&
  {Keller}}]{2014A&A...561A...3V}
{van Werkhoven}, T.~I.~M., {Brogi}, M., {Snellen}, I.~A.~G., \& {Keller}, C.~U.
  2014, \aap, 561, A3, \dodoi{10.1051/0004-6361/201322398}

\bibitem[{{Vanderbosch} {et~al.}(2020){Vanderbosch}, {Hermes}, {Dennihy},
  {Dunlap}, {Izquierdo}, {Tremblay}, {Cho}, {G{\"a}nsicke}, {Toloza}, {Bell},
  {Montgomery}, \& {Winget}}]{2020ApJ...897..171V}
{Vanderbosch}, Z., {Hermes}, J.~J., {Dennihy}, E., {et~al.} 2020, \apj, 897,
  171, \dodoi{10.3847/1538-4357/ab9649}

\bibitem[{{Vanderbosch} {et~al.}(2021){Vanderbosch}, {Rappaport}, {Guidry},
  {Gary}, {Blouin}, {Kaye}, {Weinberger}, {Melis}, {Klein}, {Zuckerman},
  {Vanderburg}, {Hermes}, {Hegedus}, {Burleigh}, {Sefako}, {Worters}, \&
  {Heintz}}]{2021arXiv210602659V}
{Vanderbosch}, Z.~P., {Rappaport}, S., {Guidry}, J.~A., {et~al.} 2021, arXiv
  e-prints, arXiv:2106.02659.
\newblock \doarXiv{2106.02659}

\bibitem[{{Vanderburg} \& {Rappaport}(2018)}]{2018haex.bookE..37V}
{Vanderburg}, A., \& {Rappaport}, S.~A. 2018, {Transiting Disintegrating
  Planetary Debris Around WD 1145+017}, ed. H.~J. {Deeg} \& J.~A. {Belmonte},
  37, \dodoi{10.1007/978-3-319-55333-7\_37}

\bibitem[{{Vanderburg} {et~al.}(2015){Vanderburg}, {Johnson}, {Rappaport},
  {Bieryla}, {Irwin}, {Lewis}, {Kipping}, {Brown}, {Dufour}, {Ciardi}, {Angus},
  {Schaefer}, {Latham}, {Charbonneau}, {Beichman}, {Eastman}, {McCrady},
  {Wittenmyer}, \& {Wright}}]{2015Natur.526..546V}
{Vanderburg}, A., {Johnson}, J.~A., {Rappaport}, S., {et~al.} 2015, \nat, 526,
  546, \dodoi{10.1038/nature15527}

\bibitem[{{Veras} {et~al.}(2020){Veras}, {McDonald}, \&
  {Makarov}}]{2020MNRAS.492.5291V}
{Veras}, D., {McDonald}, C.~H., \& {Makarov}, V.~V. 2020, \mnras, 492, 5291,
  \dodoi{10.1093/mnras/staa243}

\bibitem[{{Virtanen} {et~al.}(2020){Virtanen}, {Gommers}, {Oliphant},
  {Haberland}, {Reddy}, {Cournapeau}, {Burovski}, {Peterson}, {Weckesser},
  {Bright}, {van der Walt}, {Brett}, {Wilson}, {Jarrod Millman}, {Mayorov},
  {Nelson}, {Jones}, {Kern}, {Larson}, {Carey}, {Polat}, {Feng}, {Moore},
  {VanderPlas}, {Laxalde}, {Perktold}, {Cimrman}, {Henriksen}, {Quintero},
  {Harris}, {Archibald}, {Ribeiro}, {Pedregosa}, {van Mulbregt}, \&
  {Contributors}}]{scipy}
{Virtanen}, P., {Gommers}, R., {Oliphant}, T.~E., {et~al.} 2020, Nature
  Methods, \dodoi{https://doi.org/10.1038/s41592-019-0686-2}

\bibitem[{{Welsh} {et~al.}(2011){Welsh}, {Orosz}, {Aerts}, {Brown},
  {Brugamyer}, {Cochran}, {Gilliland}, {Guzik}, {Kurtz}, {Latham}, {Marcy},
  {Quinn}, {Zima}, {Allen}, {Batalha}, {Bryson}, {Buchhave}, {Caldwell},
  {Gautier}, {Howell}, {Kinemuchi}, {Ibrahim}, {Isaacson}, {Jenkins}, {Prsa},
  {Still}, {Street}, {Wohler}, {Koch}, \& {Borucki}}]{koi54}
{Welsh}, W.~F., {Orosz}, J.~A., {Aerts}, C., {et~al.} 2011, \apjs, 197, 4,
  \dodoi{10.1088/0067-0049/197/1/4}

\bibitem[{{Wenger} {et~al.}(2000){Wenger}, {Ochsenbein}, {Egret}, {Dubois},
  {Bonnarel}, {Borde}, {Genova}, {Jasniewicz}, {Lalo{\"e}}, {Lesteven}, \&
  {Monier}}]{2000A&AS..143....9W}
{Wenger}, M., {Ochsenbein}, F., {Egret}, D., {et~al.} 2000, \aaps, 143, 9,
  \dodoi{10.1051/aas:2000332}

\bibitem[{{Xu} {et~al.}(2016){Xu}, {Jura}, {Dufour}, \&
  {Zuckerman}}]{2016ApJ...816L..22X}
{Xu}, S., {Jura}, M., {Dufour}, P., \& {Zuckerman}, B. 2016, \apjl, 816, L22,
  \dodoi{10.3847/2041-8205/816/2/L22}

\bibitem[{{Zari} {et~al.}(2018){Zari}, {Hashemi}, {Brown}, {Jardine}, \& {de
  Zeeuw}}]{2018A&A...620A.172Z}
{Zari}, E., {Hashemi}, H., {Brown}, A.~G.~A., {Jardine}, K., \& {de Zeeuw},
  P.~T. 2018, \aap, 620, A172, \dodoi{10.1051/0004-6361/201834150}

\bibitem[{{Zhou} {et~al.}(2018){Zhou}, {Rappaport}, {Nelson}, {Huang},
  {Senhadji}, {Rodriguez}, {Vanderburg}, {Quinn}, {Johnson}, {Latham},
  {Torres}, {Gary}, {Tan}, {Johnson}, {Burt}, {Kristiansen}, {Jacobs},
  {LaCourse}, {Schwengeler}, {Terentev}, {Bieryla}, {Esquerdo}, {Berlind},
  {Calkins}, {Bento}, {Cochran}, {Karjalainen}, {Hatzes}, {Karjalainen},
  {Holden}, \& {Butler}}]{2018ApJ...854..109Z}
{Zhou}, G., {Rappaport}, S., {Nelson}, L., {et~al.} 2018, \apj, 854, 109,
  \dodoi{10.3847/1538-4357/aaa9b9}

\end{thebibliography}
\bibliographystyle{aasjournal}

\end{document}